\documentclass[12pt,epsfig]{article}
\usepackage{graphicx}

\catcode`\@=11   
\@addtoreset{equation}{section}

\def	\eqnum		#1{(\ref{#1})}       
\def	\scite		#1{$^{\cite{#1}}$}     
	\newdimen\eqskip
	\newdimen\txtskip
	\baselineskip=\txtskip
\begin{document}

\def	\be		{\begin{equation}}
\def	\ee		{\end{equation}}
\def	\ba		{\begin{eqnarray}}
\def	\ea		{\end{eqnarray}} 
\def	\nn		{\nonumber}
\def	\=		{\;=\;} 
\def	\ret		{\\[\eqskip]}
\def	\to		{\rightarrow }

\def	\ie		{{\em i.e.\/} }
\def	\eg		{{\em e.g.\/} }
\def	\q		{\mbox{$q$}}
\def	\qbar		{\bar q}
\def	\pbar		{\mbox{$\bar p$}}
\def	\ibar		{\mbox{$\bar\imath$}}
\def	\eqibar		{\mbox{${\scriptstyle \bar \imath}$}}
\def 	\jbar		{\mbox{$\bar\jmath$}}             
\def	\eqjbar		{\mbox{${\scriptstyle \bar \jmath}$}}
\def	\mc		{Monte-Carlo}
\def	\mcs		{Monte-Carlo's}
\def	\sub		{sub-amplitude}
\def	\subs		{sub-amplitudes}
\def	\amp		{\mbox{$m(1,2,\dots,n)$}}
\def	\pt		{\mbox{$p_t$}}       
\def	\et		{\mbox{$E_t$}}
\def	\w		{$W$}        
\def	\l		{\mbox{$\lambda$}}
\def	\tr		{\mbox{$tr$}}
\def	\d		#1{\mbox{d}{#1}}
\def    \Q              {q}
\def    \AQ             {\overline{q}}
\def    \L              {L}         
\def    \AL             {\overline{L}}

\def	\as		#1{\mbox{$\alpha_s^{#1}$}}
\def 	\gs		#1{\mbox{$g_s^{#1}$}}
\def	\Nc		#1{\mbox{$N^{#1}$}}
\def	\frac		#1#2{{#1 \over #2}}
\def	\lstring	#1#2{\mbox{$(\l^{#1_1}\l^{#1_2}\dots
			\l^{#1_#2})$}}   	     
\def	\istring	#1#2{\mbox{$({#1}_1,\dots,{#1}_{#2})$}} 

\def	\bra		{\langle}         
\def 	\ket		{\rangle}
\def	\sp		#1#2{\mbox{$\bra #1 #2 \ket$}}    
\def	\cp		#1#2{\mbox{$  [  #1 #2 ]$}}       
\def	\rsp		#1{\mbox{$\,\vert #1 \ket$}}      
\def	\lsp		#1{\mbox{$\bra  #1 \!\vert\,$}}   
\def	\sppr		#1#2{\mbox{$\bra #1 \!\vert #2 \ket$}}  
\def	\ps		(#1){\mbox{$\psi(#1)$}}           
\def	\psc		#1(#2){\mbox{$\psi_{#1}(#2)$}}    
\def	\eps		#1#2#3{\mbox{$\epsilon_#1^#2(#3)$}} 
\def	\epsh		#1#2{\mbox{$\hat{\epsilon}^#1(#2)$}} 
\def	\gfive	    	{\mbox{$\gamma_5$}}          
\def    \gJ             #1#2#3{\mbox{$J_{#1}(#2,\dots,#3)$}}  
\def    \qJ       #1#2#3{\mbox{$\overline{U}(#1,#2,\dots,#3)$}}    
\def    \aJ       #1#2#3{\mbox{$V(#1,\dots,#2,#3)$}} 

\def	\secint		{1 }
\def	\secthel	{2 }
\def	\sectccf	{3 }
\def	\sectsus	{4 }
\def	\sectexp	{5 }
\def	\sectsof	{6 }
\def	\secrsq		{7 }
\def	\sectapp	{10 }
                            
\def	\cffone		{1}
\def	\cfftwo		{2}
\def	\cffthree	{3}
\def	\cfffour	{4}
\def	\cfffive	{5}
\def	\exaone		{6}
\def	\exatwo		{7}
\def	\sfigone	{8}
\def	\sfigtwo	{9}
\def	\sfigthree	{10}
\def	\sfigfour	{11}
\def  	\appone 	{12}


\begin{flushright}
			FERMILAB-Pub-90/113-T \\
			May 1990
\end{flushright}

\vskip 1in                 
\begin{center}         
       	{ \Large  \bf \sc         
	Multi-Parton Amplitudes in Gauge Theories }
\vskip 0.3in

{\it Michelangelo L. MANGANO}\\[0.1in]
{\small Istituto Nazionale di Fisica Nucleare \\
Scuola Normale Superiore and    \\            
Dipartimento di Fisica, Pisa, ITALY } \\
                                      
 and  \\[0.1in]
      
{\it Stephen J. PARKE }\\[0.1in]
{\small Fermi National Accelerator Laboratory}
\footnote{Fermilab is operated by the Universities Research 
Association Inc. under contract with the United States Department of 
Energy.}\\
{\small P.O. Box 500, Batavia, IL 60510, U.S.A.} \\[0.2in]
\end{center}

\vskip 2in
\begin{abstract}
In this report we review recent developments in perturbation theory 
methods for gauge theories. 
We  present techniques and results
that are useful in the calculation of cross sections 
for processes with many final state partons which have
applications in the study
of multi-jet phenomena in high-energy Colliders.
\end{abstract}                                   

\newpage
\tableofcontents
                
                             
\newpage                            
\section{Introduction}
In high
energy collisions among hadrons and/or leptons the 
production of final states with
a large number of energetic, widely separated partons gives rise to events with
many jets in the final state
\footnote{For experimental analyses of multijet production in $e^+e^-$
collisions see, for example, \cite{bethke,bartel,tasso}. For production of
multijets in $\pbar p$ collisions, see \cite{ua2_86,afs,cdf,ua1_88,ua2_90}.
For associated production of weak gauge bosons and jets in $\pbar p$
collisions, see \cite{ua1_w,ua2_w,cdf_w}. }.
In many cases these multi-jet events offer a
potentially important probe on new physics \cite{ehlq},  \eg in the case of the
sequential decays of new heavy particles, such as a Higgs decaying to four jets
through real $W/Z$ pairs, or such as a pair of heavy gluinos decaying into a
multi-jet system through a chain-decay of the various unstable supersymmetric
particles. The possibility of using these observables to identify new phenomena
relies on our capability to predict the production rates and features of the
standard multi-jet production mechanisms which often provide a significant
background to these discovery channels.

Monte-Carlo techniques exist to describe processes with many partons in
the final state through a branching process driven by the leading logarithmic
approximation to the multiple emission probabilities \cite{webber86}. 
This approach provides a
scheme in which the number of final state particles is not fixed, and can be as
large as allowed by the relative branching probabilities. Most of the emitted
particles will be either soft or collinear to the leading ones involved in a
primary scattering process (say $gg \to gg$), because these configurations are
enhanced by the dynamics.  Inclusive energy measurements, such as calorimetric
detection of jets, are insensitive in first order to the precise details of the
structure of a collinear shower and, in the case just mentioned of a
$gg\to gg$ scattering, will usually only detect two jets at large angle.
Events with multi-jets are generated in the   
Monte-Carlo approach whenever a branching with large relative transverse
momentum takes place. In this case a new branch of the
partonic shower will arise and will independently evolve as a secondary jet. 
However, while the branching probabilities properly describe the parton 
evolution within a jet in the leading log approximation, this approximation does
not properly describe the emission of                           
partons at large relative transverse momentum. For these processes, therefore,
a full calculation of the matrix elements for the hard process involving the
many leading partons is required.

The calculation of these processes is made particularly difficult by the large
number of Feynman diagrams which appear in the perturbative expansion. 
As an example in Table~\ref{diagmult}\ we have collected from
Ref.\cite{kk89} the number
of diagrams contributing to the process {\it gluon-gluon $\to$ n-gluons}. 
                                                                          
The structure of
the non-abelian vertices, furthermore, leads to an almost uncontrollable
inflation in the number of terms which are generated, and very soon standard
techniques of numerical evaluation or algebraic symbolic manipulation become
useless.
Significant simplifications in these calculations have been achieved in 
recent years thanks to the use of 
new simple representations for vector polarizations,
a better organization of the diagrammatic expansion which fully exploits the
properties of gauge invariance, the discovery of recursive relations which
connect amplitudes with $n+1$ partons to amplitudes with $n$ ones and the use
of supersymmetric Ward identities to relate gluonic and quark-gluon amplitudes.
In spite of these advances, the results of these calculations are still often
very complicated and sometimes of limited use, even numerically, 
for systematic analysis of their phenomenological implications. In addition to
the development of these tools for the calculation of exact matrix elements,
effort has therefore also been put into finding proper approximations which
reliably simulate the exact solutions in the relevant regions of the
multi-particle phase-space and which are sufficiently simple to be handled
analytically and fast to evaluate numerically.

{\renewcommand{\arraystretch}{1.8}
    \begin{table}[tbh]
	\begin{center}
       	\begin{tabular}{|c||r|r|r|r|r|r|r|} \hline
       $ n $  &     2   &  3  &  4  &  5  &    6   &  7     &    8  
\\  \hline                                                       
$ \# $ of diagrams    
	&   4	& 25 &  220 & 2485 & 34300 & 559405 & 10525900 
\\ \hline                       
	\end{tabular} 
	\end{center}
	\caption{The number of Feynman diagrams contributing to the scattering
process $gg\to$ n $g$~.}
	\label{diagmult}   
\end{table}  }      

In this Report we collect and review these recent developments for the 
calculation of multi-parton matrix elements in non-abelian gauge 
theories. For examples of how these matrix elements can be used to 
obtain cross sections for processes in high energy colliders see EHLQ
\cite{ehlq} and references contained within.

In Section 2 we describe the helicity-amplitude technique and introduce
explicit parametrizations of the polarization vectors in terms of massless
spinors. To reach a wide an audience as possible we have chosen not 
to use the Weyl - van der Waarden formalism preferred by some researchers, 
see for example Ref.\cite{bg87}.

In Section~3 we introduce an alternative to the standard Feynman
diagram expansion, based on the equivalence between the massless sector of a
string theory and a Yang-Mills theory. This expansion groups together subsets
of Feynman diagrams for a given process in a gauge invariant way. These subsets 
are easier to evaluate than the complete set and different gauges can be used
for each subset so as to maximize the simplifications induced by a proper
choice of gauge.          
Furthermore, different subsets of diagrams are
related to one another through symmetry properties or algebraic relations
and can be obtained without further effort from the knowledge of a small number
of {\em building blocks}. This expansion can be extended to arbitrary
processes involving particles in representations other than the adjoint, and in
this Section we construct this generalization.

Section 4 describes the use of Supersymmetry Ward identities to relate
amplitudes with particles of different statistics. These relations are useful
even when dealing with non-supersymmetric theories because in many cases the
additional supersymmetric degrees of freedom decouple from the processes of
interest. In addition, if the energy of the scattering process is large with
respect to the mass splittings within supersymmetry multiplets, these relations
can be used to easily calculate the matrix elements for the production of
supersymmetric particles. 

In Section 5 we illustrate the use of these tools with the explicit calculation
of matrix elements for processes with four and five partons, and give results
for the scattering of six gluons and four gluons plus a quark-antiquark pair. 
We hope this Section is useful for the reader who wants to familiarize himself
with the details of how these calculations are performed.

In Section 6 we prove various factorization properties using a string-theoretic
approach, which provides a compact way to represent multi-parton amplitudes.
The results contained in this Section are useful for a better understanding of
the structure of multi-parton amplitudes in gauge theories. 

Section 7 introduces the Berends-Giele recursion relations, which allow to
calculate matrix elements in a recursive fashion, providing an algebraic
algorithm which can be efficiently used for numerical evaluation of higher
order processes. 

In Section 8 we collect some explicit results concerning matrix elements for
processes with an arbitrary number of particles. These expressions hold for
amplitudes with a simple helicity structure, and whose properties are fully
determined by their behaviour at the collinear and infrared poles. These
results help understanding and extending known coherence properties of the
soft radiation in non-abelian gauge theories, as will be discussed. 

In Section 9 we show how to use these techniques in the case of gauge groups
which are the product of different groups, and how to calculate in presence of
massive gauge bosons from a spontaneously broken gauge theory. As an
application, we collect the known matrix elements for the processes involving a
massive gauge boson produced in association with two gluons from the scattering
of a quark-antiquark pair. 

Section 10 describes the approximation techniques mentioned above. We review
different approaches that have been proposed and illustrate their use for
processes involving $n$-gluons or $n$-gluons plus a quark-antiquark pair. 

Finally, we collect in five Appendices various definitions, conventions and
results which are useful in performing explicitly analytic calculations or
numerical evaluations of matrix elements.        


\newpage                            
\section{Helicity Amplitudes}
The use of helicity amplitudes for the calculation of multi-parton scattering
in the high-energy (massless) limit was pioneered in papers by J.D.
Bjorken and M. Chen \cite{bj}, and by O. Reading-Henry \cite{henry},
and later further developed and fully exploited                     
by the Calkul Collaboration in a classical set of papers
\cite{calkul82a,calkul82b}. 
The application of this technique to QED processes is extensively reviewed in
the book by Gastmans and Wu \cite{wubook}.
                                          
According to this approach, one calculates matrix elements with external states
having a given assigned helicity. Since different helicity configurations do
not interfere, to obtain the full cross section it is sufficient to sum
incoherently the squares of all of the possible helicity amplitudes which can
contribute to the process. The advantage over more standard techniques is that
by choosing a definite helicity configuration one can exploit gauge
invariance and select an explicit representation for the polarization vectors
which will simplify the calculation. 

Since the polarization vectors always enter in an amplitude
contracted with a gamma matrix in QED processes, 
the Calkul group found it useful to introduce a
representation in terms of the two momenta ($q,q'$) of one of the pairs of 
external  charged  fermions in the process :
\be	\label{calkul}          
	\epsh{\pm}{p} \;\equiv\; \eps{\mu}{\pm}{p}\cdot \gamma^\mu
	  \=                           
	N [\hat{p}\hat{q}'\hat{q} (1\pm\gfive) -
	   \hat{q}'\hat{q}\hat{p} (1\mp\gfive) \mp
	   2(q\cdot q')\hat{p}\gfive ],                                     
\ee                            
where $N$
is a normalization factor,                                
\be                                                       
	N \= [16(q\cdot q')(p\cdot q)(p\cdot q')]^{-\frac{1}{2}}.
\ee                                          
With this choice for $\epsilon^\pm$, many of the terms which appear in the
diagrammatic expansion simply vanish. Because of gauge invariance, all of the
terms generated by the third piece in Eq.\eqnum{calkul}, proportional to
$\hat{p}$, will sum up to zero.
Furthermore, helicity conservation along a fermionic line guarantees that 
at least one of the two remaining terms, containing orthogonal chiral
projections, will vanish.
Finally, if the photon happens to be attached to an external fermion whose
momentum is one of the reference momenta used to define the photon
polarization, then this diagram will also vanish provided the helicities match
(see the $e^+e^-$ annihilation case in the Appendix for an explicit example).
                 
If the set of diagrams contributing to
the given matrix element can be split into the sum of gauge
invariant subsets, we can choose different reference momenta $q,q'$ for
different subsets, provided we keep track of the relative phase which can
appear in the polarizations when they are referred to different $q$'s.
As an example of a process in which this splitting is possible, we indicate
$e^+e^- \to \mu^+\mu^- \gamma$, and refer to the previously quoted papers for
the explicit calculation.

While this technique turns out to be extremely useful for pure QED
calculations, the complexity of a non-abelian theory calls for something even
simpler. In the non-abelian theory, in fact, the proliferation of diagrams is
such that the bookkeeping of the different phases becomes very complex, and the
existence of processes without external fermions calls for a different choice
of reference momenta to achieve the desired simplification. An improved version
of the Calkul representation, which is more apt to use in non-abelian theories,
was introduced by Xu, Zhang and Chang 
in Ref.\cite{xu} and, independently, in Ref.\cite{gk85b,kls85b}.
In this improved version, a vector polarization is expressed in terms of
massless spinors and just one reference momentum. Here in the following we will
give a simple derivation of this result making use of Supersymmetry
\cite{wess}. We will always work in four space-time dimensions, but the
construction could be extended in principle to higher dimensions, and possibly
to non-integer dimensions as well.

To start with, we will set our notation and will present some definitions
concerning the spinor algebra that will be extensively used in the following.
For additional details and properties, see the Appendix.

Let \ps(p)\ be a massless four-dimensional Dirac spinor, \ie:
\be
	\hat p \; \ps(p) \equiv p \cdot \gamma \ps(p) = 0 \quad\quad p^2=0.
\ee                       
We define the two helicity states of \ps(p) by the two chiral projections:
\be
	\psc{\pm}(p) \= \frac{1}{2} (1 \pm \gfive) \ps(p)
	\= \psc{\mp}(p)^c,
\ee                       
the last identity being just a conventional choice of relative phase between
opposite helicity spinors fixed by the properties under charge conjugation
($^c$):
\be                  
	\ps(p)^c \= C \ps(p)^* , \quad\quad C\gamma_\mu^* C^{-1} \= \gamma_\mu.
\ee                                                                            

Following \cite{xu}, we introduce the following notation:
\ba                                             
   	\rsp{p\pm} \= \psc{\pm}(p)  
	&\quad &
	\lsp{p\pm} \= \overline {\psc{\pm}(p)}         
	\ret                                  
   	\sp pq = \sppr{p -}{q +} = \overline {\psc{-}(p)}\psc{+}(q)
   	&\quad & 
	\cp pq = \sppr{p +}{q -} =\overline {\psc{+}(p)}\psc{-}(q) .
\ea                                                                 

The spinors are normalized as follows:
\be
	\lsp{p\pm} \gamma_\mu \rsp{p\pm} = 2 p_\mu .
\ee                                                 
From the properties of the Dirac algebra, it is straightforward to prove the
following useful identities:
\ba     \label{ids1}
      &&	
	\sppr{p +}{q +} =  \sppr{p -}{q -} = \sp pp = \cp pp = 0
	\ret
	\label{ids2}
      &&	
	\sp pq = -\sp qp ,   \quad\quad  \cp pq = -\cp qp
	\ret
	\label{ids3}
      &&
       	\vert \sp pq \vert ^2 = 2(p \cdot q)
      	\ret
	\label{ids4}
      &&           
	\lsp{A +} \gamma_\mu \rsp{B +} \lsp{C- } \gamma^\mu \rsp{D -} =
       	2 \cp AD \sp CB.
\ea                                                                 

We now turn to the description of four-dimensional massless vectors.
In four dimensions the physical Hilbert space of a massless vector is
isomorphic to the physical Hilbert space of a massless spinor (up to a $Z_2$
transformation), since they both 
lie in one-dimensional representations of $SO(2)$, the little group of
$SO(3,1)$. 
This isomorphism is realized through a linear transformation which relates
like-helicity vectors and fermions:
\ba	\label{susy}
       	\eps{\mu}{+}{p} &=& A \overline{u_{+}(p)} \gamma_\mu v,  \ret
	\eps{\mu}{-}{p} &=& (\eps{\mu}{+}{p})^*            
\ea                                                          
where \eps{\mu}{\pm}{p} is the polarization vector of an {\it outgoing} (\ie
positive-energy-) massless vector of momentum $p$, $u_{+}(p)$ is a massless
spinor as defined above,  $v_{\alpha}$ is an a priori arbitrary Dirac spinor
and $A$ is a normalization constant, needed to enforce the usual normalization
conditions:
\be                                 
	\eps{{}}{+}{p} \cdot \eps{{}}{+}{p} = 0, \quad \quad
	\eps{{}}{+}{p} \cdot \eps{{}}{-}{p}=-1.
\ee
                                                               
In this isomorphism, the gauge invariance associated with the massless vector
can be parametrized
by the arbitrariness in the choice of the spinor $ v$. Although this
parametrization does not exhaust all the possible gauge choices, nevertheless
it will turn out to be particularly useful in the following. It is easy to
check
that by properly choosing the gauge we can always select a spinor
$ v(k)$ to be used in \eqnum{susy}\ that satisfies the following properties:
\ba	\label{arbspinor1}
   &&	\hat k v(k) \equiv k \cdot  \gamma v(k)=0,    \ret
	\label{arbspinor2}
   &&	k^2=0,	\quad  k \cdot p \ne 0.
\ea                                  
We will refer to the arbitrary $k$ as to the {\em reference} momentum.
Therefore we can always write, for a proper gauge choice:                
\ba	\label{polar}
	\eps{\mu}{+}{p,k} &=& A   \; \lsp{p+} \gamma_\mu \rsp{k+}   \ret
	\eps{\mu}{-}{p,k} &=& A^* \; \lsp{p-} \gamma_\mu \rsp{k-}     
\ea                                                             

The normalization $A$ has to be chosen to give unit norm to the polarization.
Using Eq.\eqnum{ids4} we easily obtain:
\be
	\eps{{}}{+}{p,k}\cdot \eps{{}}{-}{p,k} =
	- 2 \vert A \vert ^2  p\cdot k 
\ee
and thus:
\ba	
       	\eps{\mu}{+}{p,k} &=& 
	e^{i\phi(p,k)}\frac{\lsp{p+} \gamma_\mu \rsp{k+}}{\sqrt{2} \sp kp}
   	\ret                                                           
       	\eps{\mu}{-}{p,k} &=& 
	e^{-i\phi(p,k)}\frac{\lsp{p-} \gamma_\mu \rsp{k-}}{\sqrt{2} \cp pk},
\ea                                                                         
where $\phi(p,k)$ is a phase which a priori depends on the vector momentum
$p$, and on the reference momentum $k$.
If we set this phase to zero, it is easy to show that that the change 
in the polarization vector caused by a change in the reference 
momentum is given by:
\be
	\eps{\mu}{+}{p,k} \rightarrow \eps{\mu}{+}{p,k'} - 
	\sqrt 2 \frac{\sp k{k'}}{\sp kp \sp {k'}p} p_\mu.
\ee                                              
Note that for this choice of $\phi(p,k)$ the a priori phase factor in 
front  of $\eps{\mu}{+}{p,k'}$ is equal to unity in this equation.
A similar result holds for the negative helicity vectors. Therefore the choice 
of polarization vectors used through out this review is 
\be	\label{norm}
       	\eps{\mu}{\pm}{p,k} \= \pm
	\frac{\lsp{p\pm} \gamma_\mu 
	\rsp{k\pm}}{\sqrt{2}\sppr{k\mp}{p\pm}}.
\ee                                                                         
Using this 
representation, \eqnum{norm}, for the polarization vectors 
in the calculation of a given amplitude, we can
choose not only a different reference momentum $k$ for each polarization vector
in the process, but we can also choose different reference momenta for each
gauge invariant part of the full amplitude, without having to worry about
relative phases. This property will be used extensively in the following
applications, where we will decompose each amplitude into a sum over gauge
invariant components. 

A proper assignment of reference momenta to the
different external vectors will result in significant simplifications.
As an example, by using Eqs.\eqnum{ids4},\eqnum{ids1}\ one can easily prove the
following identities:                                            
\be   \label{idpol1}                                             
	\eps{{}}{+}{p,k} \cdot \eps{{}}{+}{p',k} =
	\eps{{}}{+}{p,k} \cdot \eps{{}}{-}{k,k'} =
	0                                   
\ee                                            
These identities suggest that it is convenient to choose the reference momenta
of like-helicity vectors to be the same and to coincide with the external
momenta of some of the vectors with the opposite helicity.

The representation \eqnum{norm}\ for the polarizations is also particularly
helpful when calculating processes with external fermions in addition to the
vectors. The polarization vectors contract with the gamma
matrices in the following way:                      
\be	
	\label{epshat}
	\eps{{}}{\pm}{p,k} \cdot \gamma =  \pm
	\frac{\sqrt 2}{\sppr{k\mp}{p\pm}} 
	(\rsp{p\mp}\lsp{k\mp} + \rsp{k\pm}\lsp{p\pm}).
\ee       
An explicit example of the use of these formulas for the simple case of
$e^+e^-$ annihilation into two photons is given in the Appendix.

As a final comment, we add that the gauges generated by this choice of
polarization vectors are equivalent to axial gauges. In fact it is
straightforward to prove on the basis of the identities given here and in the
Appendix, that:
\be       
	\sum_{pol} \epsilon^{\lambda}_{\mu} \, (\epsilon^{\lambda}_{\nu})^*
	\=                                                                 
	\eps{\mu}{+}{p,k} \, \eps{\nu}{-}{p,k} \;+\;
	\eps{\mu}{-}{p,k} \, \eps{\nu}{+}{p,k}
	\= -g_{\mu\nu} \;+\; \frac{p_\mu k_\nu + p_\nu k_\mu}{p \cdot k}.
\ee
Because of this reason, we will expect these gauges to make calculations
particularly simple when studying matrix elements in the eikonal approximation.

The representation of polarization vectors in terms of spinors has 
been generalized to the case of massive particles of spin $1/2$, 1 and 
$3/2$ in Ref.\cite{passarino}.


\newpage
\def\subamp{\mbox{$m(1,2,\cdots,n)$}}
\section{The Color Form Factors}          

\subsection{Gluonic Amplitudes: Duality and Gauge Invariance}

In perturbative QCD the calculation of multi-gluon scattering  amplitudes, even
at tree level, is very challenging. The number of diagrams describing a given
process grows very quickly, and the redundancy due to the gauge invariance
leads to a rapid proliferation of terms. One way to simplify these calculations
is to divide all of the diagrams contributing to a given matrix element
into subsets of diagrams which are independently gauge invariant under
redefinition of the polarizations: $\epsilon^\mu_i(p_i) \to \epsilon_i^\mu(p_i)
+ \alpha_i p_i^\mu$, with the $\alpha_i$'s being arbitrary functions.
It might then be possible to choose different gauges for these different
subsets in such a way as to simplify the calculation as much as possible. 
By using the polarization vectors introduced in the previous Section, different
gauge choices will not change the relative phases between the different gauge
invariant pieces,  thus contributing to a further simplification.

The issue then is to find a systematic way of dividing processes into gauge
invariant components.  In this Section we will provide such a criterion based
on the work initiated in \cite{mpx87,bg87,mpx88}.  This criterion can be
applied to any gauge theory: here for simplicity we will refer to simple
unitary groups $SU(N)$, but the techniques introduced can be easily extended to
more general cases, such as products of groups, as will be shown in a later
Section. 
         
A very complete study of the relation between gauge invariance and color
structures in the context of the large-$N$ limit \cite{largen} of QCD and the
loop expansion was presented by Cvitanovi\'c and collaborators in 
Ref.\cite{cvit}.  Some of the results presented here do overlap with theirs. 

For the sake of reference, we
will often refer to the Yang-Mills gauge bosons as to {\em gluons}.
As we will prove in what follows, it turns out to be
useful to consider the space of color configurations for the given scattering
process. If we expand the amplitude with respect to  an
orthogonal basis in this space, this expansion is guaranteed to be gauge
invariant.  Therefore there are many different ways of breaking up the
amplitude into gauge invariant components. A particular choice which can be
singled out for its prompt physical interpretation and for its many important
properties is to insure that these gauge invariant components be invariant
under cyclic permutations of the external gluons. 
Consider an $SU(N)$ Yang-Mills theory; then at {\em tree level} in 
perturbation theory any vector particle scattering amplitude,
with colors $a_1,a_2 \dots a_n$, external momenta $p_1,p_2 \dots p_n$
and helicities $\epsilon_1,\epsilon_2 \dots \epsilon_n$,
can be written as 
\begin{eqnarray}
	\label{exp}
	{\cal M}_n\,&=&\,\sum_{\{1,2,\dots,n\}'}\;tr\,(\l^{a_1}\l^{a_2}
	\dots\l^{a_n})                                        
	\;m(p_1,\epsilon_1;p_2,\epsilon_2;\cdots;p_n,\epsilon_n),
\end{eqnarray}
where the sum with the {\it prime}, $\sum_{\{1,2,\dots,n\}'}$, is over all
$(n-1)!$ {\em non-cyclic}  permutations of $1,2,\dots,n$ and the $\l$'s are the
matrices of the symmetry group in the  fundamental representation, which we
choose to normalize as follows \footnote{ This normalization of the \l\
matrices differs from the usual one by a $\sqrt 2$, which we explicitly add to
the Feynman rules (see Appendix~C): this choice is purely conventional, and
just simplifies the bookkeeping of factors of 2 in the calculations.}:
\be    \label{lamnorm}
	[\l^a,\l^b] \;=\; i  f^{abc}\l^c
	\quad\quad,\quad\quad
	tr(\l^a\l^b)\;=\; \delta^{ab}.
\ee
The color structures given by the traces of $\l$ matrices do not provide a
complete basis for the possible color configurations of $n$ gluons, but
nevertheless they are sufficient to describe the tree-level scattering of
$n$-gluons, as we will show below.  It should also be pointed out that the
color structures used in equation \eqnum{exp}\ are only orthogonal at the
leading order in the expansion in powers of $N$; if $\{a\}$ and $\{b\}$ are
two permutations of the gluon color indices we have in fact (see the Appendix):
\be
	\sum_{a_i=1,N^2-1}
	\tr   \lstring{a}{n} \left [ \tr \lstring{b}{n} \right ] ^* \=
	N^{n-2} \; (N^2-1) \; (~\delta_{\{a\}\{b\}} ~+~ {\cal O}(N^{-2}) ~),
\ee                                                
where the $\delta_{\{a\}\{b\}}$ is equal to 1 if and only if the two
permutations are the same (up to cyclic re-orderings); this partial
orthogonality, nevertheless, is clearly still sufficient
to guarantee the gauge invariance of the expansion, which must hold order by
order in 1/$N$. For a different  choice of base in the color space, which is
exactly orthogonal, see the alternative approach developed by Zeppenfeld in
Ref.\cite{zepp88}.                                              
              
The proof of Equation \eqnum{exp}\ is very simple if one uses the relations
\eqnum{lamnorm}:
in any tree level Feynman 
diagram, replace the color structure function at some vertex using
$f_{abc}\,=\,-i~tr( \l^a\l^b\l^c
~-~\l^c\l^b\l^a)$. 
Now each leg attached to this vertex has a $\l$ matrix associated 
with it. At the other end of each of these legs there is either 
another vertex or this is an external leg. If there is another vertex,
use the $\l$ associated with this internal leg to write the 
color structure of this vertex $f_{cde}~\l^c$ as 
$-i~[\l^d,\l^e]$. Continue this processes until all vertices 
have been treated in this manner. Then this Feynman diagram has been 
placed in the form of Equation \eqnum{exp}. Repeating this procedure for all
Feynman  diagrams for a given process completes the proof.

The sub-amplitudes 
$m(1,2,\dots,n) ~\equiv~
m(p_1,\epsilon_1;p_2,\epsilon_2;\dots p_n,\epsilon_n)~~$ 
of Equation \eqnum{exp}\ are by construction independent of the color indices
and satisfy a number of important properties and relationships:

\begin{enumerate}
\item $m(1,2,\dots,n)$ is gauge invariant.\\
\item $m(1,2,\dots,n)$ is invariant under cyclic permutations of 
	$1,2,\dots,n$ \\
\item $m(n,n-1,\dots,1)=(-1)^n\, m(1,2,\dots,n)$\\
\item The Dual Ward Identity:
\begin{eqnarray}                                               
	m(1,2,3,\dots,n)~+~m(2,1,3,\dots,n) ~+~ m(2,3,1,\dots,n) &&
	\nonumber \\ 
	\label{ward}
	+~\cdots~+~m(2,3,\dots,1,n) &=& 0.
\end{eqnarray}                            
\item Factorization of $m(1,2,\cdots,n)$ on multi-gluon poles.\\
\item Incoherence to leading order in the number of colors:
\be
	\label{ncfact}
	\sum_{colors}\;\vert {\cal M}_n \vert^2 \=
	N^{n-2}(N^2-1)                            
	\sum_{\{1,2,\dots,n\}'} \left\lbrace \vert \subamp \vert^2 + {\cal O}
	(N^{-2})\right\rbrace.
\ee
\end{enumerate}                                               
                                                              
This set of properties for the sub-amplitudes we will refer to as duality and
the expansion in terms of these dual sub-amplitudes the dual  expansion.
Properties (1) and (2) can be seen directly from the properties  of linear
independence (to the leading order in $N$, and for arbitrary $N$) and
invariance under cyclic permutations of $tr\,(\l^1\l^2\dots\l^n)$.  Whereas (3)
and (4) follow by studying the sum of  Feynman diagrams which contribute to
each sub-amplitude.  The sum of Feynman diagrams which enter into the 
Dual Ward
Identity   is such that each diagram is paired with another with
opposite sign so that the combination  contained in Equation \eqnum{ward}\
trivially vanishes. Property (5) will be discussed in detail in Section
\sectsof and the  incoherence to leading order in the number of colors (6) was
obtained above and follows from  the color algebra of the $SU(N)$ gauge group.
                                                                               
To the string theorist this expansion and the duality properties (1) to  (6),
see \cite{jacob}, are quite familar since the string amplitude, in the zero
slope limit, reproduces the Yang-Mills amplitude on mass shell  \cite{schwarz}.
Each sub-amplitude then corresponds to the zero slope limit of a string      
diagram, and the sub-amplitude can be obtained by using the usual Koba-Nielsen
formula \cite{koba}. 
Kawai, Lewellen and Tye, Ref.\cite{klt}, have 
derived a relationship between the closed string tree amplitudes and the 
open string tree amplitudes which allows this connection to be 
explicitly extended to the heterotic string as 
well as to the closed bosonic and the 
type II superstring.
The traces of $\l$ matrices are just the Chan-Paton
factors \cite{chan}.   For the string amplitude the properties (1) through (6)
are satisfied even before the zero slope limit is taken, and in particular
Equation \eqnum{ward}\ holds as a Ward identity for correlation functions of
products of two-dimensional conformal fields. We will see later on in 
this Section and also in Section \sectsof on factorization properties,
useful examples of how to use  
the string representation to derive various properties of the Yang-Mills
amplitudes.

Which diagrams contribute to a given sub-amplitude and with which coefficients
they enter can be determined by the  procedure developed earlier in this
Section for re-writing the color   factors. It is however helpful to think in
terms of string diagrams, and to realize that the contributing Feynman diagrams
can just be obtained by pinching in all possible ways on multi-particle
poles the string diagram itself (see for example Figure~\cffone).

The relationship with the string diagrams, the possibility of choosing an {\em
ad hoc} gauge and the simple factorization properties that the dual
sub-amplitudes must satisfy, suggest that a Yang-Mills amplitude expressed as
in Equation \eqnum{exp}\  will assume a particularly simple form. That this is
in fact the case will be shown in Section~\sectexp, where we will consider some
explicit examples.
         
The gauge invariance and properties under cyclic and reverse  permutations
$(12\dots n)\to(n\dots 21)$ allow the calculation of far fewer than the
$(n-1)!$ sub-amplitudes that appear in the dual expansion. In fact the number
of sub-amplitudes that are needed is just the number of different  orderings of
positive and negative helicities around a circle. Of course some of the
sub-amplitudes vanish because of the partial  helicity conservation of tree
level Yang-Mills and others are simply  related to one another through the
properties (2) through (4). Kleiss and Kuijf in Ref.\cite{kk89} have given a 
detailed, general accounting of the minimum number of independent 
gluonic subamplitudes that are needed for the n-gluon scattering.
Their results is that $(n-2)!$ subamplitudes are independent.
                            
\subsection{Quark-Gluon Amplitudes}
In this Section we will extend the color representation introduced above to
processes with fermions in the fundamental representation of the gauge group
$SU(N)$ \cite{cvit,mlm88,dk89a}. 
We will aim at a representation which satisfies the properties of gauge
invariance and factorization to the leading order in $1/N_c$,
Eq.\eqnum{ncfact}.
As before, we will refer to the gauge bosons as to {\em gluons}, and
to the fermions as to {\em quarks}. 
We will start from processes with one quark-antiquark pair, and for the time
being we will only consider tree level diagrams.

The color structure of diagrams where all of the gluons are emitted directly
from the quark line\footnote{We will denote these diagrams as QED-type
\cite{cvit}.}
(see Figure~\cfftwo) is obtained in a straightforward way directly from the 
Feynman rules:                                                        
\be    \label{qqbarcol}                                                    
	\lstring{a}{n}_{i\eqjbar},
\ee                           
where ($i,\jbar$) are the color indices of the $q\qbar$ pair and
$a_{1,\dots,n}$ are the color indices of the gluons in the order they are
emitted. 
In order to
analyze diagrams with gluons coupling to each other, let us consider the case
in which just one gluon is emitted from the quark and develops into a tree
(Figure~\cffthree).
We can factorize the color structure of the diagram into the color
coefficient of the $q\qbar g$ vertex, namely $(\l^A)_{i\eqjbar}$, and the
color structure of the remaining gluon tree. By using the dual representation,
we can express this as the sum over traces of permutations of \l\ matrices. As 
a result, we will obtain that the color coefficient of this diagram is given by
a sum over permutations of the following expression:
\be                                              
	\sum_A (\l^A)_{i\eqjbar} \; \tr[\l^A\lstring{a}{n}] \=
	\lstring{a}{n}_{i\eqjbar} \;-\; \frac{1}{N} \delta_{i\eqjbar} 
	\tr\lstring{a}{n}.
\ee                                                                         
This identity follows from the following property of the \l\
matrices:     
\be    \label{trace}
	\sum_{a=1}^{N^2-1} (\l^a)_{i_1\eqjbar_1}
	(\l^a)_{i_2\eqjbar_2} \=
	\delta_{i_1 \eqjbar_2} \delta_{i_2 \eqjbar_1} \; - \;
	{1 \over N} \delta_{i_1 \eqjbar_1} \delta_{i_2 \eqjbar_2}
\ee
with the normalization given in Equation \eqnum{lamnorm}.

  The term proportional to \Nc{-1}\ corresponds to the subtraction
of the trace of the $U(N)$ group in which $SU(N)$ is embedded. This trace
couples to the quarks but commutes with $SU(N)$ itself, and then it doesn't
couple to the gluons. As such it must disappear after the sum over
permutations. That this is in fact the case, can be easily checked.
Since an arbitrary diagram can be factorized\footnote{Again, the factorization
we are referring to here is just a factorization of the color structure, and
not of the full amplitude, since because of gauge invariance  factorization
cannot be applied on a diagram-by-diagram basis.} 
into diagrams of the QED type and                 
diagrams with a tree evolution initiated by a single gluon, we conclude that
any diagram with a $q\qbar$ pair can be decomposed
in terms of the $n!$ permutations of the color structure given in
Eq.\eqnum{qqbarcol}. Notice that all terms of the form 
$\delta_{ij} \tr\lstring{a}{n}$ do cancel (at tree level). 
                                         
By repeated use of the factorization properties of the color coefficients one
easily arrives at the general representation in terms of which it is possible
to decompose diagrams with more than one quark pair: 
\be    \label{qgcolor}       	
	\Lambda (\{n_i\}, \{\alpha\}) \,  \= 
	\frac{(-1)^p}{\Nc{p}}\; (\l^{a_1} \dots\l^{a_{n_1}})_{i_1\alpha_1}	
	(\l^{a_{n_1+1}} \dots \l^{a_{n_2}})_{i_2\alpha_2}	    
	\dots	(\l^{a_{n_{m-1}+1}}\dots \l^{a_{n}})_{i_m\alpha_m}	
\ee 
Here $m$ is the number of $q\qbar$
pairs present in the diagram, $n$ is number of gluons emitted and the indices
$n_1,\dots,n_{m-1}$ (with $1\le n_i\le n$) correspond to an arbitrary partition
of an arbitrary permutation of the $n$ gluon indices.  A product of zero
$\l$ matrices has to be interpreted as a Kronecker delta.             The
indices $i_1,\dots,i_m$ are the color indices of the quarks and the indices
$\alpha_1,\dots,\alpha_m$ are the color indices of the antiquarks. By
convention all of the particles are outgoing, so each external quark is
connected by a fermionic line to an external antiquark. When we want to
indicate that a quark with color index $i_k$ and an antiquark with color
index $\alpha_k$ are in fact connected by a fermionic line, we identify the
index $\alpha_k$ with the index $\ibar_k$. Therefore the string 
$\{\alpha\}=\istring{\alpha}{m}$ is a generic permutation of the string
$\{\ibar\}=\istring{\ibar}{m}$.  The power $p$ is determined by the number of 
correspondences between the string $\{\alpha\}$ and the string $\{\ibar\}$, \ie
by the number of times $\alpha_k = \ibar_k$.   If  $\{\alpha\}\equiv
\{\ibar\}$, then $p=m-1$.  Contrarily to the process with only one $q\qbar$
pair, in which terms with $p \ne 0$ vanish after gauge invariant quantities are
formed, here the terms with $p \ne 0$ do not vanish. The reason for this fact
is that while one $q\qbar$ pair cannot couple via the $U(N)$-trace to a set of
gluons, the $U(N)$-trace can connect two $q\qbar$ pairs, and then it
has to be {\em explicitly} subtracted if the gauge group we want is just
$SU(N)$. These subtraction terms are exactly given by the color structures in
Eq.\eqnum{qgcolor}\ proportional to \Nc{-p}, ($p>0$). 
The negative powers of \Nc{}\ are a consequence of the coupling between a quark
and a $U(N)$-trace, which according to  the normalization chosen in
Eq.\eqnum{lamnorm}\ is given by  $1/\sqrt{N}$.
                                             
To give an example, in the case of two
quark pairs and two gluons the possible color structures are the following:
\ba        &
      	(\l^a \l^b)_{i_1 \eqibar_2} \, \delta_{i_2 \eqibar_1} 
	\quad , \quad
	(\l^a)_{i_1 \eqibar_2} (\l^b)_{i_2 \eqibar_1} \quad , 
	\quad
	\delta_{i_1 \eqibar_2} (\l^a\l^b)_{i_2 \eqibar_1} ,
	\ret &
      {1 \over N} (\l^a \l^b)_{i_1 \eqibar_1} \, 
	\delta_{i_2 \eqibar_2} \quad ,
	\quad
      {1 \over N} (\l^a)_{i_1 \eqibar_1} (\l^b)_{i_2 \eqibar_2}
	 \quad , \quad
      {1 \over N} \delta_{i_1 \eqibar_1} \,
	(\l^a \l^b)_{i_2 \eqibar_2}  ,
\ea
where $a$ and $b$ represent the color indices of the two gluons, and the six
additional color structures with $a$ and $b$ interchanged have been omitted.
                                                                       
The representation given in Eq.\eqnum{qgcolor}\ has the simple physical
description which we will now illustrate.
            
To start with, let us consider the color structure of an amplitude with quarks
only, at tree level. As before, we will take all the particles as outgoing, 
and will  assign indices \istring{i}{m}\ to the quarks and indices
\istring{\ibar}{m}\ to the
antiquarks. It is understood that the quark $i_k$ is continuously connected
through a fermionic line to the antiquark $\ibar_k$, for each $1\le k \le m$.
Helicity will be conserved along this quark line, as well as flavor, since only
gluons can be emitted. We can furthermore assume all quarks to be of different
flavor,  the case with identical quarks
being similar but more confusing. It is easy to verify that the color functions
accompanying each diagram contributing to this scattering process can be
decomposed in terms of the following color structures:
\be    \label{qcolor}                                 
       D(\{\alpha\})= 
       \frac{(-1)^p}{N^{p}} \; \delta_{i_1\alpha_1}\delta_{i_2\alpha_2}\dots
	\delta_{i_m\alpha_m}
\ee                       
where $\{\alpha\}=\istring{\alpha}{m}$ is a permutation of 
$\{\ibar\}=\istring{\ibar}{m}$. Each color structure $D(\{\alpha\})$ defines a
color-flow pattern inside the diagram.  As before, $p$ is the number of
$U(N)$-trace gluons that can appear in a given color-flow configuration, and
whose contribution has to be subtracted. In this case, $p$ is the number of
$\delta_{i_k\eqibar_k}$ appearing in the product, with the exception that  when
{\em all} of the delta functions connect quark pairs that belong to the same
fermionic line (\ie $\{\alpha\}\equiv\{\ibar\}$) then  $p=m-1$.

In Figure~\cfffour\ the case $m=2$ is shown, with the possible
color factors given by (see Eq.\eqnum{trace}):
\be                                          
	\delta_{i_1 \eqjbar_2} \delta_{i_2 \eqjbar_1} \quad , \quad
	{1 \over N} \delta_{i_1 \eqjbar_1} \delta_{i_2 \eqjbar_2} 
\ee                                                       
In Figure~\cfffour\ the propagation of the $U(N)$-trace gluon
is represented by the dashed line between the two quark lines.

The color structure in Equation~\eqnum{qgcolor}
has then a very simple physical interpretation. In fact it
corresponds to the emission of the gluons off the color-flow lines defined by
the functions $D(\{\alpha\})$. Each function $D(\{\alpha\})$ 
defines a net of color flows, as shown in Figure~\cfffive\ for the case
$m=2$. Each of these color-flow lines, specified by a pair of indices
$(i_k,\jbar_{k'})$, acts as a sort of antenna, that radiates gluons with an
associated color factor $(\l \dots \l)_{i_k \,
\eqjbar_{k'}}$ (see Figure~\cfffive).
This color factor is the one appearing in the QED-type diagrams,
\ie diagrams in which all the gauge bosons are emitted from the fermionic
line and no three- or four-vector vertices are present. 
Equation~\eqnum{qgcolor} shows that even graphs with non-abelian vertices can
be decomposed as sums   of QED-like diagrams.

Given a helicity configuration for the external states the matrix element for
$m$ quark-pair plus $n$ gluon scattering can then be expressed as: 
\be   \label{amp}
  	{\cal M}_{m,n} \;=\; \sum \, 
       	\Lambda_p (\{n_i\}, \{\alpha\}) \, 
       	\tilde m^p_{\{n_i\},\{\alpha\}}(q,h).
\ee        
where the sum is over the permutations of $(j_1,\dots,j_m)$.
$\Lambda_p (\{n_i\}, \{\alpha\}) $ are the color factors appearing in
Eq.\eqnum{qgcolor}: they depend 
upon the partition and permutation $\{n_i\}$
of the gluon indices and upon the antenna pattern determined by the permutation
of indices $\{\alpha\}$. We introduced the subscript $p$ to remind that
depending on the permutation  $\{\alpha\}$ the color factor will be
proportional to a given power \Nc{-p}.
The sub-amplitudes $\tilde m^p(q,h)$ multiplying a
given color factor are functions of the momenta $q$ and helicities $h$ 
of the external
particles. These \subs\ are obtained by summing contributions from various
different Feynman diagrams.
If some of the external states are in a given color configuration, for example
in a color singlet, the amplitude can be easily obtained by contracting
Eq.\eqnum{amp}\ with the proper projector.

To the leading order in $N$ only the terms in the sum with $\alpha_k \ne j_k$
will contribute, and the sum over colors of the amplitude squared will be the
sum of the squares of the functions $\tilde m^{p=0}_{\{n_i\},\{\alpha\}}$, the
interferences being suppressed by negative powers of $N$, as can be easily
checked using Eq.\eqnum{trace}:         
\be    \label{qgsquare}       
	\sum_{col} \vert {\cal M}_{m,n}  \vert ^2 \;=\;  
     \Nc{m+n} {\widehat{\sum_{\{n_i\}\{\alpha\}}}} \;  
	 \vert m^0_{\{n_i\},\{\alpha\}}(q,h) \vert ^2.
\ee	                                          
The `hat' restricts the sum to the permutations $\{\alpha \}$
with $\alpha_k \ne j_k$ for all $k$'s.
     
Each \sub  $~m^p_{\{n_i\},\{\alpha\}}(q,h)$ is invariant under gauge
transformations  of the gluon polarizations $\epsilon^{i}_\mu \rightarrow
\epsilon^i_\mu + \beta p^i_{\mu}$. To prove this it is sufficient the
orthogonality, to the leading order in $N$, of the color factors. We will now
prove this fact. 
                 
Let $\delta m^p_{\{n_i\},\{\alpha\}}(q,h)$ be the gauge variation
of a given \sub. Let $\{\bar n_i\},\{\bar\alpha\}$ be a given partition
and a given permutation of quark and gluon indices chosen in such a way that
$p=0$.  Then the following identity
follows:
\be
	0 = \sum_{col} \Lambda^{p=0}_{\{\bar n_i\},\{\bar \alpha\}}(q,h)
	\delta {\cal M}_{m,n} =      
  N^{n+m}\,\delta m^0_{\{\bar n_i\},\{\bar\alpha\}}(q,h)
	+ {\cal O}(1/N^2).
\ee
This shows that all the \subs\ labelled by $p=0$ are gauge-invariant,
since gauge-invariance does not depend on $N$ and variations of ${\cal
O}(1/\Nc{2})$ cannot cancel the leading piece.         
We can now select all of the \subs\ corresponding to $p=1$, and repeat the same
construction to show that they are gauge invariant too. In this way one can
continue until $p=m-1$ is reached, thus proving that each \sub\ is in fact
gauge invariant.

This gauge invariance is particularly useful for the calculation of
the sub-amplitudes, since different gauges can be chosen for different sub-sets
of gauge invariant diagrams. 
                             
To conclude this Section, we indicate how these color basis generalizes to the
case of loop amplitudes. First of all let us remind that loop amplitudes can be
obtained by applying proper dispersion relations to tree-level amplitudes,
where some of the external particles have been identified and a sum over their
possible internal quantum numbers performed. We can then obtain the color form
factors which generalize our construction to loop amplitudes by contracting
pairs of color indices in the
color representations Eqs.\eqnum{exp},\eqnum{qgcolor}. 

As an example, let us consider one-loop corrections to the $q\qbar g_1\dots
g_n$ process, whose color structure is described at tree-level by
Eq.\eqnum{qqbarcol}. For simplicity we will take $n=2$. At one-loop we can have
either a gluon contraction (from a $q\qbar$ plus four gluon tree diagram), or a
quark contraction (from a $q\qbar q'\qbar'$ plus two gluon tree diagram). Let
us study the gluon loop first: for this we need  to consider the color
structure of a $q\qbar$ plus four gluon tree-level diagram,
\lstring{a}{4}$_{i\eqibar}$. Up to permutations of the indices, we have three
possible independent color structures arising from the three inequivalent
contractions of gluons:
\ba                    
	&&  \delta_{cd} (\l^a\l^b\l^c\l^d)_{i\eqibar}   \=
	    \delta_{cd} (\l^a\l^c\l^d\l^b)_{i\eqibar}   \=
	\frac{\Nc{2}-1}{\Nc{{}}} (\l^a\l^b)_{i\eqibar}  ,
	\ret                                           
	&&   \delta_{cd} (\l^c\l^a\l^b\l^d)_{i\eqibar}   \=
	\delta_{i\eqibar} \tr (\l^a\l^b) \;-\;  
	\frac{1}{\Nc{{}}}(\l^a\l^b)_{i\eqibar} ,
        \ret
	&&  \delta_{cd} (\l^a\l^c\l^b\l^d)_{i\eqibar}   \=
	\frac{-1}{\Nc{{}}}(\l^a\l^b)_{i\eqibar}    .
\ea
Two comments are in order: first, a term of the form $\delta_{i\eqibar} 
\tr (\l^a\l^b)$, which was absent at tree-level, is now generated. 
It originates from a color configuration in which the color of the quark flows
to the antiquark through the gluon in the loop, without emitting any radiation,
while the two external gluons are emitted from the remaining color line
circulating in the loop. Since the graph of the color flow is planar and since
no trace over internal color lines appears, this term is of order \Nc{0}.
The second comment is that each of the \subs\ that correspond to the three
color structures (and their permutations):
\be \label{qqloop}                  
	\Nc{{}}(\l^a\l^b)_{i\eqibar}  ,
	\quad
	\frac{-1}{\Nc{{}}}(\l^a\l^b)_{i\eqibar}  ,
	\quad
	\delta_{i\eqibar} \tr (\l^a\l^b)  
\ee
is gauge invariant. The proof follows the one given above for the tree-level 
case. Notice that even though the first two color structures are proportional,
nevertheless they are independently gauge invariant. Graphically, they
correspond to planar and non-planar diagrams, respectively. 

From the analysis of the diagrams with a quark contraction, finally, we find
again  the color form factors given in Eq.\eqnum{qqloop}\  plus  the form
factor $(\l^a\l^b)_{i\eqjbar}$ (from pure quark-loop diagrams). 

In the general case of $n$ external gluons and $m$ external quark pairs the 
possible color form factors can be represented, in a symbolic fashion, by the
following expression:
\be
	\Nc{p} \;\; \tr(\l\dots\l) \dots \tr(\l\dots\l) \;
	\delta_{\alpha\bar\alpha} \dots \delta_{\beta\bar\beta}  \;
	(\l\dots\l)_{\gamma\bar\gamma}\dots (\l\dots\l)_{\delta\bar\delta}.
\ee
If only external gluons are present, the form factors are given by products of
traces of \l\ matrices. In general the power $p$ is an integer, determined by
the degree of non-planarity of the given color flow configuration, by the
number of closed color lines, and by the number of $U(N)$-trace subtractions.
Once again all of the \subs\ relative to a given form factor are gauge
invariant. In spite of the proliferation of form factors, which form a highly
reducible basis for the color space of a given process, the possibility of
breaking the sum of diagrams into many gauge invariant components turns out to
be an extremely efficient bookkeeping device to explicitly carry out the
calculations of complex matrix elements.


\newpage                                               
\section{Supersymmetry Relations among Amplitudes}
The properties  of the color form factors introduced in the previous Section
only depend on the representation of the gauge group to which the partons
belong and to the gauge nature of the couplings, while they are not directly
related, for example, to the particle's spin.  By this we mean that the
scattering amplitudes for scalar particles transforming as the fundamental
representation of $SU(N)$, for example, can be expanded into the same color
basis -- Equation \eqnum{qqbarcol}\ -- as the amplitudes for quarks. This
expansion will still be gauge invariant and satisfy the important properties
illustrated in the previous Section.  Likewise, amplitudes with fermions
transforming according to the adjoint representation of the gauge group (as in
a supersymmetric Yang-Mills theory) can be expanded using the dual basis,
Equation \eqnum{exp}.

In a supersymmetric theory, in which particles with different spins are
related to one another by symmetry transformations,  the relation between the
color structures extends to a relation between the sub-amplitudes as well. 
This proves to be extremely useful in simplifying the calculations for
multi-particle processes in supersymmetric theories, as different amplitudes
are connected by simple algebraic identities.  In particular, amplitudes with
scalars or fermions are much simpler to evaluate than amplitudes with gauge
bosons, as the number of diagrams and the complexity of the couplings are
smaller in the first case. 

The general properties of scattering amplitudes in supersymmetric theories
were first discussed by Grisaru et al. in Reference \cite{susy1,susy2}.  
The                                                                  
importance of these supersymmetry relations for calculations in
non-supersymmetric theories was then
pointed out in Reference
\cite{pt85}, where it was suggested the use on $N=2$ supersymmetry for the
evaluation of tree-level multi-gluon processes. As was noticed in Reference
\cite{pt85}, in fact, the diagrams contributing to multi-gluon
processes  at tree-level are exactly the same in the ordinary Yang-Mills theory
as they are in its supersymmetric extension, since neither scalar nor
fermionic particles are allowed to appear as internal propagators. The
amplitudes with only gluons can be related through supersymmetry to 
easier-to-evaluate amplitudes with scalar and fermion external states, thus
significantly simplifying the calculations. 
$N=1$ supersymmetry was also employed by Kunszt~\cite{kunszt} for the
calculation of six-parton processes in QCD. 

In this Section we will illustrate the basic features of this technique, and
we will show how to efficiently complement it with the color expansion
developed earlier.

Here we will just use $N=1$ supersymmetry, rather than $N=2$. 

One possible representation of $N=1$ supersymmetry contains a massless
vector ($g^{\pm}$) and a massless spin $1/2$ Weyl spinor ($\Lambda^{\pm}$).
The $\pm$ refers to the two possible helicity states of the vector and the
spinor.
Let  $Q(\eta)$ be the supersymmetry charge \cite{wess} with $\eta$ being the
fermionic parameter of the transformation. Then $Q(\eta)$ acts on the doublet
$(g,\Lambda)$ as follows$^{\cite{susy1,susy2}}$:
\ba     \label{susyb}
        [Q(\eta),g^{\pm}(p)]&=&\mp \; \Gamma^{\pm}(p,\eta)\;\Lambda^{\pm},
        \\[0.2in]
        \label{susyf}
        [Q(\eta),\Lambda^{\pm}(p)]&=&\mp \; \Gamma^{\mp}(p,\eta)\;g^{\pm}.
\ea
$\Gamma^{\pm}(p,\eta)$ is a complex function
linear in the anticommuting c-number components of $\eta$ and satisfies:
\be     \label{gamma}
        \Gamma^+(p,\eta)=[\Gamma^-(p,\eta)]^*=\bar{\eta}
        \; u_-(p),
\ee           
with $u_-(p)$ a negative helicity spinor
satisfying the massless Dirac
equation with momentum $p$. Because of the arbitrariness in choosing the
supersymmetry parameter $\eta$, we choose this to be a negative helicity
spinor obeying the Dirac equation with an arbitrary massless momentum $k$
times a Grassmann variable $\theta$. This variable is used to remind 
us that $\Gamma^{\pm}(p,\eta)$ anti-commutes with the fermion creation 
and annihilation operators and commutes with the bosonic operators.
If we use the notation introduced in the first Section, we then obtain:
\be
        \Gamma^+(p,k)\equiv\Gamma^+(p,\eta(k))=
         \theta \langle k+ \vert p- \rangle \equiv \theta ~ [kp].
\ee
As a notation, we choose to label the supersymmetry charge $Q(\eta)$ with
the momentum $k$ characterising the parameter $\eta$: $Q(k)=Q[\eta(k)]$.
 
Because of supersymmetry, the operator $Q(k)$ annihilates the vacuum.
It follows that the commutator of $Q(k)$ with any string of operators
creating or annihilating vectors $g^{\pm}$ and spinors $\Lambda^{\pm}$ has
a vanishing vacuum expectation value. If $z_i$ represent any of these
operators, we then obtain the following Supersymmetry Ward 
identity (SWI) \cite{susy1}:
\be   \label{swi}                                            
       0= \langle [Q,\prod_{i=1}^{n} z_i] \rangle _0 =
        \sum_{i=1}^n \langle z_1 \cdots [Q,z_i] \cdots z_n \rangle _0,
\ee
where $ \langle \dots \rangle _0$ indicates the vacuum expectation value.
If we substitute in equation~\eqnum{swi}\ the commutators, we obtain
a relation among scattering amplitudes for particles with different spin.
General features of Yang-Mills interactions, like helicity conservation
in the fermion-fermion-vector vertex guarantee the vanishing of some of
the amplitudes in \eqnum{swi}. The arbitrariness in choosing the reference
momentum $k$ for the supersymmetry parameter $\eta$ allows a further
simplification of equation~~\eqnum{swi},
by choosing $k$ to be equal to one
of the external momenta.

As was first pointed out in Reference~\cite{susy1}, these relations can be
used to prove general properties of helicity amplitudes in supersymmetric 
Yang-Mills theories; at tree level, these properties hold for the non-symmetric
theory as well. We will here prove some of these properties as an example of
the use of the supersymmetry relations. 
In particular, we will show the vanishing of all the helicity amplitudes
of the kind $(\pm\pm\dots\pm)$ and $(\mp\pm\dots\pm)$, where we assume all of
the particles as outgoing.  For the two-to-two scattering processes in
Yang-Mills and gravity, these vanishing theorems were first proved in
Reference~\cite{dewitt}.                                          
                        
Let us start applying the supersymmetry charge to the following string of
operators:
\ba    \nn
   0 &=& \langle \left[ Q , \Lambda_1^+g_2^+g_3^+\dots g_n^+ \right ] 
	\rangle \=
   -\Gamma^-(p_1,k)~A(g_1^+,g_2^+,\dots,g_n^+) ~+               
	\ret  \label{hconservI}                      
   && +\Gamma^+(p_2,k)~A(\Lambda_1^+,\Lambda_2^+,\dots,g_n^+) ~+~ \dots ~+~
     \Gamma^+(p_n,k)~A(\Lambda_1^+,g_2^+,\dots,\Lambda_n^+).               
\ea              
Since all of the couplings of fermions to vectors are helicity conserving,
all of the amplitudes with two fermions of the same helicity must vanish, and
as a consequence of the identity the first term on the right hand side of
Equation~\eqnum{hconservI}\ must vanish as well. Therefore maximal helicity
violation is forbidden in perturbation theory in 
any supersymmetric gauge theory, and at tree-level in
any gauge theory.

To prove the same theorem for the next helicity violating amplitudes, let us
consider the following identity:
\ba    \nn
   0 &=& \langle \left[ Q , \Lambda_1^+g_2^-g_3^+ \dots g_n^+ \right ] 
	\rangle \=
	\ret  \label{hconservII}                      
   && -~\Gamma^-(p_1,k)~A(g_1^+,g_2^-,\dots,g_n^+) ~-~
        \Gamma^-(p_2,k)~A(\Lambda_1^+,\Lambda_2^-,\dots,g_n^+).
\ea              
Here we have omitted all of the vanishing 
amplitudes with both fermions having the same
helicity. Equation~\eqnum{hconservII}\ must be satisfied for any choice of the
vector $k$, and in particular we can then choose $k=p_2$, proving that the
gluonic amplitude must vanish, or $k=p_1$, thus proving the vanishing of the
amplitudes with the fermion pair.             
                                                             
As a first example of non-vanishing amplitudes, 
let us now consider the helicity
amplitude 
$(g^-_{1},g^-_{2},g^+_{3},\dots,g^+_{n})$, with two negative-helicity
gluons and $n-2$ positive-helicity gluons where all of the particles
are outgoing. Through the SWI we can relate this amplitude to amplitudes
with two fermions and $n-2$ vectors. Helicity conservation
for the fermions implies that only an amplitude with one positive- and one
negative-helicity gluino can be non-vanishing.  In this way
equation~\eqnum{swi} reduces to:
\ba
       \Gamma^-(p_1,k)A(\Lambda^-_{1},g^-_{2},\Lambda^+_{3},g^+_{4},
                \dots,g^+_{n}) &+&
       \Gamma^-(p_2,k)A(g^-_{1},\Lambda^-_{2},\Lambda^+_{3},g^+_{4},
                \dots,g^+_{n})
\nn \\[0.2in]
      &-& \Gamma^-(p_3,k)A(g^-_{1},g^-_{2},g^+_{3},\dots,g^+_{n})=0.
\ea
Choosing, for example, $k=p_1$ we therefore obtain the following relation:
\be     \label{susymhv}
      A(g^-_{1},g^-_{2},g^+_{3},\dots,g^+_{n}) \=
     \frac{\sp 12}{\sp 13} A(g^-_{1},\Lambda^-_{2},\Lambda^+_{3},g^+_{4},
                \dots,g^+_{n})
\ee
As we said before, the purely gluonic amplitudes for the non-supersymmetric and
the supersymmetric theory coincide.             

It is very important to notice that the supersymmetry identities and all
the relations that can be obtained through their use -- like the  
vanishing theorems -- hold separately for each of the the sub-amplitudes in
which we can expand the full amplitude. This can be
easily proved on the basis of gauge invariance and leading order orthogonality
of the color form factors. 

Another important consequence of the expansion in terms of the color structures
described in the previous Section is the possibility of relating directly
amplitudes with a pair of quarks to amplitudes with a pair of gluinos. 
The SWI will then allow us to connect directly amplitudes 
with only gluons to amplitudes with a quark pair.
To make explicit this relation, we remind from the previous Section that
amplitudes with a quark pair (gluino pair) can be written in the following way:
\ba                          
	\label{quarks}
    	{\cal M}_{n+2}(q,g_1,\dots,g_n,\qbar) &=&
	\sum_{\{1,\dots,n\}} \;(\l^{a_1}\l^{a_2}
	\dots\l^{a_n})_{i\eqjbar}                                        
	\;m_q(q,p_1,\dots,p_n,\qbar),
\ret           
	\label{gluinos}
    	{\cal M}_{n+2}(g_1,\dots,g_n,\Lambda_{n+1},\Lambda_{n+2}) &=&
	\sum_{\{1,\dots,n+2\}'} \;tr\,(\l^{a_1}\l^{a_2}
	\dots\l^{a_{n+2}})                                        
	\;m_\Lambda(p_1,\dots,p_n,q_{n+1},q_{n+2}),
\nn \ret
\ea	           
where the momenta labeled with $q$ are fermion momenta, and where the
sub-amplitudes $m_q$ and $m_\Lambda$ can be found by summing over subsets of
Feynman diagrams obtained according to the prescriptions introduced in the
previous Section. The {\it prime} indicates that only non-cyclic permutations
have to be summed over.                  

The main difference between Equation~\eqnum{quarks}\ and
Equation~\eqnum{gluinos}\ lies in the fact that while the quark sub-amplitudes
$m_q$ always have the two fermions adjacent, there are gluino sub-amplitudes
with non-adjacent fermions.
The gluino sub-amplitudes, furthermore, satisfy the Dual Ward Identity,
Equation~\eqnum{ward}. It is easy to prove, by just studying the structure of
the relevant Feynman diagrams, that the following identity between quark and
gluino amplitudes holds:
\be
	m_q(q,p_1,\dots,p_n,\qbar)  \=
	m_\Lambda(q_{n+2},p_1,\dots,p_n,q_{n+1}),
\ee                                            
where $q=q_{n+2}$ and $\qbar=q_{n+1}$. The complete proof can be found in
Reference~\cite{mp88}.

Therefore by just calculating the gluino amplitudes we can automatically obtain
the quark amplitudes, using the previous identity, and the purely gluonic
amplitudes, by using the SWI. Even if we are only interested in the quark
amplitudes, it may still be nevertheless useful to consider the gluino
amplitudes with non-adjacent fermions as auxiliary objects,  because they
satisfy the Dual Ward Identity and can help simplifying expressions.
                                                                    
We will present a complete example of how this
works in detail in the next Section, when calculating the five parton
processes.



\newpage
\section{Explicit Results}
In this Section we will illustrate the use of the various techniques introduced
up to now with the explicit calculation of some four- and five-parton
processes in a $SU(N)$ massless gauge theory. These results were known since
the original papers
\cite{four1,four2,fiveI,fiveII,fiveIV,fiveV,keith86a}, but reproducing
them here                                                    
will show the simplicity of these techniques and their advantage over the more
standard approach. At the end of the section we will collect some results
concerning 6-parton processes as well, without going into the explicit details
of the calculations\footnote{For the details on explicit analytical derivations
see \cite{gk85a,xu} for 4-quark plus 2-gluons,
\cite{kunszt,pt87,mp88} for 2-quark plus 4-gluons and
\cite{pt86a,kunszt,gunion,bg87,mpx88} for the 6-gluon processes. For 
7-gluons see \cite{kk89} using the recursion relations of Section 
\secrsq.}. 
We hope
this Section will be helpful to the reader who wants to familiarize himself
with the explicit use of these tools.
                                     
\subsection{Four Partons}
As was mentioned in the previous Section it is generally convenient to start the
calculations from processes with a pair of fermions and to use the
supersymmetry relations to simply obtain the amplitudes of purely gluonic
processes as a by-product. The use of the polarization vectors introduced in
the first Section and of the dual color basis, however, makes the four gluon
calculation so simple by itself that it is useful to just start from it.
We will then derive the fermionic amplitude by using the SWI. In the next
subsection, when describing the five-parton processes, we will follow the
opposite route, as then the fermionic amplitude calculation is considerably 
simpler than the gluonic one.

We introduce the following notation: $q$
and $\qbar$ are the momenta of quark and antiquark, $h_q$ is the quark helicity
( the helicity of the antiquark, $h_{\qbar}$,  is fixed by helicity
conservation) and $i,\ibar$ are the color indices; $p_i,\ (i=1,2,3)$, $h_i$ and
$a_i$ will be respectively  the momenta, helicities and colors of the three
gluons.  All the particles are taken as outgoing, and therefore momentum
conservation is given by $q+\qbar+\sum p_i = 0$.               

All of the diagrams contributing to the four gluon amplitude have the following
color structure:
\be 
	f^{abX}f^{Xcd} \= - tr ([\l^a,\l^b][\l^c,\l^d]).
\ee
Here we used the normalization conventions introduced in Section~\sectccf.
The Feynman rules are given in Appendix~C.
The Feynman diagrams that enter in the calculation of a given sub-amplitude can
just be found by imposing the condition that they contain the trace of the
string of \l\ matrices in the proper permutation. For example, when calculating
$m(1,2,3,4)$ the only diagrams which will contribute are those drawn in
Fig.~\exaone.  

Notice that the first diagram will also contribute to the
subamplitudes corresponding to the permutations $(1243)$, $(2134)$ and
$(2143)$, but remember that for the calculation of the subamplitudes only the
{\em kinematical} part of the Feynman rules has to be used (see Appendix~C). 
There is only one diagram with the four-gluon coupling, and that contributes to
all 6 the permutations.
                                                                    
Before proceeding, let us classify the possible helicity configurations.
As it was shown using the supersymmetry relations in the previous Section, 
the helicity amplitudes with all of the gluons having the same helicity, and
the amplitudes with all of the gluons but one having the same helicity are
zero. We can prove this independently here by using an explicit representation
for the polarization vectors of the gluons, as given in the first Section. In
fact, assign to the gluons with the same helicity the same reference momentum,
and in the case of \subs\ of the kind $m(\mp,\pm,\dots,\pm)$ fix this
reference momentum to be the momentum of the gluon with opposite helicity from
the others. Then
it is easy to see using the identities given in Appendix~A that all of the
products $\epsilon^i \cdot \epsilon^j$ vanish. Since by the Feynman rules (or 
dimensional analysis) it follows that at tree-level each diagram will contain at
least a  factor $\epsilon^i \cdot \epsilon^j$, it follows that these amplitudes
will vanish. 
             
Therefore for four-gluon scattering the only non-zero amplitudes will be of the
form $(--++)$, up to permutations of the indices. Let us then consider the
\sub\ $m(1^-,2^-,3^+,4^+)$, with the reference momenta
for the gluons (1,2,3,4) given by the momenta of gluons (3,3,2,2), respectively.
For this choice of reference momenta 
the only non-zero $\epsilon_i \cdot \epsilon_j$ is $\epsilon_1 \cdot 
\epsilon_4$. Therefore the only non-zero diagram from Fig.~\exaone ~ is the 
first one, which gives explicitly:
\ba
	m(1^-,2^-,3^+,4^+) & = & -2i g^2  ~{\epsilon_1 \cdot \epsilon_4 ~~ 
	\epsilon_2 \cdot p_1~ \epsilon_3 \cdot p_4  \over s_{12} }
\nonumber \\ \label{eg4g}
     &=& -ig^2 ~{ {\langle 12 \rangle}^2 [34]^2 \over s_{12} ~s_{23} }
     \=  ig^2 ~ \frac {\sp 12 ^4}{\sp 12 \sp 23 \sp 34 \sp 41} ,
\ea                                                               
where various definitions and properties of the spinor dot products collected in
the Appendix  -- together with the kinematical identity $s_{12}$=$s_{34}$ --
were used. Notice that even though the diagram containing the
$t$-channel exchange vanishes as a consequence of the choice of reference
momenta (\ie the gauge choice), the $t$-pole ($1/s_{23}$) appears from the
normalization of the polarization vectors, signalling that the gauge chosen is
singular for $s_{23}=0$. This should be expected since, as we pointed out in
Section~\secthel, these gauges are light-cone gauges.  Needless to say the
result  in Eq.\eqnum{eg4g}\ is gauge invariant.

By using the Dual Ward Identity and the invariance under cyclic permutations 
we obtain the following identity, which allows
us to express the other inequivalent \sub\ $m(+-+-)$ in terms of the one we
just evaluated using Feynman diagrams:
\be
	m(1^-2^+3^-4^+) \= -m(1^-3^-2^+4^+)-m(3^-1^-2^+4^+).
\ee
Applying the Fierz identity Equation~\eqnum{aids8}\ to this DWI gives finally:
\be
	m(1^-,2^+,3^-,4^+) 
     \=  ig^2 ~ \frac {\sp 13 ^4}{\sp 12 \sp 23 \sp 34 \sp 41} ,
\ee
which generalizes to
\be     \label{gggg}
	m(1,2,3,4)            
     \=  ig^2 ~ \frac {\sp IJ ^4}{\sp 12 \sp 23 \sp 34 \sp 41} ,
\ee                           
where $I$ and $J$ are the indices of the negative helicity gluons.
The full amplitude will therefore be given by:
\ba
   	M(1,2,3,4) &=&
	ig^2 \sp IJ ^4 \sum_{perm'} tr (\l^{a_1}\l^{a_2}\l^{a_3}\l^{a_4})
	\frac {1}{\sp 12 \sp 23 \sp 34 \sp 41}                          
\nn \ret           &=&
	ig^2 \sp IJ ^4 \sum_{perm''}
	[tr (\l^{a_1}\l^{a_2}\l^{a_3}\l^{a_4})
	+tr (\l^{a_4}\l^{a_3}\l^{a_2}\l^{a_1})]
	\frac {1}{\sp 12 \sp 23 \sp 34 \sp 41}                          ,
\nn \ret
\ea
where the prime indicates sum over non-cyclic permutations, and the double
prime indicates sum over permutations up to cyclic and reverse (\ie 
$(1,2,..,n) \to (n,..,2,1)$ ) re-orderings.

In squaring the four gluon amplitude and summing over colors  the ${\cal
O}(N^{-2})$ terms in equation~\eqnum{ncfact} 
can be shown to vanish by using only the 
general properties, especially the Ward Identity, of the  \sub\ (see 
Appendix~D for the details). Therefore,
\begin{eqnarray}                        
	\sum_{colors}\;\vert {\cal M}_4 \vert^2\;=\;& &
	{N^2(N^2-1)}
	\sum_{perm^{\prime}}  \vert m(1,2,3,4) \vert^2 ,
\end{eqnarray}
and the square of each sub-amplitude is very simple because the 
spinor product is the square root of twice the dot product.
The final 
result is the standard four gluon matrix element squared:
\be   \label{pt4g}                                             
       \sum_{hel.} \sum_{colors} \vert{\cal M}_4\vert ^2 =
	N^2(N^2-1)~g^4~ \left(\sum_{i>j} s_{ij}^4 \right)
	\sum_{perm^{\prime}}{1 \over s_{12}s_{23}s_{34}s_{41}}.
\ee
Here we have not averaged over helicities or colors. 

To obtain the amplitude for two gluons plus a $q\qbar$ pair we can use the SWI
explicitly given in the previous Section for amplitudes of the kind
$M(\mp\mp\pm\dots\pm)$, Eq.\eqnum{susymhv}, and we simply get:
\be
	m(q g_1g_2 \qbar ) \=
	ig^2 \frac {\sp qI ^3 \sp \qbar{I} }
		   {\sp \qbar q \sp q1 \sp 12 \sp 2\qbar}	,
\ee                                                             
where we chose the quark and gluon $I$ ($I$=1,2) to have negative helicity, and
where \sp qI \ is a short-hand for \sp q{p_I}. For
a negative helicity anti-quark (\ie positive helicity quark) it is sufficient
to exchange $q$ with $\qbar$ in the numerator.
The full amplitude will be:
\be
	M(q g_1g_2\qbar) \=
	ig^2 {\sp qI ^3 \sp \qbar{I} }   \sum_{\{1,2\}}
	(\l^{a_1}\l^{a_2})_{i {\scriptstyle\bar\imath} }                  
	   \frac {1}{\sp \qbar q \sp q1 \sp 12 \sp 2\qbar}	
\ee                                                             
with the following square, summed over colors and
helicities:                           
\begin{eqnarray}                                         
       \sum_{hel.} \sum_{colors} \vert{\cal M}_4\vert ^2 &=&
	2 N(N^2-1)~g^4~ \left(\sum_{i=1,2} s_{qi}s_{\qbar i}^3 + 
	 s_{qi}^3s_{\qbar i} \right)
	\sum_{\{1,2\}}{1 \over s_{q\qbar}s_{12}s_{q1}s_{\qbar2}}
\nn \ret
	&+& {\cal O}(1/N^2).
\end{eqnarray}           
For the details of the squaring and the explicit form of the sub-leading piece,
see the Appendix. It is straightforward to check that these results agree with
the standard calculations.
                          
\subsection{Five Partons} 
We will begin from the calculation of the matrix elements for the scattering of
one $q\qbar$ pair and three gluons. 
First of all we classify the possible color form factors for the process.
According to Eq.\eqnum{qgcolor}\ these are given by the 6 permutations of the
expression $(\l^{a_1}\l^{a_2}\l^{a_3})_{i{\scriptstyle\bar\imath}}$. To these
six permutations there will correspond six (a priori different) \subs. We will
consider now the permutation $(1,2,3)$ of the gluon indices, and will show
afterwards how to obtain the others by using the various identities introduced
previously.  Having chosen a color factor, we need to find all of the diagrams
which contain this given color factor. For the process under study, these
diagrams are shown in Fig.~\exatwo). 
Notice that only the diagram ($a$)
contributes exclusively to this \sub. In fact it is easy to see that, for
example, diagram ($b$)  will also contribute to the \sub\ corresponding to the
permutation $(2,1,3)$,  diagram ($d$) to $(1,3,2)$, $(2,3,1)$ and $(3,2,1)$,
and diagram ($f$) will contribute to all six the permutations. According to our
technique, in the calculation of a given \sub\ we will just sum up the terms of
each diagram proportional to the corresponding permutation.

Now we have to classify the various possible helicity configurations. Up to
permutations and charge conjugation, we have four different cases: we can have
either all of the three gluons with the same helicity as the quark, or just
two, or one, or none.
As in the purely gluonic case, amplitudes of the type
$M(\mp\pm\dots\pm)$ vanish identically, as was proven using supersymmetry. 
We will now prove this by using an explicit representation for the
polarization vectors.                                         

Let us consider the case where all the gluons have the same helicity,
opposite to the helicity of the quark.
Let us choose the reference momentum of the gluon polarizations
to be 
the quark momentum. It then follows from Eqs.\eqnum{aepshat}\ and \eqnum{apol5}:
\be     \label{likehel}                     
	\lsp{q\pm} \epsh{\mp}{p_i,q} \= 0 , \quad \quad
	\epsilon^i\cdot\epsilon^j \= 0.
\ee
The {\em bra} spinor represents an outgoing quark with helicity $\pm$.
Let us then study the branch of gluons starting from the first gluon emitted by
the quark leg. The only vector quantities that can contract with the $\gamma$
matrix present at the vertex are the polarization vector of one of the external
gluons emitted by this branch, or some combination of momenta of the external
gluons themselves. In the first case the diagram is zero because of the first
equation above. In the second case, possible only if the branch has more than
one external gluon, the saturation of the indices and the dimensionality of the
couplings (\ie there can only be at most one power of momentum for each gluon
vertex) forces at least one scalar product between two polarization vectors. In
this case the diagram vanishes because of the second identity above.  This
proof of course extends to tree-level processes with a $q\qbar$ pair and an
arbitrary number of gluons, and can be easily repeated for the case with all of
the gluons having the same helicity, {\em equal} to the quark's one. 

Let us now consider the case with two gluons (say 1 and 2) having the 
helicity opposite to the quark (say +). According to the matching rule
discussed in Appendix~B, we will choose the reference momentum for the
polarizations of gluons 1 and 2 to be the quark momentum $q$, and the reference
momentum for the gluon 3 to be the antiquark momentum $\qbar$. 
We then have the following identities:
\be
	\lsp{q-} \epsh{+}{p_{1,2},q}  \= 0 , \quad
	\epsh{-}{p_3,\qbar} \rsp{\qbar -} \= 0 , \quad
	\epsilon^1 \cdot \epsilon^2 \= 0.
\ee
Using these identities it is straightforward to show that the only non
vanishing diagrams are ($d$) and ($e$). The evaluation of these two diagrams is
very simple if use is made of the various identities given in the Appendix~A,
and leads to the following result for the \sub:                   
\be    \label{qqggg}                                         
   	m(q^-,g_1^-,g_2^+,g_3^+,\qbar^+) \=
	i \, \gs{3} \, \frac{\sp q1 ^3 \sp \qbar1  }{\sp{\qbar}{q}
	\sp q1 \sp 12 \sp 23 \sp 3\qbar}.                                 
\ee

Before giving the expressions for the other permutations and 
helicity combinations, we will use Eq.\eqnum{qqggg}\  and the
supersymmetry transformation to derive the \subs\ for the five gluon process.
The supersymmetry relation that we need is the following:
\ba    \label{susy5}     
       \Gamma^-(p_1,k)m(\Lambda^-_{1},g^-_{2},\Lambda^+_{3},g^+_{4},
                g^+_{5}) &+&
       \Gamma^-(p_2,k)m(g^-_{1},\Lambda^-_{2},\Lambda^+_{3},g^+_{4},
                g^+_{5})
\nn \\[0.2in]
      &-& \Gamma^-(p_3,k)m(g^-_{1},g^-_{2},g^+_{3},g^+_{4},g^+_{5})=0.
\ea                     
with $\Gamma^-(p,k)=\sp pk$ and $\Lambda$ being the fermion field. The second
\sub\ entering this identity corresponds to the quark amplitude we just
calculated, as was mentioned in the previous Section, while the first term
is one of the fermionic \subs\ that would be necessary for the calculation of
amplitudes with fermions in the adjoint representation. By choosing
$k=p_1$ we can exclude this term, 
and using Eq.\eqnum{qqggg}\ we directly obtain: 
\be    \label{ggggg}                                         
   	m(g_1^-,g_2^-,g_3^+,g_4^+,g_5^+) \=
	i \, \gs{3} \, \frac{\sp 12 ^4}{\sp 12 \sp 23 \sp 34 \sp 45 \sp 51}.
\ee                                                                       
To get the \subs\ for the other permutations, we only need to use 
the symmetry of the \sub\ under exchange of identical bosons and the
repeated application of the Fierz relation (Eq.\eqnum{aids8}) on the Dual Ward
Identity :
\be       
	\sum_{cycl(1,..,4)} m(1,2,3,4,5) \= 0,
\ee
the sum being over the 4 cyclic permutations of $(1,2,3,4)$. One easily obtains:
\be     
   	m(g_1,g_2,g_3,g_4,g_5) \=
	i \, \gs{3} \, \frac{\sp IJ ^4}{\sp 12 \sp 23 \sp 34 \sp 45 \sp 51 },
\ee                                                                          
where $I$ and $J$ are the momenta of the two gluons with the same (negative)
helicity.  By using again now Eq.\eqnum{susy5}\ one then obtains the expression
for the general permutation of the fermionic \sub:
\be          
   	m(q^-,g_1,g_2,g_3,\qbar^+) \=
	i \, \gs{3} \, \frac{\sp qI ^3\sp \qbar I}{\sp \qbar q \sp q1 
	\sp 12 \sp 23 \sp 3\qbar },
\ee                                
where now $I$ is the index of the only gluon with negative helicity. 
Similarly, the \sub\ for the helicity configuration with one negative helicity
gluon and a negative helicity antiquark 
is given by:
\be          
   	m(q^+,g_1,g_2,g_3,\qbar^-) \=
	i \, \gs{3} \, \frac{\sp qI \sp \qbar I ^3 }{\sp \qbar q \sp q1 
	\sp 12 \sp 23 \sp 3\qbar }.             
\ee                                
All of the \subs\ for the processes with opposite helicities (\ie
$(++---)$) can be obtained from the previous expressions by replacing 
\sp{\ }{}\  products with \cp{\ }{}\  products.

Squaring the full amplitude and summing over colors and helicity
configurations, we then obtain:
\ba	\label{5gsq}
   &	\vert M(g_1,\dots,g_5) \vert ^2 \=  &
     	2 \gs{6} \Nc{3}(\Nc{2}-1) \sum_{i>j} s_{ij}^4 \;
	\sum \; \frac{1}{s_{12} s_{23} s_{34} s_{45} s_{51}}  ,
\ret
	\label{qq3gsq}
   &	\vert M(q,\qbar,g_1,g_2,g_3) \vert ^2 \= &
     	2 \gs{6} \Nc{2}(\Nc{2}-1) 
	\sum_{i} (s_{qi}^3 s_{\qbar i} \;+\; s_{qi} s_{\qbar i}^3)
	\nn \ret                    
   &&	\sum_{\{1,2,3\}} \;\frac{1}{s_{q\qbar} s_{q1} s_{12} s_{23} s_{3\qbar}}
    	\;+\;                                                                 
	O(N^{-2}).
\ea                                                                        
For the details of the squaring of the color part, see Appendix~D.
                
\subsection{Six Partons}
The six-parton processes are more complex: two independent sets of helicity
amplitudes are needed: ${\cal }M_{2-4+}$ and ${\cal M}_{3-3+}$. The first ones
are a trivial generalization of the five-parton amplitudes, and are given in
the case of two quark-four gluon and six gluons, respectively, by: 
\be    \label{q2g4I}                                               
      {\cal M}(\bar q^+_{1},q^-_{2},g^-_{3},g^+_{4},\dots,g^+_{6}) =
        i g^4 \langle 23 \rangle ^3 \langle 13 \rangle
        \sum_{\lbrace 3,4,5,6\rbrace}
        \;
        (\lambda^3\lambda^4\lambda^5\lambda^6)_{\hat2 \hat1 }
        \;
        {1 \over  \langle 12 \rangle  \langle 23 \rangle \cdots
         \langle 61 \rangle }.
\ee                           
\be    \label{g6}
      {\cal M}(g^-_{1},g^-_{2},g^+_{3},g^+_{4},\dots,g^+_{6}) =
        i g^4 \langle 12 \rangle ^4
        \sum_{\lbrace 1,2,3,4,5,6\rbrace'}
        \;                               
        tr(\l^{a_1}\l^{a_2}\dots \l^{a_6})
        \;
        {1 \over  \langle 12 \rangle  \langle 23 \rangle \cdots
         \langle 61 \rangle },
\ee
These sub-amplitudes can be shown to satisfy all of the required properties,
such as the SWI, the Dual Ward Identity and the proper soft and collinear
factorization (see Section \sectsof). 
At the leading order in $N$ the sum of these matrix elements squared, summed
over colors and over the configurations with helicities $(--++++)$ and
$(++----)$, can be easily obtained using the properties of the \l\ matrices,
giving:   
\be	
   	\vert M(g_1,\dots,g_6) \vert ^2 \=  
     	2 \gs{8} \Nc{4}(\Nc{2}-1) \sum_{i>j} s_{ij}^4 \;
	\sum \; \frac{1}{s_{12} s_{23} \cdots  s_{61}}  
    	\;+\;                                                                 
	O(N^{-2}),  \nn \\  \ \ \label{6gsq}
\ee
\ba
	\label{qq4gsq}
   	\vert M(q,\qbar,g_1,g_2,g_3,g_4) \vert ^2 \= &
     	2 \gs{8} \Nc{3}(\Nc{2}-1)                     
	\sum_{i} (s_{qi}^3 s_{\qbar i} \;+\; s_{qi} s_{\qbar i}^3)
	\nn \ret                    
   &	\sum_{\{1,2,3,4\}} \;\frac{1}{s_{q\qbar} s_{q1} s_{12} s_{23} 
	s_{34} s_{4\qbar}}
    	\;+\;                                                                 
	O(N^{-2}).
\ea                                                                        
Notice that contrarily to the 4- and 5-gluon case, here the 6-gluon amplitude
squared has a non-vanishing contribution at the sub-leading order in $N$. Its
precise form is given in Appendix~D.
Using the factorization properties of the amplitude, however, it is easy to
check that this sub-leading terms do not have collinear divergencies
\cite{mpx87}. The absence of these enhancement factors makes the numerical
value of these sub-leading terms even smaller than what one would naively expect
from the simple $1/N^2$ suppression. This fact will be discussed in more detail
in Section \sectapp, where we will illustrate some techniques to approximate
the multi-parton matrix elements. 
   
The six-parton helicity amplitudes $ {\cal M}_{3-3+}$ is described by three
distinct sub-amplitudes, characterised by three inequivalent helicity orderings:
$(+++---)$, $(++-+--)$ and $(+-+-+-)$. Because of duality,  as explained in
Ref.\cite{jacob}, all of these sub-amplitudes can be
written in the following form:
\ba
        m(1,2,\dots,6)
        &=& ig^4 \left[
        { P_1 \over t_{123} s_{12} s_{23} s_{45} s_{56}}
        ~+~{ P_2 \over t_{234} s_{23} s_{34} s_{56} s_{61}}
        \right. \nonumber \\[0.2in]
        & ~+~& {P_3 \over t_{345} s_{34} s_{45} s_{61} s_{12}}
        ~+~ \left. { P_s
         \over { s_{12} s_{23} s_{34} s_{45} s_{56} s_{61} } } ~\right].
     \label{poles.app}                                 
\ea
where $t_{ijk} \equiv (p_i+p_j+p_k)^2~=~s_{ij}+s_{jk}+s_{ki}$.
The coefficients $P_i$ will depend on the particular helicity configuration and
on the process (6-gluons or 2-quark plus four-gluons). For the purely gluonic
case, a further relation can be found between the $P$'s that will reduce
Eq.\eqnum{poles.app}\ to \cite{mpx88}:
\begin{eqnarray}             
        m_{3+3-}(g_1,g_2,\dots,g_6)
        \= ig^4 \left[       
        { \alpha^2 \over t_{123} s_{12} s_{23} s_{45} s_{56}}
        ~+~{ \beta^2 \over t_{234} s_{23} s_{34} s_{56} s_{61}}
        \right. && \nonumber \\
        ~+~ { \gamma^2 \over t_{345} s_{34} s_{45} s_{61} s_{12}}
        ~+~ \left. { {t_{123} \beta \gamma ~+~ t_{234} \gamma \alpha
        ~+~ t_{345} \alpha \beta }
         \over { s_{12} s_{23} s_{34} s_{45} s_{56} s_{61} } } ~\right], &&
        \label{six1}                                                       
\end{eqnarray}
For reference, we give the coefficients $P_i$'s and $\alpha,\beta,\gamma$ in
Table~\ref{qtable1}-\ref{qtable2} and Table~\ref{gtable}, respectively, without
derivation. Here we will just show how to relate the two sets of coefficients,
for the purely gluonic and the $q\qbar$ plus gluons case, using the various
identities introduced in the previous Sections. For simplicity we will just
work with the $(---+++)$ helicity ordering, but the same construction can be
repeated for the other orderings as well. 
                                             
Suppose we have calculated the fermionic amplitudes; then it is easy
to prove  the following identity, using a proper SWI:
\ba
        \cp 36 m(g^-_{1},g^-_{2},g^-_{3},g^+_{4},g^+_{5},g^+_{6})  &=&
     -\cp 31 m(\Lambda^+_{6},\Lambda^-_{1},g^-_{2},g^-_{3},g^+_{4},g^+_{5})
        \nn \\[0.2in]                                                      
     && -\cp 32 m(\Lambda^+_{6},g^-_{1},\Lambda^-_{2},g^-_{3}, g^+_{4},g^+_{5}).
\label{wi}                                                                     
\ea
Here by $\Lambda$ we refer to a generic fermion, $q$ or $\qbar$.
Helicity conservation has been used to cancel the two amplitudes with two
negative-helicity fermions, and the Grassmannian nature of $\Gamma^{\pm}$
was used when moving it through $\Lambda_2$.
The amplitude with the non-adjacent fermions can be extracted by using the Dual
Ward Identity obtained by moving the gluon 1 :
\ba
	m(\Lambda^+_{1},g^-_{2},\Lambda^-_{3},g^-_{4}, g^+_{5},g^+_{6})
    	&=&  \nn \ret
       -m(\Lambda^+_{1},\Lambda^-_{3},g^-_{2},g^-_{4}, g^+_{5},g^+_{6})
   &-&  m(\Lambda^+_{1},\Lambda^-_{3},g^-_{4},g^-_{2}, g^+_{5},g^+_{6})
    	\nn \ret
       -m(\Lambda^+_{1},\Lambda^-_{3},g^-_{4}, g^+_{5},g^-_{2},g^+_{6})
   &-&  m(\Lambda^+_{1},\Lambda^-_{3},g^-_{4}, g^+_{5},g^+_{6},g^-_{2}).
\ea                                                                     
Therefore the knowledge of the fermionic amplitudes is completely sufficient to
obtain the purely gluonic ones without having to calculate any additional
Feynman diagram. In particular, if one were just interested in the numerical
value of the amplitudes to calculate scattering processes, one could just use
the previous equations as operative definitions of the gluonic amplitudes,
without having to go through the algebra necessary to find explicit
expressions.

{\renewcommand{\arraystretch}{1.8}        
\begin{table}[htb]                          
        \label{qtable1}
        \begin{center}                
        \begin{tabular}{|l||c|c|c|}       
        \hline\hline
        {}
        & $P_1$ & $P_2$ &       $P_3$  \\
        {$(g_3,g_4,g_5,g_6)$}
        & $U=p_1~+~p_2~+~p_3$ & $V=p_2~+~p_3~+~p_4$ & $W=p_3~+~p_4~+~p_5$
\\ \hline\hline
         $ (-,-,+,+)_{(I)} $
        & $ {}[56]^2 \langle 13 \rangle  \langle 23 \rangle
                 \langle 1 \vert U \vert 4 \rangle ^2 $
        & 0
        & $ -[16][26]\langle 34 \rangle ^2 \langle 5 \vert W \vert 2 \rangle ^2$
\\ \hline
         $ (+,+,-,-)_{(II)} $
        & $ -[13][23]\langle 56 \rangle ^2 \langle 4 \vert U \vert 2 \rangle ^2$
        & 0
        & $ {}[34]^2 \langle 16 \rangle \langle 26 \rangle
                \langle 1 \vert W \vert 5  \rangle ^2 $
\\ \hline
         $ (-,+,+,-)_{(III)} $
        & $ {}[45]^2 \langle 13 \rangle \langle 23 \rangle
                \langle 1 \vert  U \vert 6 \rangle ^2 $
        & $ {}[15] \langle 23 \rangle  \langle 5 \vert V \vert 3 \rangle
                \langle 4 \vert  V \vert 6 \rangle ^2 $
        & $ {}[45]^2 \langle 16 \rangle  \langle 26 \rangle
                 \langle 1 \vert W \vert 3 \rangle ^2 $
\\ \hline
         $ (+,-,-,+)_{(IV)} $
        & $ -\langle 45 \rangle ^2 [ 13] [ 23]
                \langle 6 \vert  U \vert 2 \rangle ^2 $
        & $ {}[16] \langle 24 \rangle  \langle 6 \vert V \vert 4 \rangle
                \langle 3 \vert  V \vert 5 \rangle ^2 $
        & $ -\langle 45 \rangle ^2 [16]  [26]
                 \langle 3 \vert W \vert 2 \rangle ^2 $
\\ \hline
         $ (-,+,-,+,)_{(V)} $
        & $ {}[46]^2 \langle 13 \rangle  \langle 23 \rangle
                 \langle 1 \vert U \vert 5 \rangle ^2 $
        & $ {}[16] \langle 23 \rangle  \langle 6 \vert V \vert 3 \rangle
                 \langle 4 \vert V \vert 5 \rangle ^2 $
        & $ -[16]{}[26] \langle 35 \rangle ^2
                 \langle 4 \vert W \vert 2 \rangle ^2 $
\\ \hline
         $ (+,-,+,-,)_{(VI)} $
        & $ -[13][23] \langle 46 \rangle ^2
                 \langle 5 \vert U \vert 2 \rangle ^2 $
        & $ {}[15] \langle 24 \rangle \langle 5 \vert V \vert 4 \rangle
                 \langle 3 \vert V \vert 6 \rangle ^2 $
        & $ {}[35]^2 \langle 16 \rangle  \langle 26 \rangle
                 \langle 1 \vert W \vert 4 \rangle ^2 $
\\ \hline\hline
        \end{tabular}
        \caption{The numerator functions $P_i$ for
        $m(\bar{q}^+_1,q^-_2,g_3,g_4,g_5,g_6)$.
        The left column contains
        the helicity orderings of the gluons we define
         $\langle I \vert K \vert J \rangle           
        \equiv \langle I+ \vert K \cdot \gamma \vert J+ \rangle$.}
\vspace{0.5in}
        \end{center}
\end{table}    }      
 
\newpage
{\renewcommand{\arraystretch}{1.8}
\begin{table}[htb]
        \label{qtable2}
        \begin{center}
{\small
        \begin{tabular}{|l||c|} \hline\hline
        {$(g_3,g_4,g_5,g_6)$}   & $P_s$
\\ \hline\hline
         $ (-,-,+,+)_{(I)} $
        & $ \langle 23 \rangle  \langle 34 \rangle [56][61] \left(
         \langle 1 \vert U \vert 4 \rangle  \langle 2 \vert V \vert 1 \rangle
         \langle 5 \vert W \vert 2 \rangle
        - S_{56}[23] \langle 34 \rangle  \langle 5 \vert W \vert 2 \rangle
        - S_{34}[56] \langle 61 \rangle  \langle 1 \vert U \vert 4 \rangle
        \right) $
\\ \hline
        $ (+,+,-,-)_{(II)} $
        & $ [23][34] \langle 56 \rangle  \langle 61 \rangle
         \langle 4 \vert U \vert 2 \rangle  \langle 1 \vert V \vert 2 \rangle
         \langle 1 \vert W \vert 5 \rangle $
\\ \hline
         $ (-,+,+,-)_{(III)} $
        & $ -t_{123}[15][45] \langle 13 \rangle  \langle 26 \rangle
         \langle 4 \vert V \vert 6 \rangle  \langle 1 \vert W \vert 3 \rangle
            -t_{234}[45]^2 \langle 13 \rangle  \langle 26 \rangle
         \langle 1 \vert U \vert 6 \rangle  \langle 1 \vert W \vert 3 \rangle
        $
\\ &
        $     +t_{345}[15][45] \langle 13 \rangle  \langle 23 \rangle
        \langle 1 \vert U \vert 6 \rangle  \langle 4 \vert V \vert 6 \rangle
            +[45][56] \langle 12 \rangle  \langle 36 \rangle
         \langle 1 \vert U \vert 6 \rangle  \langle 4 \vert V \vert 6 \rangle
         \langle 1 \vert W \vert 3 \rangle  $
\\ \hline
         $ (+,-,-,+)_{(IV)} $
        & $ t_{123}[16][26] \langle 24 \rangle  \langle 45 \rangle
         \langle 3 \vert V \vert 5 \rangle  \langle 3 \vert W \vert 2 \rangle
            +t_{234}[13][26] \langle 45 \rangle ^2
         \langle 6 \vert U \vert 2 \rangle  \langle 3 \vert W \vert 2 \rangle
        $
\\ &
        $     -t_{345}[13][26] \langle 24 \rangle  \langle 45 \rangle
        \langle 6 \vert U \vert 2 \rangle  \langle 3 \vert V \vert 5 \rangle
            +[12][36] \langle 34 \rangle  \langle 45 \rangle
         \langle 6 \vert U \vert 2 \rangle  \langle 3 \vert V \vert 5 \rangle
         \langle 3 \vert W \vert 2 \rangle  $
\\ \hline
        $ (-,+,-,+)_{(V)}  $
        & $ -t_{123}[16] \langle 35 \rangle  \langle 6 \vert V \vert 3 \rangle
         \langle 4 \vert V \vert 5 \rangle  \langle 4 \vert W \vert 2 \rangle
          -t_{234}[46] \langle 35 \rangle  \langle 6 \vert V \vert 3 \rangle
         \langle 1 \vert U \vert 5 \rangle  \langle 4 \vert W \vert 2 \rangle$
\\ &
        $  +t_{345}[46] \langle 23 \rangle  \langle 6 \vert V \vert 3 \rangle
         \langle 1 \vert U \vert 5 \rangle \langle 4 \vert V \vert 5 \rangle
        -[46][56] \langle 34 \rangle  \langle 35 \rangle
        \langle 1 \vert U \vert 5 \rangle
         \langle 4 \vert V \vert 5 \rangle  \langle 4 \vert W \vert 2 \rangle
         $
\\ \hline
        $ (+,-,+,-)_{(VI)} $
        & $ [12][23][15][35] \langle 14 \rangle  \langle 24 \rangle
             \langle 26 \rangle  \langle 56 \rangle
             \langle 5 \vert U \vert 2 \rangle
          +  (S_{12}S_{23}-S_{12}S_{45})[15][35]
              \langle 24 \rangle  \langle 46 \rangle
               \langle 3 \vert V \vert 6 \rangle
$ \\ & $
          -  S_{23}S_{16}[35]^2 \langle 26 \rangle  \langle 46 \rangle
              \langle 1 \vert W \vert 4 \rangle
          -  [15]^2[23][34] \langle 12 \rangle  \langle 16 \rangle
              \langle 24 \rangle  \langle 46 \rangle
              \langle 1 \vert W \vert 4 \rangle
$ \\ & $
          +  S_{23}S_{16}[15][35] \langle 46 \rangle ^2
              \langle 3 \vert 1+5 \vert 2 \rangle
          +  S_{12}S_{15}[15][23][35]
             \langle 24 \rangle  \langle 26 \rangle  \langle 46 \rangle $
\\ \hline\hline
        \end{tabular}
			}
        \caption{The numerator functions $P_s$ for
         $m(\bar{q}^+_1,q^-_2,g_3,g_4,g_5,g_6)$ with the same
        notation as Table~1.}
\vspace{0.5in}             
        \end{center}
\end{table}  }        
 
The squaring of the amplitudes is independent of the particular helicity
configurations, and the explicit formulas for leading and sub-leading terms are
given in Appendix~D. The same consideration concerning the collinear finiteness
of the sub-leading piece made before for the helicities $(--++++)$ also holds
here. 

{\renewcommand{\arraystretch}{1.8}
    \begin{table}[htb]  \label{gtable}
        \begin{center}           
        \begin{tabular}{|l||c|c|c|} \hline\hline
      &  $1^+2^+3^+4^-5^-6^-$ & $1^+2^+3^-4^+5^-6^-$ &
        $1^+2^-3^+4^-5^+6^-$ \\
         & $X=p_1+p_2+p_3$& $Y=p_1+p_2+p_4$&
         $Z=p_1+p_3+p_5$ \\ \hline\hline
        { $\bf \alpha$} &
        0 &
        $-[12]\langle 56 \rangle \langle 4 \vert Y \vert 3 \rangle $&
        $[13]\langle 46 \rangle \langle 5 \vert  Z \vert 2 \rangle $
         \\ \hline
        {$\bf \beta$ } &
        $[23]\langle 56 \rangle \langle 1 \vert X \vert 4 \rangle$ &
        $[24]\langle 56 \rangle \langle 1 \vert Y \vert 3 \rangle $&
        $[51]\langle 24 \rangle \langle 3 \vert Z \vert 6  \rangle$ \\ \hline
 
        {$\bf \gamma$} &
        $[12]\langle 45 \rangle \langle 3 \vert  X \vert 6 \rangle $ &
        $[12]\langle 35 \rangle \langle 4 \vert  Y \vert 6 \rangle$ &
        $[35]\langle 62 \rangle \langle 1 \vert  Z \vert 4 \rangle $ \\
        \hline\hline
        \end{tabular}
        \caption{Coefficients for the
        $m_{3+3-}(g_1,g_2,g_3,g_4,g_5,g_6)$ sub-amplitudes. We define
         $\langle I \vert K \vert J \rangle                          
        \equiv \langle I+ \vert K \cdot \gamma \vert J+ \rangle$}
\vspace{0.5in}
        \end{center}
\end{table}   }
 


\newpage
\section{Factorization Properties of Dual Amplitudes}
One of the most important properties of the dual amplitudes, which partly
accounts for the relative simplicity of their explicit expressions, is their
factorizability on multi-particle poles. The residues at these poles are
determined by unitarity, and can be expressed in terms of dual amplitudes for
processes with a smaller number of external particles. The possibility of
factorizing these amplitudes into products of amplitudes and near-the-pole
propagators, puts such severe constraints on the amplitudes themselves that
often it is possible to deduce their explicit form by just imposing unitarity
and Lorentz invariance. Subtle cancellations which usually are made explicit
only at the matrix element square level for the full amplitude, here are made
manifest at the matrix element level for each single dual amplitude. From the
technical point of view, the constraints imposed by factorizability provide
furthermore a powerful check all along the way while performing complex
calculations.                                                     

A very simple and instructive way to prove these factorization properties
\cite{uppsala} is by using the Koba-Nielsen representation for the amplitudes
\cite{koba,schwarz}. While this representation may not be too helpful in
carrying out explicit calculations\footnote{
The calculation of the five gluon amplitudes has however been carried out
explicitly using the Koba-Nielsen representation \cite{mpx88,dk88}.},
this compact symbolic representation                                
provides a powerful tool for deriving general properties of the amplitudes. 
It was used independently by Lipatov in Ref. \cite{lipatov88a} to study the
emission of soft gluons and gravitons, in Ref. \cite{lipatov88b} to study the
production of gluons in tachyon-tachyon scattering and by Fadin and Lipatov 
in Ref. \cite{fadin89} to describe multi-gluon production in a
quasi-multi-Regge kinematics, in which all the pairs of final state paticles
except one have large invariant mass and fixed transverse momentum.
                                                                   
The                        
following factorization properties can also be proved in a simple and effective
way \cite{bg89} by using the recursive relations introduced by Berends and
Giele \cite{bg88}  and reviewed here in the following Section 
(see also Ref.\cite{dk89} for a derivation of the recursive relations using the
Koba-Nielsen representation of the amplitudes. and further applications of this
approach).

For the sake of definiteness we will deal in this Section with gluonic
amplitudes only. As was mentioned previously, an $n$-gluon dual amplitude can be
represented by considering the terms with lowest momentum dimensionality 
($[P]^{4-n}$) in the expansion of the following expression\footnote{For
simplicity in this Section we will omit the coupling constant.}:
\ba   \nn                                                
	m_n(k_1,\epsilon_1;k_2,\epsilon_2;\dots;k_n,\epsilon_n) &=&
	\int_{z_1<z_2<\dots<z_n} \; \prod_{i=3}^{n-1} \d{z_i}            
	\; \mu_{KN} \; \prod_{n \ge i > j \ge 1} \; (z_i-z_j)^{k_i k_j} \;
	\ret
	\label{kneq}
	&& 
	\exp \left\{ \sum_{i \ne j} \frac 12 \,
	\frac{\epsilon_i\epsilon_j}{(z_i-z_j)^2} \;+\;
	\frac{k_i \epsilon_j}{(z_i-z_j)} \right\} , 
\ea
where  $\mu_{KN}=(z_2-z_1)(z_n-z_1)(z_n-z_2)$ is the measure that makes the
integral invariant under Moebius transformations. The values of $z_{1,2,n}$ can
be chosen arbitrarily, but usually as follows: $z_1=0$, $z_2=1$ and
$z_n=\infty$.  The gluon amplitude is given by the terms in the expansion of
the Koba-Nielsen expression which are multi-linear in the polarization vectors
$\epsilon_i$. 

The singularities of the matrix elements arise from the regions of integration
where two or more $z$'s coalesce. This follows easily from Eq.\eqnum{kneq}; for
example, it is easy to check that poles like $1/(k_i+k_j+\dots+k_l)^2$ arise
from the region of integration $z_i \sim z_j \sim \dots \sim z_l$.  From this
it follows that for a given dual amplitude, represented by a determined
permutation of the indices, the only singularities that can appear are
multi-particle poles in which the indices of the momenta have to appear
consequently within the given permutation.                                 

\subsection{Soft Gluon Factorization}
Let us start from the simplest kind of singularities, \ie those due to the
emission of soft gluons. We want to show that when one of the gluons becomes
soft (\ie $p\to 0$) the dual amplitude can be written as the product of a dual
amplitude describing the process involving the remaining gluons times an
overall factor. 

Let us introduce the following conventions: we will indicate with $w$ the 
$z$ coordinate of the soft gluon, with $p$ its momentum 
and with $\zeta$ its polarization. We will
take the permutation in which the soft gluon is, by convention, inserted
between gluon 1 and gluon 2. We will furthermore fix the values of $z_{1,2,n}$
as given above, and therefore will integrate the soft gluon 'coordinate' $w$ in
the range $z_1=0 \le w \le z_2=1$.  

It is then simple to prove that in the $p\to 0$ approximation the Koba-Nielsen
formula becomes:
\ba \nn
        &&
    	m_{n+1}(k_1,\epsilon_1;p,\zeta;k_2,\epsilon_2;\dots;k_n,\epsilon_n) \=
	\int_0^1 \d{w} \; w^{pk_1}(1-w)^{pk_2}   
	\int_{z_1<w<z_2<\dots<z_n} \;  \bar Z  \; \bar E  \;
\ret   	\label{kneq2}
	&& 
	\exp \left\{  \frac{\epsilon_1 \zeta}{w^2} +     
	\frac{\epsilon_2 \zeta}{(1-w)^2} -
	\frac{k_1 \zeta}{w} +              
	\frac{k_2 \zeta}{(1-w)} +
	\zeta \cdot \sum_{i>2} ( \frac{\epsilon_i }{(z_i-w)^2} -
	\frac{k_i }{(z_i-w)}) \right\} ,                        
\ea                          
where	
\ba  
   	\bar Z &=& 	\prod_{i=3}^{n-1} \d{z_i}                    
	\; \mu_{KN} \; \prod_{n \ge i > j \ge 1} \; (z_i-z_j)^{k_i k_j} \;
\ret
	\bar E &=&
	\exp \left\{ \sum_{i \ne j} \frac 12 \,
	\frac{\epsilon_i\epsilon_j}{(z_i-z_j)^2} \;+\;
	\frac{k_i \epsilon_j}{(z_i-z_j)} \right\} . 
\ea                                                
The momentum $p$ was kept only in those terms which can give rise to
singularities. By expanding at the linear level in the polarizations, we will
now find integrals in $w$ of the following form:
\ba    \nn
	I(a,b) &=& \int_0^1 \d{w} \; w^{pk_1} (1-w)^{pk_2}\; w^{-a} (1-w)^{-b}
\ret                                                                     
	&=& \frac{\Gamma(-a+1+pk_1)\Gamma(-b+1+pk_2)}{\Gamma(pk_1+pk_2+2-a-b)}.
\ea
where the pair $a,b$ can take the following values: $(a=0,1,2 ; b=0)$ or 
$(a=0 ; b=0,1,2)$. 
The only integrals which give the leading infrared singularities are 
$I(1,0)$ and $I(0,1)$, which in the soft limit behave, respectively, like 
$1/pk_1$ and $1/pk_2$. Therefore the dual amplitude corresponding to the
emission of a soft gluon takes the following form:
\ba   \nn              
    	m_{n+1}(k_1,\epsilon_1;p,\zeta;k_2,\epsilon_2;\dots;k_n,\epsilon_n) 
	&=&
	\left[ \zeta \cdot (\frac{k_2}{pk_2} - \frac{k_1}{pk_1}) \right]
    	m_{n}(k_1,\epsilon_1;k_2,\epsilon_2;\dots;k_n,\epsilon_n)      
\ret  \label{softfact}
	& \equiv &  \zeta \cdot j_{eik} ~m_{n},
\ea                                
where $j_{eik}$ is the classical gauge invariant eikonal current. Because of
gauge invariance, we can use the spinorial representation of the polarization
$\zeta$ with an arbitrary reference momentum -- say $k_2$. For a
positive-helicity soft gluon, we find the following result:
\be
	\zeta \cdot j_{eik} \= \frac{ \lsp{p\,} k_2 \rsp{k_1} }{\sqrt 2 \;
	\sp{k_1}{p} (pk_2) } \=  \sqrt 2 \frac{\sp 12}{\sp 1p \sp p2},
\ee                                                                     
which is the square root of the usual eikonal factor:
\be
     \vert \zeta \cdot j_{eik} \vert ^2 \=  \frac{(12)}{(p1)(p2)}.
\ee
For the emission of a negative-helicity gluon we just have to change 
the \sp{i}{j}\  products with \cp{j}{i}\ products.
The factorization of the sub-amplitude, Eq.\eqnum{softfact}, does not imply the
eikonalization of the full matrix element, as is the case in QED, because of the
convolution with the color Chan-Paton factors: the interference of gluons
in the non-abelian theory persists in the soft limit.  Repeated applications of 
Eq.\eqnum{softfact}\ lead to the multi-gluon amplitudes introduced in
\cite{pino}. The properties of the sub-amplitudes in presence of soft-gluon
emission were also studied in detail by Berends and Giele in Ref.\cite{bg89}.
Here expressions were given for the case of multiple soft emission, in the 
case were soft gluons are strongly ordered in energy $(E_1>>E_2>>\dots)$, and
in the case in which energies are not strongly ordered $(E_1 \sim E_2 \dots
\sim E_k << E_{k+1} \dots)$. We refer to that paper for the details. 
                              
\subsection{Factorization of Collinear Poles}
In a similar fashion one can analyze the factorization properties of the
amplitudes near a collinear singularity by studying the residues of the
appropriate poles in the  Koba-Nielsen variables. To this end, we will assume
that the collinear pair is formed by the first two gluons, and will label the
variables in the following fashion: the first two gluons will have momenta
$p_1$ and $p_2$, respectively, and polarizations $\zeta_{1,2}$. Their
Koba-Nielsen variables will be $w_1$ and $w_2$. For the remaining $n$ gluons we
will use the notation $k_i$, $\epsilon_i$ and $z_i$ for momentum, polarization
and Koba-Nielsen coordinate, respectively. Furthermore we will fix the range of
the Koba-Nielsen integration as follows:
\be
	w_1=0 \; < \; w_2=w \; < \; z_1=1 \; < \; \dots \; < \; z_n \to \infty.
\ee
The collinear singularity -- $1/(p_1p_2)$ -- will arise from the region 
$w \to 0$. To isolate the leading contributions, therefore, we will expand the
KN integral in a Laurent series in $w$, keeping only the singular  part:
\ba   \nn                                
        &
    	m_{n+2}(p_1,p_2,k_1,\dots,k_n) \=
	\int_0^1 \d{w} \; w^{p_1p_2} \prod_{i=1}^n z_i^{Pk_i} \;
	\left ( 1 - w \sum \frac{p_2k_i}{z_i} \right ) \;
        \exp (-P \sum \frac{\epsilon_i}{z_i})
\ret                                       
	& 
	\int_{i=2,n-1} \d{z_i} \;  \bar Z  \bar E  \;
	\left\{ \frac{\zeta_1\zeta_2}{w^2} \
	- \frac{1}{w} \left[ (\zeta_1\zeta_2) p_2^\mu +
	(p_1\zeta_2)\zeta_1^\mu - (p_2\zeta_1)\zeta_2^\mu \right]
	      \sum                                                
	(\frac{\epsilon_i^\mu}{z_i^2} + \frac{k_i^\mu}{z_i} )
        \right\}  ,                    
\ea                          
where $\bar E$ and $\bar Z$ were defined above, and where $P=p_1+p_2$. We left
out terms like $(p_1\zeta_2)(p_2\zeta_1)$ because they have higher dimension
(\ie they would disappear in the zero slope limit, in the string theory
language).  The integrals in $w$ can be regularized by introducing a factor
$(1-w)^\epsilon$, which allows them to be defined in terms of Euler functions
by analytic continuation, and then taking the $\epsilon \to 0$ limit. In this
way only the integral in $w^{(p_1p_2 -1)}$ contributes to the leading behaviour.

By performing the integrations and keeping only the leading terms, we
obtain the following expression:
\ba   \nn
        &
    	m_{n+2}(p_1,p_2,k_1,\dots,k_n) \=
	\int \prod_{i=2,n-1} \d{z_i} \;  \bar Z  \bar E  \;
	\prod_{i=1}^n z_i^{Pk_i} \;
        \frac{-1}{2(p_1p_2)} \;
        \exp (-P \sum \frac{\epsilon_i}{z_i})
\ret                                       
	& 
	\left\{ 
	(\zeta_1\zeta_2) P_\mu               
	      \sum                                                
	(\frac{\epsilon_i^\mu}{z_i^2} + \frac{k_i^\mu}{z_i} )
	+ \left[ (\zeta_1\zeta_2) Q^\mu +
	2(p_1\zeta_2)\zeta_1^\mu - 2(p_2\zeta_1)\zeta_2^\mu \right]
	      \sum                                                
	(\frac{\epsilon_i^\mu}{z_i^2} + \frac{k_i^\mu}{z_i} )
        \right\}  ,
\ea                          
where $Q=p_2-p_1$.  The term proportional to $	(\zeta_1\zeta_2) P_\mu$ 
corresponds to the coupling of a gluon with polarization proportional to its
momentum. By gauge invariance, after we integrate over the remaining
Koba-Nielsen variables it will be proportional to $P^2$, and will only
contribute to finite terms, so in the leading pole approximation we can drop
it.  What is left can be written in the following way:
\be
    	m_{n+2}(p_1,p_2,k_1,\dots,k_n) \=
        \frac{1}{2(p_1p_2)} \;              
	V_\mu
    	\frac{\partial}{\partial \zeta_\mu} m_{n+1}(P,k_1,\dots,k_n) ,
\ee                                                                   
where :                                                               
\be
	V_\mu \= 	\left[ (\zeta_1\zeta_2) Q^\mu +
	2(p_1\zeta_2)\zeta_1^\mu - 2(p_2\zeta_1)\zeta_2^\mu \right]
\ee                                                                 
is the usual three-gluon vertex,
and $\zeta$ is an 'auxiliary' polarization assigned to the gluon of momentum
$P$.

If we select an explicit representation for the helicities, and reintroduce the
coupling constant (using the normalization conventions given in the Appendix)
we obtain the following relations:   
\begin{eqnarray}
m(1^+,2^+,3, \dots ) 
& \stackrel{1^+~\parallel~2^+}{\longrightarrow} &
\left\lbrace {{i g   ~[12]} \over \sqrt{z(1-z)} }
\right\rbrace ~{i \over s_{12} }~ m(P^+,3, \dots ) \\
m(1^+,2^-,3, \dots ) 
& \stackrel{1^+~\parallel~2^-}{\longrightarrow} &
\left\lbrace {{-i g   ~
 z^2 \langle 12 \rangle} \over \sqrt{z(1-z)} } \right\rbrace
~{i \over s_{12} }~ m(P^+,3, \dots ) \\ \nonumber
& & ~~~~~+~ \left\lbrace 
{{ i g  ~ (1-z)^2~[12]} \over \sqrt{z(1-z)} }
\right\rbrace ~{i \over s_{12} }~ m(P^-,3, \dots ) \\
m(1^-,2^-,3, \dots ) 
& \stackrel{1^-~\parallel~2^-}{\longrightarrow} &
\left\lbrace {{ -i g  ~ \langle 12 
\rangle}          
 \over \sqrt{z(1-z)} } \right\rbrace
~{i \over  s_{12} }~ m(P^-,3, \dots ),
\end{eqnarray}                         
where $z$ is the momentum fraction carried by the first gluons.
One can easily check that all of the subamplitudes given explicitly in the
previous Section do satisfy these relations in the collinear limit.
  
This Equation shows the collinear factorization of the kinematical part of the
dual amplitude. As for the color part, factorization can be easily verified by
noticing that in the collinear approximation:                
\be                                                          
    	m_{n+2}(p_1,p_2,k_1,\dots,k_n) \=
    	-m_{n+2}(p_2,p_1,k_1,\dots,k_n) 
\ee
and that 
\be
	\tr (\l_1 \l_2 \dots \l_n) - \tr (\l_2 \l_1 \dots \l_n) =
	i  f_{12c} \; \tr (\l_c \dots \l_n),
\ee                                                   
which is in fact the product of the color factor of the three-gluon vertex
times the color factor of an $(n-1)$-gluon dual amplitude.

The general factorization properties of the gluon subamplitudes are
given by
\ba
m(1,2,\dots,n) 
& \stackrel{P^2 \rightarrow 0}{\longrightarrow} &
\sum_{\lambda=\pm} 
m(1,2,\dots,k,-P^{\lambda}) ~{i \over P^2} 
~m(P^{-\lambda},k+1,\dots,n) 
\ea
where $P=\sum_{i=1}^{k}~p_i$. Of course the full amplitude, including the 
color factor, must factorize. But the color factors introduced in 
section~\sectccf only factorize to leading order in the number of 
colors,
that is,
\be
tr(\lambda^1 \lambda^2 \cdots \lambda^n) = \sum_{a_x}
tr(\lambda^1 \cdots \lambda^k \lambda^{a_x}) 
tr(\lambda^{a_x} \lambda^{k+1} \cdots \lambda^n) 
~+~ {1 \over N} 
tr(\lambda^1 \lambda^2 \cdots \lambda^k) 
tr(\lambda^{k+1} \cdots \lambda^n) .
\ee
However, the 1/N terms in the full amplitude cancel at the pole because of the 
Dual Ward Identities for the gluon subamplitudes.

Similar factorization properties also exist for 
subamplitudes involving quark-antiquark pairs.


\newpage
\section{Recursive Relations}

The color structure for purely gluonic and processes involving gluons 
and a quark-antiquark pair defined in previous sections allows for the 
reorganization of the perturbation theory in a efficient and straight 
forward manner. The building blocks are color ordered vector and 
spinorial currents defined with a gluon off mass shell, or a quark or 
antiquark off mass shell, with all other particles on mass shell.
If you have calculated these building blocks for $n$ on mass shell legs 
then there are recursion relationships, the Berends-Giele recursion 
relations, ref.\cite{bg88}, 
which allow you to simply evaluate these currents with
$(n+1)$ on mass shell legs. This allows for computer evaluation of
processes with a large number of external particles\scite{bgk89b}.
A detailed and self-contained description of the use of recursive relations in
the calculation of multi-parton processes can be found in Giele's thesis
\cite{gielephd}.                                           

\subsection{Color Ordered Gluon Currents}

From the set of color truncated 
Feynman diagrams that make up the subamplitude,
\amp, one can form a color ordered gluonic current  by replacing
the polarization vector of the $n-th$ gluon with the propagator 
and allowing the momentum of 
this gluon to be off mass shell but still retain momentum 
conservation. This color ordered gluonic current will be represented
by Fig. \sfigone, where the dotted line represents the gluon which is
off mass shell. 
This current will be written as \gJ{\mu}{1}{n-1} and the subamplitude
can be reconstructed from this current by multiplying by the inverse 
propagator and contracting with a 
polarization vector and allowing the momentum of this gluon to be on 
mass shell,

\be
	\amp ~= \lbrace \epsilon^{\mu}(p_n)
		~i[P(1,n-1)]^2
		~\gJ{\mu}{1}{n-1} \rbrace ~|_{P(1,n-1)=-p_n} ,
\ee
where, $P(1,n) \equiv \sum_1^n p_i$.

Of course these currents, $J_{\mu}$,
are not gauge invariant and do depend on the 
choice of reference momenta chosen for the $(n-1)$ on mass shell 
gluons. Also they depend on the helicity of the on mass shell gluons.
However these color ordered gluonic currents can be used
as building blocks for gluonic currents with more external on mass
shell legs.
	
Consider a gluonic current with $n$ on mass shell gluons.
Then the off mass shell gluon is attached to the
rest of the gluons either through a three or a four point color
ordered gluon 
coupling. At these vertices the other legs are attached to
color ordered gluonic currents with fewer 
than $n$ on mass shell gluons. This can be seen diagrammatically in 
Fig. \sfigtwo .
Hence, the color ordered gluonic current with 
$n$ on mass shell gluons can be written in terms of gluonic currents 
with less than $n$  on mass shell gluons. This is the Berends-Giele
recursion relation\scite{bg88}
 for gluonic color ordered currents and algebraically
it is written as

\ba
\gJ{\mu}{1}{n} &=&  {-i \over P(1,n)^2}
\lbrace
\sum_{i=1}^{n-1} V3^{\mu\nu\rho}(P(1,i),P(i+1,n))
~\gJ{\nu}{1}{i}~\gJ{\rho}{i+1}{n} \nonumber \\  & & 
+ ~\sum_{j=i+1}^{n-1}  ~\sum_{i=1}^{n-2} 
V4^{\mu\nu\rho\sigma}
~~\gJ{\nu}{1}{i}~\gJ{\rho}{i+1}{j}~\gJ{\sigma}{j+1}{n}
\rbrace
\ea

where the color ordered three and four gluon vertices are, see 
Appendix~C,

\ba
V3^{\mu\nu\rho}(P,Q) & = & i\frac{g}{\sqrt 2}~(g^{\nu \rho}~(P-Q)^{\mu} 
~+~ 2g^{\rho \mu}~Q^{\nu} ~-~ 2g^{\mu \nu}~P^{\rho}),
\nonumber \\
V4^{\mu\nu\rho\sigma} & = &
i\frac{g^2}{2}~
(2g^{\mu \rho}~g^{\nu \sigma} ~-~ g^{\mu \nu}~g^{\rho \sigma} ~-~ 
g^{\mu \sigma}~g^{\nu \rho})  .
\ea

The current with one on mass shell gluon is defined as
\be
J_{\mu}(1) ~\equiv~ \epsilon_{\mu}(p_1).
\ee

The gluonic currents, \gJ{\mu}{1}{n},
satisfy properties that are similar to the
gluon subamplitude, \amp.

\begin{enumerate}
\item Dual Ward identity:
\be
J_{\mu}(1,2,3,\dots,n) ~+~ J_{\mu}(2,1,3,\dots,n) \cdots ~+~ 
J_{\mu}(2,3,\dots,n,1) ~=0.
\ee
\item Reflectivity:
\be
\gJ{\mu}{1}{n} ~=~ (-1)^{n+1} ~\gJ{\mu}{n}{1}
\ee
\item
\gJ{\mu}{1}{n} is conserved:
\be
P(1,n)^{\mu}~\gJ{\mu}{1}{n} ~=~ 0
\ee
\end{enumerate}

There are simple analytical expressions for the color ordered gluonic 
currents if all the helicities are the same or if one is different 
from the others. Of course we must define the reference momentum for 
the gluons. Here the symbol $i$ for the gluons must be expanded to 
$i^{\lambda}_k$ where the $i$-th gluon has helicity $\lambda$ and reference 
light-like momentum $k$. Then

\be
J_{\mu}(1^+_k,2^+_k,\dots,n^+_k) ~=~ g^{n-1}
{ { \lsp{k-}~\gamma_{\mu}~\hat{P}(1,n)~\rsp{k+}}
\over
{\sqrt{2}~\sp{k}{1}~\sp{1}{2}\cdots\sp{n-1}{n}~\sp{n}{k}}  }
\ee

and 

\be
J_{\mu}(1^-_k,2^-_k,\dots,n^-_k) ~=~ (-1)^n ~g^{n-1}
{ { \lsp{k+}~\gamma_{\mu}~\hat{P}(1,n)~\rsp{k-}}
\over
{\sqrt{2}~\cp{k}{1}~\cp{1}{2}\cdots\cp{n-1}{n}~\cp{n}{k}}  }.
\ee

Berends and Giele, ref.\cite{bg88}, give compact expressions for 
$J_{\mu}(1^{\mp},2^{\pm},\dots,n^{\pm})$ for a given choice of 
reference momenta. Also, Kosower, ref.\cite{dk89},
has given a light-cone formulation of these recursion relation
to derive the sub-amplitudes $m(-,-,-,+,\cdots,+)$.
Kleiss and Kuijf, Ref.\cite{kk89}, have used these recursion relations 
to calculate the 7-gluon amplitudes numerically.

\subsection{Color Ordered Quark Currents}

For the subamplitudes involving a quark-antiquark pair and gluons
one can define a Quark and Antiquark color ordered spinorial 
current, see Fig. \sfigthree, in a way similar to 
the gluon currents that were defined in the last section.
We will write the Quark current as $\qJ{\Q}{1}{n}$ and the Antiquark
current as $\aJ{1}{n}{\AQ}$. 
The quark-antiquark pair
plus gluon subamplitudes can be obtained from these currents as follows:
\ba
m(\Q,1,\dots,n,\AQ) & & =  \lsp{\Q~} (+i)(\hat{\AQ}+\hat{P}(1,n))
\aJ{1}{n}{\AQ}~|_{\AQ+P(1,n)=-\Q} \nonumber \\[0.2in]
		      & & =  \qJ{\Q}{1}{n}
(-i)(\hat{\Q}+\hat{P}(1,n)) \rsp{\AQ}~|_{\Q+P(1,n)=-\AQ} 
\ea

In manner similar to the gluon current, a recursion relation can be written 
for this color ordered Quark current \cite{bg88}, see Fig. \sfigfour 
and Appendix~C,
\be                                                                 
\label{qrec}
\qJ{\Q}{1}{n} ~=~ \sum_{m=0}^{n-1} \qJ{\Q}{1}{m} \frac{ig}{\sqrt 2}
\gamma^{\mu} 
\gJ{\mu}{m+1}{n} {i \over (\hat{\Q}~+~\hat{P}(1,n)) }
\ee
and for the Anti-quark current\scite{bg88}
\be
\label{aqrec}
\aJ{1}{n}{\AQ}= \sum_{m=1}^{n} 
{-i \over (\hat{\AQ}+\hat{P}(1,n)) }
\frac{ig}{\sqrt 2} \gamma^{\mu} \gJ{\mu}{1}{m}
\aJ{m+1}{n}{\AQ}
\ee
and
where the spinor currents for the zero gluon case are defined to be
\be 
	\overline{U}(\Q)~\equiv~\overline{u}(\Q), \quad \quad   \quad
	V(\AQ)~\equiv~v(\AQ)
\ee
in Bjorken and Drell notation.

These color ordered spinor currents can be defined for massive or 
massless quarks.
For massive quarks the propagators in the recursion relations 
Eqs. (\ref{qrec},\ref{aqrec})
must be modified by adding the appropriate mass term. 
For massless quarks these spinor currents carry a chirality such that
\be 
	(1~\pm~\gamma_5)~\aJ{1}{n}{\AQ^{\pm}}~=0,
\quad \quad \quad 
	\qJ{\Q^{\pm}}{1}{n}(1~\pm~\gamma_5)~=0.
\ee
Also for the massless case the zero gluon currents are simply
\be
	\overline{U}(\Q^{\pm})~\equiv~\lsp{\Q\pm}, \quad \quad \quad 
         V(\AQ^{\pm})~\equiv~\rsp{\AQ\mp}.
\ee

Again there are simple analytic expressions for these color ordered 
spinor currents when all the gluons have the same helicity as the 
fermion,

\be
\qJ{\Q^+}{1^+_k}{n^+_k} ~=~
-g^n~
{\lsp{k-}(\hat{\Q}+\hat{P}(1,n)) \over \sp{\Q}{1} \sp{1}{2} \cdots 
\sp{n}{k}  },
\ee

\be
\qJ{\Q^-}{1^-_k}{n^-_k} ~=~
-(-g)^n ~
{\lsp{k+}(\hat{\Q}+\hat{P}(1,n)) \over \cp{\Q}{1} \cp{1}{2} \cdots 
\cp{n}{k}  },
\ee

\be
\aJ{1^+_k}{n^+_k}{\AQ^+} ~=~
-g^n~
{ (\hat{\AQ}+\hat{P}(1,n))\rsp{k+}
\over
\sp{k}{1} \sp{1}{2} \cdots \sp{n}{\AQ} },
\ee
and
\be
\aJ{1^-_k}{n^-_k}{\AQ^-} ~=~
-(-g)^n~
{ (\hat{\AQ}+\hat{P}(1,n))\rsp{k-}
\over
\cp{k}{1} \cp{1}{2} \cdots \cp{n}{\AQ} }.
\ee

If there is one gluon with 
opposite helicity to that of the fermion, the spinorial currents are

\be
\overline{U}(\Q^+,1^-_k) ~=~
-g~ { {\cp{\Q}{k} \lsp{\Q+}} \over {\cp{\Q}{1} \cp{1}{k} }},
\ee

\be
\overline{U}(\Q^-,1^+_k) ~=~
g~ { {\sp{\Q}{k} \lsp{\Q-}} \over {\sp{\Q}{1} \sp{1}{k} }},
\ee

\be
V(1^-_k,\AQ^+) ~=~
-g~
{ { \rsp{\AQ-} \cp{k}{\AQ} }
\over
{\cp{k}{1} \cp{1}{\AQ} } }
\ee

and
\be
V(1^+_k,\AQ^-) ~=~
g~
{ { \rsp{\AQ+} \sp{k}{\AQ} }
\over
{\sp{k}{1} \sp{1}{\AQ} } }.
\ee

Finally, for two gluons with opposite helicity, we have the following 
spinorial currents,
\be
\overline{U}(\Q^-,1^+_2,2^-_1) ~=~ { -g^2 ~\sp{\Q}{2}^2 
\over \sp{\Q}{1} S_{12} (\Q+1+2)^2}  \lsp{1+}
(\hat{\Q}+\hat{1}+\hat{2}),
\ee

\be
\overline{U}(\Q^-,1^-_2,2^+_1) ~=~ { g^2 ~\cp{2}{\Q} \sp{\Q}{1} 
\over \cp{\Q}{1} S_{12} (\Q+1+2)^2}  \lsp{2+}
(\hat{\Q}+\hat{1}+\hat{2}),
\ee

\be
V(1^+_2,2^-_1,\AQ^+) ~=~ (\hat{1}+\hat{2}+\hat{\AQ})
 \rsp{2+} { -g^2 ~\cp{1}{\AQ}^2
\over  (1+2+\AQ)^2 S_{12} \cp{2}{\AQ} 
}
\ee

and
\be
V(1^-_2,2^+_1,\AQ^+) ~=~ (\hat{1}+\hat{2}+\hat{\AQ})
 \rsp{1+} { g^2 ~\cp{2}{\AQ}
\sp{\AQ}{1} \over  (1+2+\AQ)^2 
S_{12} \sp{2}{\AQ} }.
\ee

A straight forward example using these currents 
is to calculate the sub-amplitude
for $(\Q^-,1^+,2^-,\AQ^+)$ process,
\ba
m(\Q^-,1^+,2^-,\AQ^+)  
&=&
\overline{U}(\Q^-)
~i(\hat{1}+\hat{2}+\hat{\AQ})
~ V(1^+_2,2^-_1,\AQ^+) 
\nonumber \\
&=& { ig^2 ~\sp{\Q}{2}^3 \sp{\AQ}{2} \over
\sp{\Q}{1} \sp{1}{2} \sp{2}{\AQ} \sp{\AQ}{\Q} },
\ea
which is the previously obtained result. The spinorial currents 
defined here can be used to derive many of the results of other 
section, especially the section involving multiple gauge groups.


\newpage
\section{Exact Results for n-Parton Amplitudes}
The interest in exact matrix elements for $n$-parton processes in QCD 
was started in 1986 by Parke and Taylor \cite{pt86b} who realized that 
certain non-trivial helicity amplitudes in pure Yang-Mills 
theory
could be written in a simple closed form. These same helicity 
amplitudes have been extended to processes including 
quarks,in Ref.\cite{uppsala,mp88}, 
and Vector Bosons, in Ref.\cite{bg88,mlm88}, and are also used as 
the starting point for approximation schemes, in 
Ref.\cite{ks88,maxwell87}, for 
processes involving a large number of partons. These Generalized Parke 
and Taylor amplitudes are the subject of this section.
                                                      
The Berends and Giele recursive relations presented in the previous Section are
an extremely powerful tool to derive expressions for processes with a large
number of partons. They have been used up to now to obtain the amplitudes for
7- and 8-gluon processes \cite{bgk89b,bgk89c} and for processes with a
color-singlet vector boson and up to 5 colored partons \cite{bgk89a}, in this
case confirming the results independently obtained  by Hagiwara and Zeppenfeld
in Ref.\cite{zep89}.  Unfortunately most of the resulting formulae are very
complex and hard to interpret. It is very interesting, however, that for some
special helicity configurations the matrix elements conserve a very simple
universal structure independently of the number of particles involved. 

In this Section we present a collection of exact results which hold for a
specific set of helicity amplitudes with an arbitrary number of particles. The
interest in these expressions is not just academic, as these results can be
used as the starting point for the development of very powerful approximation
techniques that will be described in a following Section. Furthermore they 
help clarifying the dynamical features of hard multi-particle processes and
shed additional  light on the structure of quantum coherence in the radiation
of abelian and non-abelian radiation, as we will discuss.
                                                         
\subsection{Helicity Violating Amplitudes} 
Let us consider processes with the following
helicity structure: 
\be                         
	(-,-,+\dots,+) \quad \mbox{or} \quad  (+,+,-,\dots,-).
\ee                                                           
We will call these helicity structures Maximally Helicity Violating (MHV)
\footnote{Amplitudes with helicity configuration
$(+,\dots,+)$ and $(-,+,\dots,+)$ are equal to zero, see Section \sectsus.}. 
It is straightforward to prove that the only singularities the corresponding
amplitudes may have are soft and collinear poles of the form $1/\sqrt (p_i
\cdot p_j)$.  Because of factorization, the residues at these poles are fixed,
and given by the product of a proper Altarelli-Parisi splitting function times
an $(n-1)$-parton MHV amplitude:
\be
	m^{(n)}(-,-,+\dots,+) \to \frac{1}{\sqrt (p_i p_j)} \sqrt {f(z)} 
	m^{(n-1)}(-,-,+\dots,+).
\ee
The A-P function that will appear depends on the helicities of the collinear
partons.

As an example, we can take the 5-gluon amplitude, Eq.~\eqnum{ggggg}:
\be                                                              
   	m(g_1^-,g_2^-,g_3^+,g_4^+,g_5^+) \=
	i \, \gs{3} \, \frac{\sp 12 ^4}{\sp 12 \sp 23 \sp 34 \sp 45 \sp 51}.
\ee                                                                       
If we study the behaviour of this sub-amplitude near the three inequivalent
poles $s_{23} \to 0$, $s_{45} \to 0$ and $s_{51} \to 0$, we obtain the
following factorization relations:
\ba     \label{pole23}
   	m^{(5)} & \to &
	\frac{1}{\sp 23} \; g \; \sqrt{ \frac {z^4}{z(1-z)} } ~
         ig^2 ~ \frac {\sp 1P ^4}{\sp 1P \sp P4 \sp 45 \sp 51}    ,
\ret     \label{pole45}
   	m^{(5)} & \to &
	\frac{1}{\sp 45} \; g \; \sqrt{ \frac {1}{z(1-z)} }  ~
         ig^2 ~ \frac {\sp 12 ^4}{\sp 12 \sp 23 \sp 3P \sp P1}   ,
\ret       \label{pole51}
   	m^{(5)} & \to &
	\frac{1}{\sp 51} \; g \; \sqrt{ \frac {(1-z)^4}{z(1-z)} }    ~
         ig^2 ~ \frac {\sp P2 ^4}{\sp P2 \sp 23 \sp 34 \sp 4P}  ,
\ea                          
with :
\be
	i \= zP + {\cal O}(\sqrt {(i\cdot j)}) , \quad
	j \= (1-z)P + {\cal O}(\sqrt {(i\cdot j)}) ,
\ee
$i$ and $j$ being the two collinear momenta.
As expected, the first terms on the right-hand side of
Equations~\eqnum{pole23}-\eqnum{pole51}\ are the four-gluon \subs,
Equation~\eqnum{gggg}, and the second terms are the square roots of the
polarized A-P functions.

Lorentz invariance and the factorization properties uniquely fix the form of
the amplitudes at tree level (their squares are just rational functions), and
for the MHV amplitudes these constraints can be easily solved explicitly
with simple 'educated guesses' of what the amplitude might be.  More formally,
these amplitudes can be derived by solving the Berends and Giele recursive
relations, which for these helicity configurations turn out to be particularly
simple \cite{bg88,mpx88}.

For purely gluonic processes, the MHV amplitudes are given by the obvious
generalizations of Equations~\eqnum{gggg}\ and \eqnum{ggggg} \footnote{In this
Section we will only consider the $(-,-,+,\dots,+)$ helicities; the 
$(+,+,-,\dots,-)$ ones can be obtained by replacing \sp{i}{j}\  products with
\cp{j}{i}\  products.} :                                        
\be
        M(g^-_{1},g^-_{2},g^+_{3},\dots,g^+_{n})=
        i g^{n-2} \langle 12 \rangle ^4
        \sum_{\{1,2,\dots\}'}\;tr(\lambda_1\lambda_2\cdots\lambda_n)
        {1 \over  \langle 12 \rangle  \langle 23 \rangle \cdots
         \langle n1 \rangle },
\ee
where the sum is taken over the $(n-1)!$ non-cyclic permutations of the
indices.                                                           
It is easy to show that this Equation satisfies all the required factorization
properties and the Dual Ward Identities.

At the leading order in \Nc\ the square of these gluonic matrix elements,
summed over colors, and over all the MHV configurations, 
gives the so called Parke and Taylor Amplitudes
\cite{pt86b}:                                     
\be  \label{ptamp} 
   	\vert M(g_1,\dots,g_n) \vert ^2 \=  
     	2 \gs{2n-4} \Nc{n-2}(\Nc{2}-1) \sum_{i>j} s_{ij}^4 \;
	\sum_{\{1,2,\dots,n\}'}\; \frac{1}{s_{12} s_{23} s_{34} \dots s_{n1}}  .
\ee                               
The overall factor of 2, coming from the sum over $(--++\dots +)$ and 
$(++--\dots -)$ configurations, is clearly absent for $n=4$
                                   
Recently a universal form was found \cite{dk89} also for a specific set of
gluonic non-MHV sub-amplitudes, namely for helicity configurations of the
following  form and in the indicated order:  $(---+++\dots +)$. This form is
not as simple as the Parke and Taylor expression, therefore we will not display
it here and we refer to the original paper for the explicit result.
                                     
By using the Supersymmetry Ward Identity given in Equation~\eqnum{susymhv}\
we can now derive the MHV amplitudes for processes with a pair of (massless)
gluinos or a $q\qbar$ pair:                                             
\ba   
   &&
   M(g^-_{1},\Lambda^-_{2},\Lambda^+_{3},g^+_{4},\dots,g^+_{n}) =
       ig^{n-2}  \langle 12 \rangle ^3 \langle 13 \rangle
        \sum_{\{1,2,\dots,n\}'}\; tr(\lambda_1\lambda_2\cdots\lambda_n)
        {1 \over  \langle 12 \rangle  \langle 23 \rangle \cdots       
         \langle n1 \rangle },    \nn
  \\ && \label{gluino}
\ret   
   &&              
   M(q^-,g^-_{1},g^+_{2},\dots,g^+_{n},\qbar^+) =
       ig^{n}  
	\frac {\langle q1 \rangle ^3 \langle \qbar1 \rangle}{\sp \qbar q}
        \sum_{\{1,2,\dots,n\}}\; (\lambda_1\lambda_2\cdots\lambda_n)_{ij}     
        {1 \over  \langle q1 \rangle  \langle 12 \rangle \cdots          
         \langle n\qbar \rangle }.                       \nn
   \\  && \label{qqng}                       
\ea
We can once again square these matrix elements at the leading order in 
\Nc{}, and obtain:             
\ba  
 \vert M(g_{1},\Lambda_{2},\Lambda_{3},g_{4},\dots,g_{n}) \vert ^2 \= 
       & 	2 \gs{2n-4} \Nc{n-2}(\Nc{2}-1)                                          
       \sum_{i\ne 2,3} (s_{2i}^3s_{3i} + s_{2i}s_{3i}^3) \;
  \nn \ret \label{gluinosq} 
       & \times
	\sum_{\{1,2,\dots,n\}'}\; \frac{1}{s_{12} s_{23} s_{34} \dots s_{n1}}  ,
\ea                                                             
\ba  
   	\vert M(q,g_1,\dots,g_n,\qbar) \vert ^2 \=  
     	& 2 \gs{2n} \Nc{n-1}(\Nc{2}-1) 
	\sum_{i=1}^n (s_{qi}^3s_{\qbar i} + s_{qi}s_{\qbar i}^3) \;
   \nn \ret \label{mp} 
	& \times \frac{1}{s_{q\qbar}}
	\sum_{\{1,2,\dots,n\}}\; \frac{1}{s_{q1} s_{12} \dots s_{n\qbar}}  .
\ea                                                    
\vskip 0.5cm
Even though massless supersymmetric particles do not exist, the $m=0$
approximation might turn out to be useful if they were discovered to be
relatively light on the scale of the future hadronic supercolliders, where
their properties would be studied in detail.

Let us now take the amplitude
$M(\Lambda^+_{1},\Lambda^-_{2},\Lambda^+_{3},
g^-_{4},g^+_{5},\dots,g^+_{n})$. By commuting with the supersymmetry operator
and properly choosing the reference momentum $k$ we obtain the following SWI:
\be
        M_{\tilde g}(\Lambda^+_{1},\Lambda^+_{2},\Lambda^-_{3},\Lambda^-_{4},
        g^+_{5},\dots,g^+_{n})=
        { \langle 12 \rangle  \over  \langle 24 \rangle }
        M_{\tilde g}(g^+_{1},\Lambda^+_{2},\Lambda^-_{3},g^-_{4},g^+_{5},
                                                                \dots,g^+_{n}).
\ee
By using Equation~\eqnum{gluino} we get:
\be    \label{4gluino}
        M_{\tilde g}(\Lambda^+_{1},\Lambda^+_{2},\Lambda^-_{3},\Lambda^-_{4}
        ,g^+_{5},\dots,g^+_{n})=  i g^{n-2}
        \langle 12 \rangle  \langle 34 \rangle ^3
        \sum_{perm'}\;tr(\lambda_1\lambda_2\cdots\lambda_n)
        {1 \over  \langle 12 \rangle  \langle 23 \rangle \cdots
         \langle n1 \rangle }.
\ee
The MHV amplitude for
the scattering of gluons and a pair of massless scalar-quarks is
obtained from the SWI and
the supersymmetry transformations of a chiral superfield$^{\cite{susy1,susy2}}$:
\be
      M(\bar{\phi}^+_{1},\phi^-_{2},g^-_{3},g^+_{4},\dots,g^+_{n}) =
         i g^{n-2}\langle 23 \rangle ^2 \langle 13 \rangle^2
        \sum_{\lbrace 3,\dots,n\rbrace}
	\;(\lambda_3\lambda_4\cdots\lambda_n)_{i_2 i_1}\;
        {1 \over  \langle 12 \rangle  \langle 23 \rangle \cdots
         \langle n1 \rangle }.
\ee
$\phi^{\pm}$ are the supersymmetry partners of the two helicity states of
the quark. The two combinations $\phi^+ \pm \phi^-$ transform respectively
as a scalar and a pseudoscalar under the Lorentz group.
For $n=4,5$ these are the only independent non-vanishing
helicity amplitudes for this process.

It is interesting to notice, even though the authors do not have a clear
understanding of the deep reasons for this result, that all of these exact
matrix elements for the MHV amplitudes with gluons, fermions and scalars can be
generated in a straightforward way as correlation functions of fields of a
two-dimensional Wess-Zumino-Witten gauge model with $N=4$ supersymmetry
\cite{nair}.    

\subsection{Color Coherence}
Let us go back now to the $q\qbar$ plus gluons process, Equation~\eqnum{qqng}:
if we put the
color factors $(\lambda^{a_1}\dots\lambda^{a_n})_{ij}$ equal  to 1
for each permutation of 1 through $n$, then Equation~\eqnum{qqng}\ gives
rise to the QED result for the amplitude with one quark-pair and $n$ photons.
This can be easily proved diagrammatically by observing that diagrams with
non-abelian gluon vertices entering the graph expansion for the \subs\ cancel
in pairs when we perform the sum over permutations. In this way the only
diagrams left are the QED-type diagrams, with the common trivial abelian
color structure. This result is independent of the helicity
configuration, and for the helicities considered above we then obtain:
\be     \label{photon}
	M^{n}_{\gamma}(h_q,h_\gamma) =     i \, (\sqrt 2 e)^n
	{ \{p\gamma\}^3 \{\pbar\gamma\}  \over \{p\pbar\} } \, 
	\sum_{\{1,2,\dots,n\} }
       	{ 1 \over \{p1\}\{12\}\dots\{n\pbar\} },
\ee                                          
where $\gamma$ is the momentum of the photon with helicity different from
the others, $h_\gamma$ is its helicity, and where the curly brackets stand for 
the spinor products or their complex conjugates, depending on the helicity of
the photon:
\be    
	\{ij\}_{h_\gamma=-}= \langle ij \rangle 
	\quad , \quad \{ij\}_{h_\gamma=+}= [ij] 
\ee                                                     
                               
The following remarkable identity holds:
\be    \label{eik_id}
	\sum_{\{1,2,\dots,n\} }
       	{ \{p\pbar\} \over \{p1\}\{12\}\dots\{n\pbar\} }    \;=\;
	\prod_{i=1}^n \; { \{p\pbar\}  \over  \{pi\}\{i\pbar\} }.
\ee
Equation~(\ref{eik_id}) can be proved by iteratively using the Fierz identity:
\be     \label{fierz}
	\{ p\pbar \} \{ q\qbar \} \;=\;
	\{ p\qbar \} \{ q\pbar \} + \{ pq  \} \{ \pbar\qbar\} .
\ee
Equation~(\ref{eik_id})  can be thought of as a sort of `square root' of the
eikonal identity. It allows us to put equation~(\ref{photon}) into the 
eikonalized form:
\be     \label{photon_eik}
	M^{n}_{\gamma}(h_q,h_\gamma) =     i \, (\sqrt 2 e)^n
   	\delta_{ij}
	{ \{p\gamma\}^3 \{\pbar\gamma\}  \over \{p\pbar\}^2 } \, 
	\prod_{i=1}^n \; { \{p\pbar\}  \over  \{pi\}\{i\pbar\} }.
\ee
Equations~(\ref{qqng}), (\ref{photon}) and (\ref{photon_eik}) offer
a nice example of the difference between the properties of the non-abelian
radiation as opposed to the abelian radiation. Let us take, in fact, the
square of these three expressions, summed over the colors of the quarks
and of the gluons (when present):
\ba    
	&&
  	\sum_{col} \vert M^{n}(h_q,h_g) \vert ^2 = g^{2n} N^{n+1}
	{ (pg)^3(\pbar g) \over (p\pbar) } \, \sum_{\{1,2,\dots,n\} }
	{1 \over (p1)(12)\dots (n\pbar)} \;+\;
	{1 \over N^2}(interf.),   
        \nn \\   &&   
	\label{square1}              
	\ret        
	&&
  	\sum_{col} \vert M^{(n)}_\gamma (h_q,h_\gamma)  \vert ^2 = (\sqrt 2
	e)^{2n} N
	{ (p\gamma)^3(\pbar\gamma) \over (p\pbar) } \, \sum_{\{1,2,\dots,n\} }
	{1 \over (p1)(12)\dots (n\pbar)} \;+\;
	(interf.),                                  
        \nn \\  &&
	\label{square2}              
	\ret   &&                      
	\label{square3}              
  	\sum_{col} \vert M^{(n)}_\gamma(h_q,h_\gamma)  \vert ^2 = (\sqrt 2 e)^{2n} N
	{ (p\gamma)^3(\pbar\gamma) \over (p\pbar)^2 } \, 
	\prod_{i=1}^n
	{ (p\pbar) \over (pi)(i\pbar) },
\ea                                     
where:
\be
	 (ij)=2 \; i\cdot j.                             
\ee
Equations~(\ref{square2}) and (\ref{square3}) are identical, thanks to the
eikonal identity, but we wrote them in the two different ways to establish a
connection with the expression for the gluon emission. Equation(\ref{square3})
shows that the photon emission is incoherent: the photons only know about
their source, \ie the quark line, but they do not know about each other.
Up to the overall factor in front, the probability for the emission of $n$
photons is just the product of the probabilities for the independent emission
of each of them. 
\footnote{This result, which is exact for this specific helicity
configuration, also holds for any other helicity configuration in the limit
of soft-photon emission. The reason why it cannot hold for an arbitrary
helicity configuration is that in general the amplitude will have poles
of the kind $1/(p+k+k')^2$, $k$ and $k'$ being arbitrary photon momenta.}

On the contrary, if we now look at equation~(\ref{square1}) we see that
the gluon emission is not incoherent: gluons know of each other's presence,
and the full probability is not a product of probabilities. The interference
terms coming from the product of different permutations are suppressed by
a factor of $1/N^2$; this suppression originates from the interferences of the
color factors. For the photon emission, vice versa, we can see from
equation~(\ref{square2}) that the interferences among
different permutations are not suppressed and they conspire to cancel the
coherence apparent into the sum of squares, giving rise to the factorized
expression given in~(\ref{square3}). 

The phenomenological consequences of this coherence effects have been explored
experimentally \cite{string} (the {\it string effect}) and theoretically
(see for example \cite{azetal,lund,pino,mw84a,emw87,muller87,mw88}).

\subsection{$q\qbar q\qbar$ plus gluons}
In general the factorization of the color structure  exhibited in
equation~(\ref{qgcolor}) does not imply a similar factorization of the
kinematical part of the amplitude. In other words, the sub-amplitude that
multiplies a given color factor does not factorize into products of terms that
only depend upon the kinematical variables (helicities and momenta) of the
particles belonging to the same antenna. 
One remarkable non-trivial exception
to this general feature is given by the amplitude for a process with two quark
pairs and an arbitrary number of like-helicity gluons (all the particles are
outgoing). Up to an overall factor that only depends
upon the helicity configuration
each sub-amplitude factorizes into the product of two terms that 
only depend upon the momenta of the gluons emitted by one or the other of the
two antennas: 
{\samepage    
\ba 	\label{two pairs}                     
 	M(h_p,h_q,h_g) &&=\;i\,g^{n+2}A_0(h_p,h_q,h_g) \;\;\cdot \nn \\{}\nn\\
	 \sum && { \{ p \bar q \}  
	\over \{pa_1\}\{a_1a_2\} \cdots \{ a_k \bar q \} } 
	{\{ q \bar p \}  \over \{qb_1\}\{b_1 b_2\} \cdots \{b_{k'} \bar p\}} 
	(\lambda^{a_1} \cdots \lambda^{a_k})_{i_1j_2} 
	(\lambda^{b_1} \cdots \lambda^{b_{k'}})_{i_2 j_1}  
	\nn \\ {} \nn \\ 
	-{1 \over N}&& {\{ p \bar p \}  \over 
	\{pa_1\}\{a_1a_2\} \cdots \{a_k\bar p\}} 
	{\{ q \bar q \} \over \{q b_1\}\{b_1 b_2\}\cdots 
	\{b_{k'} \bar q\}} 
	(\lambda^{a_1} \cdots \lambda^{a_k})_{i_1 j_1} 
	(\lambda^{b_1} \cdots \lambda^{b_{k'}})_{i_2 j_2} .                  
	\nn \\ {} 
\ea  }  \\
The helicity structure of this amplitude 
uniquely determines the pole structure and the residues
of these poles, through unitarity: Equation(\ref{two pairs}) is the
only Lorentz invariant amplitude that gives rise to the right poles and
the right residues. Alternatively, one can prove Equation~\eqnum{two pairs}\
by using the appropriate form of the
Berends and Giele recursive relations \cite{bg88}.
                                                  
If the gluons have all positive helicity, then $\{ij\}=\sp ij$, otherwise
$\{ij\}=\cp ij$. 
The indices $p$ and $q$ ($\bar p$ and $\bar q$) refer to the quarks (antiquarks)
and the indices $a_\alpha,b_\beta$ refer to the gluons. The arguments $h$
represent the helicities of the two quarks and of the gluons; the helicities
of the anti-quarks are fixed by helicity conservation along the fermion
lines. The sum is over
all the partitions of the $n$ gluons ($k+k'=n, k=0,1,\dots,n $) and over the
permutations of the gluon indices. 
When  $k=(0,n)$ the product of zero $\lambda$ matrices becomes a Kronecker delta
and one of the two kinematical factors is equal to one. 
The overall factor $A_0$ can be written as follows:
\be
	A_0(h_p,h_q,h_g) \;=\; { a_0(h_p,h_q,h_g) \over (p+\pbar)^2(q+\qbar)^2 }.
\ee
The functions $a_0$ are given in Table~\ref{utable}, where helicity
configurations obtained by permuting the quark helicities have been omitted.
The functions $a_0$
are universal, in the sense that they only depend upon the spin-1/2 nature
of the quarks. As we will see later, they also enter in processes like deep
inelastic scattering or $e^+e^-$ annihilation.
{\renewcommand{\arraystretch}{1.8}
    \begin{table}
	\begin{center}
       	\begin{tabular}{|l||c|} \hline\hline
       $(h_p,h_q,h_g)$ & $a_0(h_p,h_q,h_g)$  \\ 
          \hline\hline                      
	$(+,+,+)$                                     
		 & $ \langle \pbar \qbar \rangle ^2  
		      [p\pbar][q\qbar]  $
	\\ \hline		
	$(+,+,-)$
	       	 & $ [ pq ] ^2  
		      \langle  p \pbar \rangle  \langle q \qbar \rangle$
	\\ \hline
	$(+,-,+)$                                     
		 & $ \langle \pbar q     \rangle ^2  
		      [p\pbar][q\qbar]  $
	\\ \hline		
	$(+,-,-)$
	       	 & $ [ p \qbar ] ^2  
		      \langle  p \pbar \rangle  \langle q \qbar \rangle$
	\\ \hline\hline
	\end{tabular} 
	\end{center}
	\caption{The universal functions $a_0(h_p,h_q,h_g)$. 
	}
	\label{utable}   
\end{table}  \\ }    

To the leading order in $N$, the amplitude squared summed over colors is
furthermore given by:
{\samepage
\ba                                                       
	\sum_{col} \vert M(h_p,h_q,h_g)  \vert ^2 & =&
      	g^{2n+4}N^n(N^2-1) \vert A_0(h_p,h_q,h_g) \vert ^2 \nn \\  {}\nn \\
	\label{square}    
       & \sum & {( p \bar q )  \over (pa_1)(a_1a_2) \cdots (a_k \bar q)}\,
	{( q \bar p )  \over (q b_1)(b_1 b_2) \cdots (b_{k'} \bar p)}.
\ea                         }
If the quarks are identical we must add
the contribution from the crossed channel $p \leftrightarrow q$. 

As in the case with one quark pair, we can here compare the properties of
photon radiation with those of gluon radiation. A reasoning similar to the
one used in the previous section allows us to write the amplitude for the
emission of $n$ like-helicity photons off two quark-pairs:     
\ba     
   	M(h_p,h_q,h_g) &=& i\,g^2 (\sqrt 2 e)^{n}A_0(h_p,h_q,h_g) \;\;\cdot 
	\nn \\ {}\nn\\
	 \sum &&  \!\!\!\!
   	 {\{ p \bar p \}  \over 
	\{pa_1\}\{a_1a_2\} \cdots \{a_k\bar p\}} 
	{\{ q \bar q \} \over \{q b_1\}\{b_1 b_2\}\cdots 
	\{b_{k'} \bar q\}} 
	(\delta_{i_1j_2}\delta_{i_2j_1} 
 	-{1 \over N}\delta_{i_1j_1}\delta_{i_2j_2}). 
	\nn \\ 
	&&     \label{two_pair_ph}
\ea                           \\
Only the contribution from gluon exchange is shown. The effect of photon
exchange between the two quark-pairs can be easily added.
A repeated use of the Fierz identity, equation~(\ref{fierz}), leads then
to the following form of equation~(\ref{two_pair_ph}):
{\samepage
\ba     \label{two_pair_eik}
   	M(h_p,h_q,h_g) &=& i\,g^2 (\sqrt 2 e)^{n}A_0(h_p,h_q,h_g) \;\;\cdot \nn\\{}\nn\\
   	&\prod_{i=1}^n &
	( { \{ p \bar p \}  \over \{pi\} \{i \bar p \} } +
	  { \{ q \bar q \}  \over \{qi\} \{i  \bar q\} } )
	(\delta_{i_1j_2}\delta_{i_2j_1}              
	-{1 \over N}\delta_{i_1j_1}\delta_{i_2j_2}).
\ea   }                          \\
This expression shows that photons are emitted independently. Once
again we expect this result to hold for an arbitrary helicity configuration
in the soft-photon limit.

If we substitute the color factor in equation(\ref{two_pair_eik}) with
the Abelian one, $\delta_{i_1j_1}\delta_{i_2j_2}$, and if we put $g$=$\sqrt2 e$,
then we obtain the amplitude for the process $e^+e^-\mu^+\mu^-\ photons$,    
as given in reference~\cite{eemm,xu}.

\subsection{$e^+e^-$ and DIS}
It is easy to derive expressions analogous to (\ref{two pairs}) and
(\ref{square})
for the
process $l\bar l q \bar q \  gluons$, where $l\bar l$ is a lepton-antilepton
pair (for example $e^+e^-$ or $e^-\bar\nu$) \cite{bg88,mlm88}. Again the gluons
have all the same helicity\footnote{In Ref.\cite{mlm88} the term $\{q\qbar\}$
was inadvertently omitted from these equations.}:
\be   \label{epem_amp}
   	M^{l}(h_l,h_q,h_g) \;=\;  i \, g^n \sum_{V=\gamma,Z,W}
	M^{l}_V(h_l,h_q,h_g) \, \sum_{\{1,2,\dots,n\}}
	(\lambda^{a_1} \dots \lambda^{a_n})_{ij}\,
	{ \{ q\qbar\} \over \{q1\}\{12\}\dots\{n\qbar\} },
\ee                                                                        
\\
\be   \label{epem_sq}
	\sum_{col}  \vert M^{l}(h_l,h_q,h_g)  \vert ^2	
	\;=\; g^{2n} N^{n-1}(N^2-1)  \; 
	 \vert \sum_{V=\gamma,Z,W} M^{l}_V(h_l,h_q,h_g)  \vert ^2
	\, \sum_{\{1,\dots,n\} } {(q\qbar)\over (q1)(12)\dots(n\qbar) }.
\ee                                  
$q$ and $\qbar$ are the quark momenta and 1 through $n$ are the gluon momenta.
The contributions from photon,$W$ and $Z$ exchange are explicitly
exhibited. 
The functions $M^{l}_V(h_l,h_q,h_g)$ are given by:
\be              
	M_V(h_l,h_q,h_g)= { Q_V(h_l)Q_V(h_q) \over (q+\qbar)^2(s - M^2_V) }
			 \, a_0(h_l,h_q,h_g)
\ee           
$Q_V(h_l)$ \ ($Q_V(h_q)$) is the charge corresponding to the interaction
of a lepton (quark) of helicity $h_l$ ($h_q$) with the vector $V$. \ 
Furthermore $s=(p+\pbar)^2$, with $p$ and $\pbar$ being the lepton momenta,
and $M_V^2$ is the mass squared of the vector boson $V$.
The universal functions   $a_0(h_l,h_q,h_g)$ coincide with those given in
Table~\ref{utable}.
        
For $e^+e^-$ scattering the effect of photon radiation (both from the initial
and the final state) can be easily incorporated into equation~(\ref{epem_amp})
by using equation~(\ref{two_pair_eik}). Here we will display directly the
result for the square of the amplitude with $n$ gluons and $m$ photons, 
to the leading order in $N$:
{\samepage                        
\ba     \label{string_eff}
	\sum_{col}  \vert M^{e^+e^-}(h_l,h_q,h_g)  \vert ^2	
	\= (2 e^2)^{m}g^{2n} N^{n-1}(N^2-1)  \; 
	 \vert \sum_{V=\gamma,Z} M^{l}_V(h_l,h_q,h_g)  \vert ^2 \times &&
	\nn \ret
	 \prod_{i=1}^m 
	 \vert 	( { \{ p \bar p \}  \over \{pk_i\} \{k_i \bar p \} } +
	  { \{ q \bar q \}  \over \{qk_i\} \{k_i  \bar q\} } ) \vert ^2
	\sum_{\{1,2,\dots,n\} } { (q\qbar) \over (q1)(12)\dots(n\qbar) }.&&
\ea                }
\\                         
The $k_i$'s are the momenta of the photons.
Once again this result is only exact if all the gluons {\em and} the photons
have the same helicity, but this is the behaviour of all the other helicity
configurations in the case of soft emission. 

Exact expressions for the full calculation of the processes $e^+e^- \to
4$ partons were given in \cite{ert} (for the complete $O(\alpha^2)$
calculation) and in \cite{ali} (tree level) and for the processes $e^+e^- \to
5$ partons in \cite{zep89,bgk89a} (tree level).



\newpage
\section{Multiple Gauge Groups}

In a theory in which the fermions are coupled to a direct product 
gauge group, say $SU(M) \times SU(N)$, then the techniques described 
earlier in this report can also be implemented. In fact the 
subamplitudes can be easily obtained from the subamplitudes given 
earlier by adding various combinations of `color' orderings together.
This can also be extended to $U(1)$ gauge groups, {\em e.g.} 
electromagnetism. For spontaneously broken gauge groups, only 
processes involving a single vector boson are easily obtained from 
previous calculations. In this section we have set all coupling 
constants to unity, but the reader can easily insert the appropriate 
couplings.

\subsection{$SU(M) \times SU(N)$ }

Consider a fermion which is in the fundamental representation of 
both $SU(M)$ and $SU(N)$, then the amplitude for scattering of
a fermion-antifermion with $m$ vectors from $SU(M)$ and $\bar{n}$ 
vectors of $SU(N)$ can be written as 
\ba
\label{qqmn}
	M(\Q,1,2,\dots,m;\bar{1},\bar{2},\dots,\bar{n},\AQ) 
&=&
\sum_{P,\overline{P}} ~~(\Lambda^1 \Lambda^2 \dots \Lambda^m)_{i~j}
     ~~(\lambda^{\bar{1}} \lambda^{\bar{2}} \dots \lambda^{\bar{n}} 
)_{\bar{i}~\bar{j} }~~~~~~~~~
\nonumber \\
& & 
m_{M,N}
(\Q,1,2,\dots,m,\AQ;~\Q,\bar{1},\bar{2},\dots,\bar{n},\AQ).
\ea
$\Lambda$ and $\lambda$ are the fundamental matrices of $SU(M)$ and 
$SU(N)$ respectively. Also $ij$ and $\bar{i}\bar{j}$ are the $SU(M)$ 
and $SU(N)$ `color' indices of the fermions and the sum is over 
all permutations of $1,2,\dots,m$ and $\bar{1},\bar{2},\dots,\bar{n}$.

The subamplitudes defined by Eq. (\ref{qqmn}) can be easily obtained from 
the subamplitudes obtained earlier, Eq. (\ref{quarks}),
\be
\label{subqqmn}
m_{M,N}
(\Q,1,2,\dots,m,\AQ;~\Q,\bar{1},\bar{2},\dots,\bar{n},\AQ)
~=~
\sum_I m(\Q,1,2,\dots,m,\bar{1},\bar{2},\dots,\bar{n},\AQ)
\ee
where the $\sum_I$ is over all ways the barred numbers can be interspersed
within the unbarred numbers maintaining the order of both the barred and 
unbarred numbers. This sum causes all Feynman diagrams which connect 
directly the vectors of $SU(M)$ with $SU(N)$ to be cancelled.

As an example, consider the scattering in which the fermion has 
negative helicity and the $\alpha$ vector boson of either gauge group
has negative helicity and all particles have positive helicity. Then,
\ba
m_{M,N}
(\Q,1,\dots,m,\AQ;\Q,\bar{1},\dots,\bar{n},\AQ)
&=&                        
i~ { {\sp{\Q}{\alpha}^3 \sp{\AQ}{\alpha}} \over 
\sp{\AQ}{\Q}^2 }
\sum_I
~~{ {\sp{\AQ}{\Q} }
\over {\sp{\Q}{1} \sp{1}{2} 
\cdots 
\sp{m}{\bar{1}} \sp{\bar{1}}{\bar{2}} 
\cdots
\sp{\bar{n}}{\AQ} }}
\nonumber \\
&=&
i~ { {\sp{\Q}{\alpha}^3 \sp{\AQ}{\alpha}} \over 
\sp{\AQ}{\Q}^2 }
~{\sp{\AQ}{\Q}
 \over {\sp{\Q}{1} \sp{1}{2} \cdots
\sp{m}{\AQ}} }
~{\sp{\AQ}{\Q}
 \over {\sp{\Q}{\bar{1}} \sp{\bar{1}}{\bar{2}} \cdots
\sp{\bar{n}}{\AQ}} }.
\ea

\subsection{$U(1)$}

For completeness we include here a discussion of the $U(1)$ gauge group 
even though it was extensively discussed in the previous section.
The $U(1)$ gauge group results can be obtained by replacing
$\lambda$ in Eq. (\ref{quarks}) with $\delta$ or by iterating 
Eq. (\ref{subqqmn}) of the previous section. Thus for QED the amplitudes
are
\be
A_{QED}(\Q,1,2,\dots,m,\AQ) = \sum_P m(\Q,1,2,\dots,m,\AQ).
\ee
The sum over all permutations causes all non-abelian Feynman diagrams 
that naively appear to be cancelled, thus leaving only the QED Feynman 
diagrams.

Again as an example consider the case of fermion-antifermion $m$ 
photon scattering in which the fermion and the $\alpha$ photon have 
negative helicity and all other particles have positive helicity.
Then,
\ba
A_{QED}(\Q,1,2,\dots,m,\AQ) &=&
i{ {\sp{\Q}{\alpha}^3 \sp{\AQ}{\alpha}} \over 
\sp{\AQ}{\Q}^2 }
 \sum_P { { \sp{\AQ}{\Q}}
\over
{\sp{\Q}{1} \sp{1}{2} \cdots \sp{m}{\AQ} 
}} \nonumber \\
&=& 
i{ {\sp{\Q}{\alpha}^3 \sp{\AQ}{\alpha}} \over 
\sp{\AQ}{\Q}^2 }
~\prod_{i}{\sp{\AQ}{\Q} \over {\sp{\Q}{i}
\sp{i}{\AQ}} }.
\ea

Combining this example with that of the previous subsection, the 
subamplitude in an $SU(M) \times SU(N) \times U(1)$ theory for a 
fermion and the $\alpha$ vector particle with negative helicity and 
all other particles of positive helicity is
\ba
& & m_{M,N,QED}(\Q,1,\dots,n,\AQ;
\Q,\bar{1},\dots,\bar{n},\AQ;
\Q,\hat{1},\dots,\hat{n},\AQ)  ~=~ \nonumber \\
& & ~~~~~
i~{ {\sp{\Q}{\alpha}^3 \sp{\AQ}{\alpha}} \over 
\sp{\AQ}{\Q}^2 }
~{\sp{\AQ}{\Q}
 \over {\sp{\Q}{1} \sp{1}{2} \cdots
\sp{m}{\AQ}} }
~{\sp{\AQ}{\Q}
 \over {\sp{\Q}{\bar{1}} \sp{\bar{1}}{\bar{2}} \cdots
\sp{\bar{n}}{\AQ}} }
~\prod_{\hat{i}} {\sp{\AQ}{\Q} \over {\sp{\Q}{\hat{i}} 
\sp{\hat{i}}{\AQ}} }.
\ea

\subsection{The Insertion of a W or Z}

A spontaneously broken gauge group does not have a simple 
generalization of the previous subsections. However, the insertion of
one such massive vector particle, W or Z, can be easily incorporated.
Consider the scattering of a quark-antiquark $n$ gluons and a W 
vector boson. Then the amplitude for this process is written as
\be
A(\Q,1,\dots,n,\AQ;W) = \sum_P
( \lambda^1 \cdots \lambda^n )_{i\bar{j}} ~m(\Q,1,\dots,n,\AQ;W)
\ee
where the subamplitude can be written as
\be
\label{qqWng}
m(\Q,1,\dots,n,\AQ;W)
= i~ \epsilon^{\mu}_W
~\sum^n_{i=0} \overline{U}(\Q^-,1,\dots,i) \gamma_{\mu} 
{ (1 - \gamma_5) \over 2 }  V(i+1,\dots,n,\AQ^+).
\ee
Here the recursion techniques of section \secrsq have been 
explicitly used.

This expression can be used in one of two ways: either one can square
it directly or allow the W boson to decay into another 
fermion-antifermion pair.
If one squares this expression directly the relationship
\be
\sum_{pol} \epsilon^{\mu}_W {\epsilon^{\nu}_W}^*
~=~ -g^{\mu \nu} + { W^{\mu} W^{\nu} \over M_W^2 }
\ee
can be employed.

The other alternative is to replace the polarization of the W vector 
boson by the amplitude for it to decay into a lepton-antilepton 
pair\footnote{Working at the amplitude level one could also use the 
representation for the heavy vector polarizations given in 
Ref.\cite{passarino}.}. Then Eq. (\ref{qqWng}) is written as
\ba
m(\Q,1,\dots,n,\AQ;\L,\AL)
&=& -i~ \overline{U}(\L^-) \gamma_{\mu} {(1-\gamma_5) \over 2 } V(\AL^+)
\nonumber \\
& \times & {(-g_{\mu \nu} + {W_{\mu}W_{\nu} \over M_W^2 })
	\over (W^2-M_W^2+iM_W\Gamma_W)} \nonumber \\
& \times & \sum^n_{i=0} \overline{U}(\Q^-,1,\dots,i) \gamma_{\nu} 
{ (1 - \gamma_5) \over 2 }  V(i+1,\dots,n,\AQ^+).
\ea
If we use the fact that the charged lepton is effectively massless,
compared to $M_W$, the Fierz rearrangement gives
\ba
m(\Q,1,\dots,n,\AQ;\L,\AL)
&=&  -2i \sum^n_{i=0} {{ ~\overline{U}(\Q^-,1,\dots,i) \rsp{L^+} 
~ \lsp{\AL^+} V(i+1,\dots,n,\AQ^+) }
\over 
{ (W^2-M_W^2+iM_W\Gamma_W)} } .
\ea

Using the results from the recursion relation section of this report,
the sub-amplitude for the process 
$\Q\AQ \rightarrow ~W~ \rightarrow \L\AL$ 
is
\ba 
m_W(\Q^-,\AQ^+;\L^-,\AL^+)
&=&   
 { {-2i~\cp{\AL}{~\AQ} \sp{\Q}{\L} } \over {(W^2-M^2_W + iM_W\Gamma_W) }} 
\nonumber \\
&=&
{ {2i~\sp{\Q}{\L}^2 ~ \cp{\AL}{\L} } \over 
{ \sp{\Q}{\AQ}~(W^2-M^2_W + iM_W\Gamma_W) } }
\nonumber \\
&=&
{ {2i~\cp{\AQ}{~\AL}^2 ~ \sp{\AL}{\L} } \over 
{ \cp{\Q}{\AQ}~(W^2-M^2_W + iM_W\Gamma_W) } }.
\ea

Adding $n$ gluons with the same helicity to this process, gives 
\ba 
m_W(\Q^-,g_1^+,\cdots,g_n^+,\AQ^+;\L^-,\AL^+)
&=&   
{ {2i~\sp{\Q}{\L}^2 ~ \cp{\AL}{\L} } \over 
{ \sp{\Q}{1} \sp{1}{2} \cdots \sp{n}{\AQ}~(W^2-M^2_W + iM_W\Gamma_W) } 
},
\ea

\ba 
m_W(\Q^-,g_1^-,\cdots,g_n^-,\AQ^+;\L^-,\AL^+)
&=&   
{ {(-1)^n~2i~\cp{\AQ}{~\AL}^2 ~ \sp{\AL}{\L} } \over 
{ \cp{\Q}{1} \cp{1}{2} \cdots \cp{n}{\AQ}~(W^2-M^2_W + iM_W\Gamma_W) } 
}.
\ea

If we add two gluons of opposite helicity, then the sub-amplitudes are
\ba 
	m_W(\Q^-,g_1^+,g_2^-,\AQ^+;\L^-,\AL^+)
	&=&     
	{{ -2i } \over 
	{ (W^2-M^2_W + iM_W\Gamma_W) } } \nonumber \\
	\bigl(  { \sp{\Q}{\L} \lsp{\AL+} \hat{\Q}+\hat{W} \rsp{2+}
	\cp{1}{\AQ}^2 \over S_{\Q W} ~S_{12} ~\cp{2}{\AQ} } 
	 & + & ~{\sp{\Q}{2} \sp{\Q}{\L} \cp{\AL~}{\AQ} \cp{1}{\AQ}
	\over \sp{\Q}{1} ~S_{12} ~\cp{2}{\AQ} } 
\nonumber \\                                            
	& + & ~{\sp{\Q}{2}^2 \lsp{1+} \hat{\AQ}+\hat{W} \rsp{\L+} \cp{\AL~}{\AQ}
	\over       
	\sp{\Q}{1} ~S_{12} S_{\AQ W}  }  \bigr)
\ea

and
\ba 
	m_W(\Q^-,g_1^-,g_2^+,\AQ^+;\L^-,\AL^+)
	&=&     
	{{ 2i } \over 
	{ (W^2-M^2_W + iM_W\Gamma_W) } } \nonumber \\
	\bigl( {\sp{\Q}{\L} 
	~\lsp{\AL+}\hat{\Q}+\hat{W} \rsp{1+} 
	~\cp{2}{\AQ} ~\sp{1}{\AQ}
	\over   S_{\Q W} ~S_{12} ~\sp{2}{\AQ} } 
	& +  & ~ { \lsp{2+} \hat{\AQ} + \hat{W} \rsp{\L+}
	~\lsp{\AL+}\hat{\Q}+\hat{W} \rsp{1+}
	\over \cp{\Q}{1} ~S_{12} ~\sp{2}{\AQ}~ } \nonumber \\
	& + & ~ { \cp{\Q}{2} ~\sp{\Q}{1} 
	~\lsp{2+} \hat{\AQ} + \hat{W} \rsp{\L+} \cp{\AL~}{\AQ}
	\over \cp{\Q}{1} ~S_{12} ~S_{\AQ W} }.
	\bigr)
\ea
These expressions reproduce the results first obtained in 
\cite{gk85b,kls85b,keith86b}. The full set of radiative corrections (to order
$\alpha_s^2$) to this process was recently calculated in
\cite{hallsie,gonsalvez}.

The previous discussion can be extended to include the Z boson by 
decomposing the coupling of the Z to the quarks and leptons into its 
left and right handed parts and then proceeding as with the W boson.
For a complete discussion of the calculation for these processes,
including the complete results for processes including a W boson plus 
five partons see Berends, Giele and Kuijf, ref.\cite{bgk89a}. These results
agree with those independently obtained by
Hagiwara and Zeppenfeld, ref.\cite{zep89}.



\newpage
\section{Approximate Matrix Elements}
The techniques described in the previous Sections provide very powerful tools
to calculate the matrix elements of very complex processes. As an example, the
Berends and Giele recursive relations were recently used for the calculation of
8-gluon scattering \cite{bgk89c}. The resulting expressions, however, prove very
slow to evaluate numerically because of their complexity, thus making it almost
impossible to generate a number of events large enough to perform relevant
physics studies.

These considerations, and the importance of having fast event generators to
simulate multi-jet processes at high-energy hadron colliders, where these
processes will provide important backgrounds to many possible new physics
signals, justify the study of approximate expressions which describe
sufficiently well the exact matrix elements throughout phase-space and at the
same time are simple enough to allow very fast simulations.

Kunszt and Stirling \cite{ks88} and Maxwell \cite{maxwell87} were the first to
realize that the Parke and Taylor amplitudes, Equation~\eqnum{ptamp}, can be
properly fudged in a systematic way so as to reproduce the full sum over all
the allowed helicity amplitudes for gluonic processes. 
This idea was later generalized to other processes in which at least one set of
helicity amplitudes is known in both hadronic 
\cite{maxwell87,mp89a,maxwell89} and $e^+e^-$ \cite{maxwellep}
multi-jet production.  An
alternative scheme based on the non-abelian version of the eikonal
approximation was also introduced in \cite{bg89}. In this Section we will
describe these various approximation schemes, referring the reader to the
original literature for numerical comparisons between them.

\subsection{The Kunszt and Stirling Approximation}
We will start from the simplest scheme, namely that of Kunszt and Stirling (KS,
see Ref.\cite{ks88}).
It amounts to assuming that all of the helicity amplitudes have 'on average'
the same value, and therefore the full amplitude can just be obtained
by multiplying the Parke and Taylor (PT) expressions by a proper weight,
representing the ratio between the number of non-zero helicity configurations
and the number of the Maximum Helicity Violating (MHV) configurations whose
matrix-elements are described by the PT formula. 

This approximation becomes particularly simple when neglecting sub-leading
terms in $1/N$. This is justified because the sub-leading terms have softer
collinear singularities than the leading ones, and therefore do not contribute
substantially to the numerical value of the matrix elements. In particular, for
$n=6$ the sub-leading terms are finite \cite{mpx87}, and only contribute of the
order of few percent to the full square. For generic $n$ it was proven in
Reference~\cite{fpr88} that the sub-leading terms are also finite in the strong
energy-ordering kinematical domain. This important result strongly justifies
neglecting the sub-leading terms at the level of precision given by these
tree-level calculations.                                      

For an $n$-gluon process the number of MHV amplitudes is $n(n-1)$ if $n>4$ and
$n(n-1)/2$ if $n=4$.  The total number of non-zero helicity amplitudes is
instead $2^n-2(n+1)$. For $n=4,5$ these multiplicities coincide, and the PT
formula describes the exact results, as is well known. For $n$ larger than 5,
the KS approximation gives:
\be  \label{ksapp}
	d\sigma^{gg\to g\dots g}_{KS} \=
	\frac{2^n - 2(n+1)}{n(n-1)} \; d\sigma^{gg\to g\dots g}_{PT}
\ee
For $n=6,7$ , for example, the fudge factor is 5/3 and 8/3, respectively.

To describe processes with initial state quarks, KS suggest the use of the so
called effective structure function approximation \cite{ua1sf1,ua1sf2}, which
gives a good description of the two-to-two QCD processes. According         to
this approximation in most of the relevant phase-space the differential
cross-sections for processes initiated by $gg$, by $qg$ and by $qq$ or $q\bar
q$ stand in a constant ratio:
\be                                               
	\mbox{d}\sigma_{gg} \;:\; \mbox{d}\sigma_{gq} \;:\; \mbox{d}\sigma_{qq}
	\;=\;             1 \;:\; 4/9 \;:\; (4/9)^2 .
\ee          
In this way the full differential cross-section, weighted by the appropriate
structure functions, reads:                 
\ba                                          
	\mbox{d}\sigma_{tot} \;=\; F(x_1)F(x_2) \mbox{d}\sigma_{gg}, 
	\\
	F(x)=g(x) + 4/9 \, (q(x)+\bar q(x)),
\ea
$g(x)$ and $q(x)$ being the gluon and quark structure functions.
For $d\sigma_{gg}$, finally, one takes Eq.\eqnum{ksapp}.

The KS approximation scheme tends to overestimate the exact results and the
effective structure function approximation is works less and less for an
increasing number of partons in the final state;
nevertheless the KS approximation is an extremely
useful tool for simple but significant estimates of multi-jet rates and
distributions. For
comparisons of this scheme with exact calculations, see for example References
\cite{ks88,mp89a,maxwell89,bgk89b}.
                                 
\subsection{The Infrared Reduction Technique}
It is well known that in the limit in which two partons (say $i$ and $j$)
become collinear, a given process can be described in the 
Weissz\"{a}ker-Williams (W-W) approximation:
\be   \label{wwapp}
	d\sigma^{(n)} \= \frac{1}{2(p_ip_j)} \; f(z) \; d\sigma^{(n-1)}
\ee
where $f(z)$ is an appropriate function of the fraction of momentum carried by
one of the two partons becoming collinear, and $d\sigma^{(n-1)}$ is the
partonic cross-section for the effective $(n-1)$-particle process in which the
two collinear partons are replaced by the single one into which they merge.
On the pole the W-W approximation is nothing but the factorization of the
amplitude, discussed in various occasions in the previous Sections. The
functions $f(z)$, in the case of a QCD process, are just the Altarelli-Parisi
(AP) \cite{ap} splitting functions.                                    

The {\em infrared reduction} technique introduced by Maxwell \cite{maxwell87}\
improves the W-W approximation by using the exact matrix elements for some
simple helicity configurations, and derives the other helicity configurations
by approximating their relative weights at the closest collinear pole.
                                                                 
Next to a collinear pole (say $p_1\cdot p_2 \to 0$) each of the non-vanishing
helicity amplitudes will factorize in the following way:
\be
	d\sigma^{(n)}_h \= \frac{1}{2(p_1p_2)} \sum_{h'}
	\; f_{h'}(z) \; d\sigma^{(n-1)}_{h'} \;+\; \mbox{finite}
\ee                                    
where $h'$ are the various helicity configurations which can contribute to the
factorization, and $f_{h'}(z)$ are the corresponding polarized
AP splitting functions, depending on the variable $z=E_1/(E_1+E_2)$.
For the time being we will restrict our attention to gluon scattering.
For the full process, factorization is described by Equation~\eqnum{wwapp},
with $f(z)$ given by:
\be
	f(z) \= g^2 \; N \; \frac{1+z^4+(1-z)^4}{z(1-z)}
\ee                         
If we just sum over the PT amplitudes, instead, we obtain:
\be    \label{irprel}
	d\sigma^{(n)}_{PT} \= \frac{1}{2(p_1p_2)} 
	\; f_{PT}(z,s_{ij}) \; d\sigma^{(n-1)}_{PT} 
\ee                                           
where $d\sigma^{(n)}_{PT}$ is the sum over all the MHV amplitudes, and 
$f_{PT}(z,s_{ij})$ is given by:
\ba                
&&     	f_{PT}(z,s_{ij}) \=
  	g^2 \, N \, \frac{R+z^4+(1-z)^4}{z(1-z)}
\ret                
&&	R \= \frac{\sum_{i>j} s_{ij}^4}{\sum_{i} s_{Pi}^4},
\ea
the indices $i$ and $j$ being different from the collinear particles, and $P$
being the sum of the collinear momenta. 
                                        
Equation~\eqnum{irprel}\ can also be rewritten in the following fashion:
\be    \label{irprel2}
	d\sigma^{(n)}_{PT} \= \chi^{-1} \; \frac{1}{2(p_1p_2)} 
	\; f_{AP}(z) \; d\sigma^{(n-1)}_{PT}
\ee                       
with:
\be    \label{chired}            
     	\chi(z,s_{ij}) \=
  	\frac{(1+R)(1+z^4+(1-z)^4}{R+z^4+(1-z)^4}
\ee
By equating Equations~\eqnum{irprel2}\ and \eqnum{wwapp}\ we therefore obtain:
\be     \label{irred}
	d\sigma^{(n)}_{full} \= d\sigma^{(n)}_{PT}  \; \chi(z,s_{ij})
	\; \frac{d\sigma^{(n-1)}_{full} }{d\sigma^{(n-1)}_{PT} }     
\ee
Maxwell suggested that while the W-W approximation is not very good unless
we are very close to a collinear pole, Equation~\eqnum{irred}\ is rather good
throughout phase-space, provided we perform the factorization considering the
pair of partons with the minimum $s_{ij}$.  In other words, while the value of
the full differential cross section is not well reproduced in the W-W
approximation away from the collinear poles, what is well approximated is the
relative weight of different helicity amplitudes.  

Since $d\sigma^{(5)}_{full} = d\sigma^{(5)}_{PT}$, for $n=6$ we obtain:
\be
   	d\sigma^{(6)}_{full} \= d\sigma^{(6)}_{PT} \chi(z,s_{ij})
\ee
while for larger $n$ the infrared reduction can be iterated, giving:
\be
   	d\sigma^{(n)}_{full} \= d\sigma^{(n)}_{PT} 
	\prod_{k=6}^{n} \chi_k(z_k,s_{ij}),
\ee                                        
with an obvious notation.
                         
If the two partons which minimize $\vert s_{ij} \vert$ belong to initial and
final state, we can still use Equations~\eqnum{irred}\ and \eqnum{chired}\
provided we keep all of the momenta as outgoing (which implies that the
energies of the initial state particles will be negative) and define:
\be
	z \= \frac{E_i}{E_i+E_j}
\ee
The $z$ defined in this way cannot be interpreted directly as the fraction of
momentum anymore, since it will not satisfy the constraint $0<z<1$. In
particular, if $i$ is the final state parton then $z<0$, while if $i$ is the
initial state, then $z>1$. However it can be easily checked that with this
prescription Equations~\eqnum{irred}\ and \eqnum{chired}\ reproduce the desired
factorization properties. 

This technique can be applied whenever we have exact expressions for some sets
of helicity amplitudes. In particular, it applies to $q\qbar g\dots g$
processes \cite{mp89a,maxwell89}\ and to $e^+e^- \to q\qbar g \dots g$ and
$eq \to eqg \dots g$ 
\cite{maxwellep}.  We will here summarize the main results concerning the 
quark-gluon processes. Similarly to the purely gluonic case, the infrared
reduction technique leads to the following relation:
\be
   	d\sigma_{full}(q\qbar \mbox{n}g) \= 
	d\sigma_{MHV}(q\qbar \mbox{n}g) 
	\prod_{k=4}^{n} \chi_k(z_k,s_{ij}),
\ee                                        
where the relevant MHV amplitudes were given in the previous Section,
Equation~\eqnum{mp}. For these processes the factors $\chi$ depend on the
nature of the partons that have the minimum $\vert s_{ij} \vert$. If these are
both gluons (with indices $\alpha$ and $\beta$), then we have:
\be
\chi_{gg}  \=   { (1+R)~(1+z^4+(1-z)^4) \over (R +z^4 + (1-z)^4) } 
\ee
as before, but with
\begin{eqnarray}
z  = { {p^0_{\alpha}} \over { P^0 } }  & , &
P  \equiv  p_{\alpha} + p_{\beta}  \nonumber \\
R  & = &  { { \sum_{i\neq \alpha , \beta }  
( s^3_{qi} s_{\bar{q}i} +  s_{qi} s^3_{\bar{q}i} )  } \over 
{( s^3_{qP} s_{\bar{q}P} +  s_{qP} s^3_{\bar{q}P} )}  } .
\end{eqnarray}
If the pair with the minimum dot product contains a quark and a gluon
then                                              
\begin{eqnarray}
\chi_{qg}  \=   { (1+R)~(1+z^2_q) \over (1+Rz^2_q ) }  
\end{eqnarray}
where
\begin{eqnarray}
z_q = { {q^0 } \over { Q^0 } }  & , &
Q  \equiv  p_{\alpha} + q \nonumber \\
R & = & { {\sum_{i \neq \alpha } s^3_{Qi} s_{\bar{q}i} }  \over 
{ \sum_{i \neq \alpha } s_{Qi} s^3_{\bar{q}i} }  } .
\end{eqnarray}
The result for an antiquark-gluon pair is 
the same as the above quark-gluon pair but with each fermion 
momentum replaced by the appropriate anti-fermion momentum.

For the situation in which the minimum $|s_{ij}|$ pair is made up of
a quark and an antiquark the multiplication factor is
\begin{eqnarray}
\chi_{ q \bar{q} }  \=   (1+R) 
\end{eqnarray}
where 
\begin{eqnarray}
G & \equiv & q + \bar{q} \nonumber \\
R & = & { { \sum_{ i<j } s^4_{ij} }  \over 
{ \sum_{i} s^4_{Gi} } }.
\end{eqnarray}

In all of these cases we assume the prescription given above when the collinear
partons belong to initial and final state.

The Maxwell approximation scheme has been checked against exact matrix elements
for various multi-parton processes, and has proved to be extremely accurate, in
addition to being numerically more efficient by 2 or 3 orders of magnitude,
depending on the number of IR reduction steps.


\newpage
\section{Conclusions}
While most of the explicit techniques summarized here 
are limited in their application to tree-level processes, 
it is auspicable that one day they can be extended for use in loop calculations
as well. For example, the color structures introduced in Section \sectccf
provide a gauge invariant decomposition of the amplitude at any order in
perturbation theory.  The spinor representation for the polarization vectors
introduced in Section \secthel and used throughout this report is however
specific to four dimensions and could not be used in a dimensional
regularization scheme.  We believe it can be extended to non-integer
dimensions in a scheme which preserves the dimensional relations between
spinors and vectors, such as supersymmetry. However such a scheme (dimensional
reduction) may not provide a consistent regularization of loop amplitudes
beyond one loop. 

Perhaps some of the beautiful features of amplitudes with
simple helicity configurations (the Parke and Taylor amplitudes) can be shown
to persist at higher orders in the loop expansion. The structure of
multi-parton amplitudes unveiled by the approaches described in this report
will hopefully lead to a better understanding of perturbation theory for
non-abelian gauge theories. 

The intriguing connection between the Parke and
Taylor amplitudes and correlation functions in a two-dimensional
Wess-Zumino-Witten model with $N=4$ supersymmetry discovered by Nair
suggests the existence of new generating functionals for these
amplitudes, and perhaps more fundamental structures underlying the perturbative
expansion in a gauge theory.                                       
                                        
For purely phenomenological purposes, the production rates
obtained by these calculations at tree level are extremely valuable
for reliable estimates of important processes. 
These complex processes with many partons in the final state, are now 
being experimentially probed by current hadron colliders
and will become more important at the next generation of hadron 
colliders currently under construction. 
A detailed understanding of 
these QCD/Electro-Weak
background processes will be fundamental for the detection of signals 
of new physics which will contain many jets in the final
state.                            

\begin{center}{\bf Acknowledgements} \end{center}
We wish to thank all our colleagues and friends with whom we have 
discussed many issues in perturbative QCD / Electro-Weak 
physics over the years. 
In particular we give special thanks to our collaborators,
T.~Taylor and Z.~Xu. Also we would like to explicitly thank
F.~Berends, J.~Bjorken, E.~Eichten, K.~Ellis, W.~Giele, Z.~Kunszt,
P.~Marchesini, C.~Maxwell, G.~Paffuti, R.~Pisarski,
C.~Quigg, J.~Stirling, L.~Trentadue and B.~Webber
for contributing to our understanding, through their comments and their
research.                                                             
We are also very grateful to W. Giele and Z. Kunszt for a careful 
reading of this report.


\newpage
\appendix
\section{Appendix: Polarization Vectors and Spinor Properties}
We will use notations and conventions as in \cite{xu}, and will summarize them
here for ease of reference.                                         
Let \ps(p)\ be a massless Dirac spinor. We will denote its chiral projections
as follows:
\be	\label{defspin}
	\rsp{p\pm} \= \psc{\pm}(p) \= \frac{1}{2} (1\pm\gfive) \ps(p)
	\quad\quad
	\lsp{p\pm} \= \overline{\psc{\pm}(p)}
\ee                
By convention, we will choose the spinor phases so as to satisfy the following
identities:
\be    \label{aconj}
	\rsp{p\pm} \= \rsp{p\mp}^c \quad\quad \lsp{p\pm} \= ^c\! \lsp{p\mp},
\ee                                                                  
where the suffix $c$ stands for the charge conjugation operation:
\ba
      &&\rsp{\ }^c \= C \rsp{\ }^*,  \quad\quad ^c\lsp{\ } \= -^*\lsp{\ }C^{-1}
	\ret                                                            
      && C \gamma_\mu^* C^{-1} \= \gamma_\mu,
	\ret
      && C \= C^\dagger \= C^{-1} \= C^* \= C^T.
\ea	
We will also introduce the following notation:
\be
	\sp pq = \sppr{p -}{q +} \quad\quad  \cp pq = \sppr{p +}{q -}
\ee
The spinors are normalized as follows:
\be
	\bra p \vert \gamma_\mu \vert p \ket = 2 p_\mu
\ee
From the properties of the Dirac algebra, it is straightforward to prove the
following useful identities:
\ba     \label{aids1}
      &&	
	\sppr{p +}{q +} =  \sppr{p -}{q -} = \sp pp = \cp pp = 0
\ret \label{aids2}
      &&	
	\sp pq = -\sp qp ,   \quad\quad  \cp pq = -\cp qp
\ret \label{aids3}
	&&                
 	2\rsp{p\pm}\lsp{q\pm} \= \frac{1}{2} (1\pm\gfive) 
	\gamma^\mu \lsp{q\pm} \gamma_\mu \rsp{p\pm}  ,
\ret \label{aids4}                                   
      &&
       	\sp pq ^* = - sign(p\cdot q)\cp pq = sign(p\cdot q) \cp qp
\ret \label{aids5}
      &&
       	\vert \sp pq \vert ^2 = 2(p \cdot q),
\ret \label{aids6}                           
	&&
	\lsp{p\pm} \gamma_{\mu_1}\dots\gamma_{\mu_{2n+1}} \rsp{q\pm}  \=
	\lsp{q\mp} \gamma_{\mu_{2n+1}}\dots\gamma_{\mu_1} \rsp{p\mp} ,
\ret \label{aids7}
	&&
	\lsp{p\pm} \gamma_{\mu_1}\dots\gamma_{\mu_{2n}} \rsp{q\mp}  \=
      -	\lsp{q\pm} \gamma_{\mu_{2n}}\dots\gamma_{\mu_1} \rsp{p\mp} ,
\ret \label{aids8}
      &&        
	\sp AB \sp CD = \sp AD \sp CB  +  \sp AC  \sp BD
\ret \label{aids9}                        
      &&         
	\lsp{A +} \gamma_\mu \rsp{B +} \lsp{C- } \gamma^\mu \rsp{D -} =
	2 \cp AD \sp CB.
\ea                                                                 
In the identity Eq.\eqnum{aids4} the possibility of having spinors with energies
of different sign is considered. This will be important in the following, since
for simplicity we will always carry out the calculations of the matrix elements
assuming all of the particles as being outgoing. Energy-momentum conservation
will then force the energy of some of the particles to be negative.
                                                                   
Notice that the following equations hold for generic chiral spinors (not
necessarily solutions of a Dirac equation) which
satisfy Eq.\eqnum{aconj}: \eqnum{aids2},  \eqnum{aids3},  \eqnum{aids6}, 
\eqnum{aids7},  \eqnum{aids8},  \eqnum{aids9}.

The polarizations for vectors with momentum $p$, as defined in the text:
\ba  \label{aepsmu}                                
       	\eps{\mu}{\pm}{p,k} &=&  \pm
	\frac{\lsp{p\pm} \gamma_\mu \rsp{k\pm}}{\sqrt{2} \sppr{k\mp}{p\pm}},
\ret  \label{aepshat}                                                      
	\eps{{}}{\pm}{p,k} \cdot \gamma &=&
	\pm \frac{\sqrt 2}{\sppr{k\mp}{p\pm}}
	(\rsp{p\mp}\lsp{k\mp} + \rsp{k\pm}\lsp{p\pm}),
\ea
enjoy the following properties:
\ba
     \label{apol1}
&&	\eps{\mu}{\pm}{p,k} \= (\eps{\mu}{\mp}{p,k} )^*  ,
\ret \label{apol2}
&&	\eps{{}}{\pm}{p,k} \cdot p \= \eps{{}}{\pm}{p,k} \cdot k \= 0 ,
\ret \label{apol3}                                 
&&	\eps{{}}{\pm}{p,k} \cdot \eps{{}}{\pm}{p,k'} \= 0 ,
\ret \label{apol4}                                  
&&	\eps{{}}{\pm}{p,k} \cdot \eps{{}}{\mp}{p,k'} \= -1 ,
\ret \label{apol5}                                  
&&	\eps{{}}{\pm}{p,k} \cdot \eps{{}}{\pm}{p',k} \= 0 ,
\ret \label{apol6}                                  
&&	\eps{{}}{\pm}{p,k} \cdot \eps{{}}{\mp}{k,k'} \= 0 ,
\ret \label{apol7}                                  
&&	\eps{\mu}{+}{p,k} \, \eps{\nu}{-}{p,k} \;+\;
	\eps{\mu}{-}{p,k} \, \eps{\nu}{+}{p,k} \= 
  	-g_{\mu\nu} \;+\; \frac{p_\mu k_\nu+p_\nu k_\mu}{p\cdot k}.
\ea

\section{Appendix: A QED Example}
In this Appendix we will describe as a simple application of the Helicity
Amplitude technique the process of electron-positron annihilation into a photon
pair. Two diagrams contribute to the process -- $t$-channel and $u$-channel
fermion exchange (see Figure~\appone). 
If $q,\qbar$ are the momenta of the incoming electron and positron,
$p_{1,2}$ are the momenta of the two outgoing 
photons and $h_{1,2}$ are their helicities,
the contributions of the two diagrams are the following:
\ba
	\label{texch}
	M_t &=& \frac{ie^2}{(\qbar - p_1)^2}
	\lsp{\qbar\pm} \epsh{{h_1}}{p_1,k_1} (\hat{\qbar} - \hat p_1)
             \epsh{{h_2}}{p_2,k_2} \rsp{q\pm},
\ret \label{uexch}
	M_u &=& \frac{ie^2}{(\qbar - p_2)^2}
	\lsp{\qbar\pm} \epsh{{h_2}}{p_2,k_2} (\hat{\qbar} - \hat p_2)
             \epsh{{h_1}}{p_1,k_1} \rsp{q\pm}.
\ea                                                  
It is straightforward to check, using Eq.\eqnum{aepshat}, that if both photons
have the same helicity then $M=M_u+M_t=0$. In fact by choosing the reference
momenta to be equal to $q$ ($\qbar$) when the common photon helicities are the
same as (opposite to) the electron helicity, we easily find that both $M_t$ and
$M_t$ identically vanish. 
Therefore the only processes which contribute have photons with opposite
helicity. 
By using the above criterion for the assignment of the reference momentum, and
by assuming for definiteness that photon 1 has the same helicity as the
electron, we can easily transform Eqs.\eqnum{texch}\ and \eqnum{uexch}\ into
the following expressions:
\ba
	\label{texch2}           
	M_t &=& \frac{-ie^2}{2(\qbar p_1)}
	\frac{(\pm)\sqrt 2}{\sppr{q\mp}{p_1\pm}}
	\frac{(\mp)\sqrt 2}{\sppr{\qbar\pm}{p_2\mp}}
	\sppr{\qbar\pm}{p_1\mp}
	\lsp{q\mp}  (\hat{\qbar}-\hat p_1)  \rsp{\qbar\mp}
	\sppr{p_2\mp}{q\pm} 
\ret \nn                
	&=&  -2ie^2 
\frac{\sppr{p_1\pm}{\qbar\mp}\sppr{q\mp}{p_2\pm}}{
		\sppr{p_1\mp}{\qbar\pm}\sppr{\qbar\pm}{p_2\mp}}  
	\= 2e^2 e^{i\phi} \sqrt{\frac{(qp_2)}{(\qbar p_2)}}  \quad ,
\ret \label{uexch2}                                                    
	M_u &=& \frac{-ie^2}{2(\qbar p_2)}
	\frac{(\pm)\sqrt 2}{\sppr{q\mp}{p_1\pm}}
	\frac{(\mp)\sqrt 2}{\sppr{\qbar\pm}{p_2\mp}}
	\sppr{\qbar\pm}{\qbar\mp}
	\lsp{p_2\mp} (\hat{\qbar}-\hat p_2) \rsp{p_1\mp}
	\sppr{q\mp}{q\pm} \equiv 0.
\ea                                                  
Notice that even though in this gauge only the $t$-channel Feynman
diagram is different from zero, it nevertheless contributes to the $u$-pole 
because of the singularity present in the definition of the polarization
vector. This is a common feature of gauges defined in terms of arbitrary
vectors, such as axial gauges.

Squaring and summing over the various non vanishing helicity configurations, we
finally obtain:             
\be    
	\sum_{pol} \vert M(e^+e^-\to\gamma\gamma) \vert ^2  \=
	8 e^4 \frac{u^2+t^2}{ut}.
\ee


\section{Appendix: Feynman Rules and the $SU(N)$ Algebra}
Our Feynman rules follow the Bjorken and Drell conventions. In this Appendix we
will collect them in the form which is most appropriate for their use with the
helicity amplitude and the dual color expansion.

By convention we will always assume all of the particles as outgoing and we
will order the indices clock-wise. Positive-
and negative-helicity quarks are represented by {\em bra} spinors:
\be
	\lsp{q\pm},
\ee           
while positive- and negative antiquarks are represented by {\em ket} spinors:
\be
	\rsp{\qbar\mp}.
\ee                  
Positive- and negative-helicity gluons are given by the the positive- and
negative polarization vectors introduced in the text and in Appendix~A.

The fermion and vector propagators are given, respectively, by:
\be
	i \; \frac{\hat q}{q^2} \; \delta^{ij}
	 \quad , \quad 
	-\frac{i}{p^2} \; g_{\mu\nu} \; \delta^{ab} ,
\ee                                          
where the indices $i,j$ and $a,b$ are the color indices of the fermion and
adjoint representations, respectively.
Since we will always be calculating helicity amplitudes, therefore dealing
with physical external gluons, the choice of Feynman gauge is technically
equivalent to any other choice. 

The fermion-antifermion-gluon vertex is given by:
\be
	\frac{i}{\sqrt 2} \; g \; (\l^a)_{ij} \; \gamma_\mu
\ee
The $\sqrt 2$ is a consequence of our choice of normalization for the \l\
matrices and their algebra:
\ba    \label{sun1}
	[\l^a,\l^b] \= i f^{abc} \l^c   \quad &,& \quad
	\tr (\l^a\l^b) \= \delta^{ab}
\ret   \label{sun2}
	\sum_a \; (\l^a)_{ij}(\l^a)_{kl} &=&
	\delta_{il}\delta_{jk} \;-\; \frac{1}{N} \delta_{ij}\delta_{kl}.
\ea                 
This is not the usual normalization, but this choice prevents the proliferation
of powers of $\sqrt 2$ which would otherwise appear in the calculation of the
matrix elements and, independently, in the squaring of the color structures. All
of these factors of 2 eventually cancel out, and our convention enforces these
cancellations since the start.
The three-gluon vertex is given by:
\ba
	V(p_1,p_2,p_3)_{\mu_1\mu_2\mu_3} &=&
   	-\frac{1}{\sqrt 2} \; g \; f^{a_1a_2a_3} \;
	F(p_1,p_2,p_3)_{\mu_1\mu_2\mu_3},
\ret                            
	F(p_1,p_2,p_3)_{\mu_1\mu_2\mu_3} &=&
	[(p_1-p_2)_{\mu_3} g_{\mu_1 \mu_2} +
	 (p_2-p_3)_{\mu_1} g_{\mu_2 \mu_3} +	
	 (p_3-p_1)_{\mu_2} g_{\mu_3 \mu_1} ].
\ea                                          
The $\sqrt 2$ is a consequence of our normalization of the $SU(N)$
algebra.
The three-gluon vertex breaks up into two pieces, corresponding to the two
possible color structures this vertex can contribute to:
\be                                                          
   	\frac{i}{\sqrt 2} \; g \; 
	[ \tr (\l^{a_1}\l^{a_2}\l^{a_3})  \; - \;
	  \tr (\l^{a_3}\l^{a_2}\l^{a_1})  ]
	 \; F(p_1,p_2,p_3)_{\mu_1\mu_2\mu_3}.
\ee                       
In the calculation of a dual amplitude, where the diagrams are determined by a
specified ordering of the gluons, we should use only the color structure
corresponding to the assigned ordering:
\be                                                          
   	\frac{i}{\sqrt 2} \; g \; 
	 \tr (\l^{a_1}\l^{a_2}\l^{a_3})  
	 \; F(p_1,p_2,p_3)_{\mu_1\mu_2\mu_3}.
\ee                       
For example, the $s$-channel diagram contributing to the dual amplitude
$m(1,2,3,4)$ would be given by:
\ba                                                          
   	& (\frac{i}{\sqrt 2} g )^2 & \sum_b \;
	 \tr (\l^{a_1}\l^{a_2}\l^{b})            
	 \; F(p_1,p_2,P)_{\mu_1\mu_2\nu} \; [-\frac{i}{P^2}] \;
	 \tr (\l^{a_3}\l^{a_4}\l^{b})  
	 \; F(p_3,p_4,P)_{\mu_3\mu_4\nu}
\nn	\ret
   =   & (\frac{i}{\sqrt 2} g )^2 &
	 \tr (\l^{a_1}\l^{a_2}\l^{a_3}\l^{a_4}) [-\frac{i}{P^2}] \;
	 \; F(p_1,p_2,P)_{\mu_1\mu_2\nu}                                   
	 \; F(p_3,p_4,P)_{\mu_3\mu_4\nu} ,
\ea	                                  
with $P=p_1+p_2$ and where a term proportional to $\tr
(\l^{a_1}\l^{a_2})\tr(\l^{a_3}\l^{a_4})$ vanishes because of the anti-symmetry
of the functions $F$.
                                                        
In an analogous way, and using the standard four-gluon vertex Feynman rule, one
can write the four-gluon {\em dual} vertex, corresponding to the permutation
$(1,2,3,4)$:
\be
	i \frac{g^2}{2}  \; \tr (\l^{a_1}\l^{a_2}\l^{a_3}\l^{a_4}) \;
	( 2 g_{\mu_1\mu_3}g_{\mu_2\mu_4} \;-\; g_{\mu_1\mu_4}g_{\mu_2\mu_3} 
	\;-\; g_{\mu_1\mu_2}g_{\mu_3\mu_4} )
\ee
The sum of this vertex plus the other 5 obtained by permuting the indices gives
the standard four-gluon vertex, as can be easily checked.                      

\newpage
\subsection{Summary of Feynman Rules}
Here we summarize the Color truncated Feynman rules, where all the 
vertices are cyclically ordered and all momenta are outgoing.
Demonstrating that the subamplitudes $m(g_1,g_2,g_3,g_4)$ 
and $m(\q,g_1,g_2,\qbar)$ are 
gauge invariant is an easy way to check the consistency of our 
conventions.
{\renewcommand{\arraystretch}{1.8}
        \begin{itemize}
\item External, outgoing fermion, $F$,  helicity $\pm$:
	\be 	 \lsp{F \pm}.                       \ee
\item External, outgoing anti-fermion, $\overline{F}$, helicity $\pm$:
	\be   \rsp{\overline{F} \mp}.  \ee
\item External, outgoing vector, momentum $p$, reference $k$, helicity 
$\pm$:
	\ba  & \eps{\mu}{\pm}{p,k} =\pm                      
                            	   \frac{\lsp{p\pm} \gamma_\mu \rsp{k\pm}}
					{\sqrt{2} \sppr{k\mp}{p\pm}} ,
	 \\  
	 & \eps{{}}{\pm}{p,k} \cdot \gamma =                                 
					\pm \frac{\sqrt 2}{\sppr{k\mp}{p\pm}}
				(\rsp{p\mp}\lsp{k\mp} + \rsp{k\pm}\lsp{p\pm}).
	\ea
\item Fermion propagator, momentum $q$, in the direction of the fermion 
arrow:
	\be  i \; \frac{\hat q}{q^2}. 	 \ee
\item Vector propagator, momentum $p$:
	\be  -i \; \frac{g_{\mu\nu} }{p^2}.  \ee
\item Fermion-vector-antifermion vertex, order $(F V \overline{F})$:
	\be   i \; \frac{g}{\sqrt 2} \; \gamma_\mu.   \ee
\item Tri-Vector vertex, order $(123)$, all momenta outgoing from vertex:
	\be   i \; \frac{g}{\sqrt 2} \; 
	[(p_1-p_2)_{\mu_3} g_{\mu_1 \mu_2} +
	 (p_2-p_3)_{\mu_1} g_{\mu_2 \mu_3} +	
	 (p_3-p_1)_{\mu_2} g_{\mu_3 \mu_1} ].   \ee
\item Quartic-Vector vertex, order $(1234)$:
	\be  i \frac{g^2}{2}  \; 
	( 2 g_{\mu_1\mu_3}g_{\mu_2\mu_4} \;-\; g_{\mu_1\mu_4}g_{\mu_2\mu_3} 
	\;-\; g_{\mu_1\mu_2}g_{\mu_3\mu_4} ).  \ee
        \end{itemize}
}
 
\newpage

\section{Appendix: Squares and Color Sums}
For the sake of definiteness we will choose the fermion color representation to
be the fundamental representation. With our conventions the \l\ matrices are
hermitian and are normalized by 
\be    
	tr(\l^a\l^b)\;=\; \delta^{ab},
\ee
and satisfy the following identities which 
are useful in reducing the color sums of 
products of traces:
\ba    
	\sum_{a=1}^{N^2-1} (\l^a)_{i_1 j_1}
	(\l^a)_{i_2 j_2} &=&
	\delta_{i_1 j_2} \delta_{i_2 j_1} \; - \;
	{1 \over N} \delta_{i_1 j_1} \delta_{i_2 j_2} \\
	\left[ \tr (\l^{a_1} \l^{a_2} \dots \l^{a_n})\right]^{*}
	&=&
	\tr (\l^{a_n} \dots \l^{a_2}\l^{a_1}) .
\ea     
As a short-hand notation, we will define:
\be                   
       	(a_1a_2\dots a_n) \= \tr \lstring{a}{n}
\ee
In squaring the gluon amplitudes and summing over colors, we will have to carry
out sums of the following form:
\be
	\sum_{a_1,\dots,a_n=1}^{N^2-1}
       	(a_1a_2\dots a_n) (b_1b_2\dots b_n)^*
\ee                                          
where $\{b\}$ is a permutation of $\{a\}$.  Using the cyclic property of the trace,
the properties under complex conjugation and Equation~\eqnum{sun2}, we can   
always rearrange the permutation $\{b\}$ so as to reduce the sum in the 
following way:                                                         
\ba            
   	\sum_{a_1,\dots,a_{n-1}=1}^{N^2-1}
	\sum_{a_n}
	(a_1a_2\dots a_n) (a_n a_{m_{n-1}} \dots a_{m_1}) &=&
	\sum_{a_1,\dots,a_{n-1}=1}^{N^2-1}            
	[ (a_1 \dots a_{n-1} a_{m_{n-1}} \dots a_{m_1}) \nn \ret      
	&-& \frac{1}{N}(a_1 \dots a_{n-1})(a_{m_{n-1}} \dots a_{m_1}) ].
\ea                                                                     
The first term can have either of the following forms:
\be                                   
	\sum_{a_{n-1}} \; (\Lambda_1 a_{n-1}a_{n-1} \Lambda_2) \=
	\frac{(N^2-1)}{N}(\Lambda_1\Lambda_2) ,
\ee
or:
\be
	\sum_{a_{n-1}} \; (\Lambda_1 a_{n-1} \Lambda_2 a_{n-1} ) \=
	(\Lambda_1)(\Lambda_2) \;-\; \frac{1}{N}(\Lambda_1\Lambda_2) .
\ee                                                                   
In either case the sum over colors can be iterated using the formulas given so
far, until all of the terms that will develop are proportional to $(a_1a_1) =
(N^2-1)$. The leading contributions in $N$ come from $\{b\}=\{a\}$, in which
case, to the leading order in $N$:
\be
	\sum_{a_1,\dots,a_n=1}^{N^2-1}
	(a_1a_2\dots a_n) ( a_1 a_2\dots  a_n)^*  \=
	N^{n-2}(N^2-1) \; ( \; 1 \; + \; {\cal O}(1/N^2) \;)
\ee                       

For the readers convenience we will collect here some useful formulas involving
traces and squares of \l\ matrices:
\ba
	(ab) &=& \delta_{ab}  \\
	\sum_a (\l^a\l^a)_{ij} &=& \frac{N^2-1}{N} \delta_{ij} \\
	\sum_{ab} (ab)(ab) &=& N^2-1 \\
	\sum_{abc}(abc)(cba) &=& (N^2-2) \left( \frac{N^2-1}{N} \right) \\
	\sum_{abc}(abc)(abc) &=& -2 \left( \frac{N^2-1}{N} \right) \\
	\sum_{abc} f^{abc}f^{abc} &=& 2N(N^2-1)
\ea
    
As an explicit example we will now calculate the square of the 4-gluon
amplitude:                                          
\be
	M^{(4)} \= \sum_{\{2,3,4\}} (a_1a_2a_3a_4) \; m(1,2,3,4)
	\= \sum_{\{2,3,4\}'} [a_1a_2a_3a_4] \; m(1,2,3,4),
\ee                          
where the prime indicates that only permutations inequivalent under reflection
( $(1234) \to (4321)$ ) should be considered, and where we introduced the
following notation:
\ba
	[a_1a_2\dots a_n]   &=& (a_1a_2\dots a_n) + (-1)^n (a_n\dots a_2a_1) ,
\ret
	[a_1a_2\dots a_n]^* &=& (-1)^n [a_n\dots a_2a_1] .
\ea                                                       
The sum over colors of the squared 4-gluon amplitude can be then written as:
\ba
	\sum_{col} \vert M^{(4)} \vert ^2 &=&
	\sum_{\{2,3,4\}'} m(1,2,3,4) \sum_{col} 
	[a_1a_2a_3a_4] \; \left( [a_1a_2a_3a_4]^* \; m^*(1,2,3,4) \right.
+ \nn \ret
      &&  	\left.
	 [a_1a_3a_2a_4]^* \; m^*(1,3,2,4) 
	+[a_1a_2a_4a_3]^* \; m^*(1,2,4,3) \right).
\ea                                                                        
It is simple to prove the following equations:
\ba
    \sum_{col} [a_1a_2a_3a_4] [a_1a_2a_3a_4]^* &=& \sum_{col} [a_1a_2a_3a_4]^2
\ret 
    \sum_{col} [a_1a_2a_3a_4] [a_1a_2a_4a_3]^* &=& \sum_{col} [a_1a_2a_3a_4]^2
	+ N [a_1a_2a_3]^2
\ret                                      
    \sum_{col} [a_1a_2a_3a_4] [a_1a_3a_2a_4]^* &=& \sum_{col} [a_1a_2a_3a_4]^2
	+ N [a_1a_2a_3]^2              
\ret 
	\sum_{col} N [a_1a_2a_3]^2 &=& -2 N^2(N^2-1).
\ea                                       
By using these equations and the Dual Ward Identity, Equation~\eqnum{ward}, we
therefore obtain:                                                         
\ba                          
	\sum_{col} \vert M^{(4)} \vert ^2 &=&
     	\sum_{\{2,3,4\}'} \vert m(1,2,3,4) \vert ^2 (- \sum_{col} 
	N [a_1a_2a_3][a_1a_2a_3] ) \nn \ret
     &=&                           
       2 N^2(N^2-1) \sum_{\{2,3,4\}'} \vert m(1,2,3,4) \vert ^2 
				\nn \ret                        
     &=&                           
     	N^2(N^2-1) \sum_{\{2,3,4\}} \vert m(1,2,3,4) \vert ^2 .
\ea
Summing over the different helicity configurations, finally, gives
Equation~\eqnum{pt4g}.

In analogous way one can show the vanishing of the subleading terms in the
square of the 5-gluon process. As for the 6-gluon case, by using the DWI one
can show that the square takes the following form:
\ba
	\sum_{col} \vert M^{(6)} \vert ^2 &=& N^4(N^2-1) 
	\sum_{\{23456\}}  \left ( \vert m(123456) \vert ^2  \right.
\nn \ret                                                         
	&+&        \left.
	\frac{2}{N^2} \; m(123456)^* \times 
	[m(135264) + m(153624) + m(136425)] \right ),
\ea
where the expressions of the dual amplitudes $m(123456)$ for the various
helicity configurations were given in Section~\sectexp.

Finally, we give here 
the structure of the amplitude squared for the
processes $(\bar{q}qgg)$, $(\bar{q}qggg)$ and $(\bar{q}qgggg)$.
To keep the following formulae as simple as possible,
we introduce the following notation for the quark \subs:
\be
        m(q,g_I,g_J,\dots,g_L,\qbar) = (I,J,\dots,L),
\ee
where $(I,J,\dots,L)$ is an arbitrary permutation of $(1,2,\dots,4)$.
From the expansion of the amplitude in the usual color basis,
\be                                                          
         M(q,g_1,\dots,g_n,\qbar)=
        \sum_{\lbrace 1,\dots,n\rbrace}\;
        ( \lambda^1\lambda^2\dots\lambda^n )_{ij}
      \; (1,2,\dots,n),
\ee
we obtain the following expression:
\be
        \sum_{colors}\;\vert  M(q,g_1,\dots,g_n,\qbar)\vert^2
        \;=\;
        {(N^2-1) \over N^{n-1} }
        \sum_{j=0}^{n-1} N^{2j}\;
        \sum_{\lbrace 1,\dots,n\rbrace} H_j(1,2,\dots,n).
\ee
For $n=2,3,4$ the functions $H_j$ are given by:
\begin{itemize}
        \item $n=2$
\ba     &&
        H_1(1,2) =  \vert (1,2) \vert ^2
        \\[0.1in] &&
        H_0(1,2) =  - (1,2)^*[(1,2)+(2,1)]
\ea
        \item $n=3$
\ba     &&
        H_2(1,2,3) =  \vert (1,2,3) \vert ^2
        \\[0.1in] &&
        H_1(1,2,3) = - (1,2,3)^* [2\;(1,2,3)+(1,3,2)+(2,1,3)-(3,2,1)]
        \\[0.1in] &&                                         
        H_0(1,2,3) = (1,2,3)^*\sum_{\lbrace I,J,K \rbrace} (I,J,K)
\ea
        \item $n=4$
\ba     &&
        H_3(1,2,3,4)=  \vert (1,2,3,4) \vert ^2,
        \\[0.1in] &&
        H_2(1,2,3,4)= (1,2,3,4)^*  [
         -3\;(1,2,3,4) - (1,2,4,3) - (1,3,2,4) \nn
        \\ &&
\quad\quad\quad\quad\quad
- (2,1,3,4)
+ (1,4,3,2) + (3,2,1,4) + (3,4,1,2) \nn
        \\ &&
\quad\quad\quad\quad\quad  + (3,4,2,1) + (4,2,3,1) + (4,3,1,2) ],
        \\[0.1in] &&
        H_1(1,2,3,4)=  (1,2,3,4)^*  [ M(1,2,3,4) - M(4,3,2,1) ]
        \\ &&
        M(1,2,3,4) = 3\;(1,2,3,4) + 2\;(1,2,4,3) + 2\;(1,3,2,4)
         + 2\;(2,1,3,4) \nn
        \\ &&
\quad\quad\quad\quad  + (1,3,4,2) + (1,4,2,3) + (2,1,4,3)
                + (2,3,1,4) + (3,1,2,4) ,
        \\[0.1in]  &&
        H_0(1,2,3,4)=  -(1,2,3,4)^*  \sum_{\lbrace I,J,K,L\rbrace}(I,J,K,L).
\ea
\end{itemize}
 
The formulae for $n=2,3$ can be used to compare our results with
the expressions already known. In doing this it is useful to apply the DWI to
the functions $H_j$ and use the gluino \subs\
with non-adjacent fermions
as auxiliary functions.

The \subs\ for the various helicity configurations were given in
Section~\sectexp.

\section{Appendix: Numerical Evaluation of the Spinor Products}

To calculate the matrix element squared for a given process, it is 
frequently easier to evaluate the sub-amplitudes as complex numbers
and then form the appropriate square using the color algebra of the 
previous Appendix. For all the processes discussed in this review,
the sub-amplitudes can be calculated from sums of products of spinor
products. Therefore, we need an algorithm for evaluating these spinor 
products. Given that both $ \langle ij \rangle $ 
and $ [ij] $ are complex square roots of the Lorentz invariant
$ S_{ij} ~\equiv~ (p_i~+~p_j)^2 $, 
\ba
&& \langle ij \rangle \equiv \sqrt{|S_{ij}|} ~\exp{(i\phi_{ij})},
\\    {}\nn \\
&& [ij] \equiv \sqrt{|S_{ij}|} ~\exp{(i\tilde{\phi}_{ij})} .
\ea
If both momenta having positive energy, the phase factor $\phi_{ij}$
is defined, in a popular representation of 
the gamma matrices, by

\ba
\cos \phi_{ij} & = & { (p^1_i p^+_j ~-~ p^1_j p^+_i) \over 
                     \sqrt{ p^+_i p^+_j S_{ij} } } 
\nn \\    {}\nn \\
\sin \phi_{ij} & = & { (p^2_i p^+_j ~-~ p^2_j p^+_i) \over 
                     \sqrt{ p^+_i p^+_j S_{ij} } }.
\ea
Where $p^{\pm} = (p^0 ~\pm~ p^3)$ and since $p^2_i = 0$, the spinor product
for this representation of gamma matrices 
are  undefined for a momentum vector in the minus 3 direction. 
If one or more of the momenta in $ \langle ij \rangle $ have negative 
energy, $\phi_{ij}$ is calculated with minus the momenta with negative 
energy and then $ n \pi /2 $ is added to $\phi_{ij}$ where $n$ is the 
number of negative momenta in the spinor product.  The 
associated phase factor, $\tilde{\phi}_{ij}$ ,
for $[ij]$ can be 
calculated from S$_{ij}$ using the identity $S_{ij} \equiv \langle 
ij \rangle ~ [ji] $. The above identities can be used to evaluate the 
spinor products with approximately the same amount of computational
effort as the evaluation of $\sqrt{S_{ij}}$.

Similarly, the matrix element squared can usually be written 
simply as a sum of traces of momentum vectors, see \cite{mpx88}. 
If these traces are expanded, the                            
resulting expressions are extremely cumbersome. However, the phase 
factors defined above can be used to evaluate these traces in an 
efficient manner.
Consider the trace of a large string of light-like momentum vectors
with all vectors having positive energy, then
\ba
Tr(\hat{P}_1 \hat{P}_2 \hat{P}_3 \cdots \hat{P}_{2n} )
& = & 
 \cp{1}{2} \sp{2}{3} \cdots \sp{2n}{1} 
					~+~ \sp{1}{2} \cp{2}{3} \cdots \cp{2n}{1} \nonumber \\
& = & 2~ \sqrt{S_{12}S_{23} \dots S_{2n1}}
~\cos(\phi_{12}-\phi_{32}+\phi_{34}-  \dots -\phi_{1~2n} ).
\ea
Where the identity for positive energy spinor products,
$\tilde{\phi}_{ij}=-\phi_{ji}$, has been used.
For traces involving $\gamma_5$, the corresponding identity is
\ba
 Tr(\hat{P}_1 \hat{P}_2 \hat{P}_3 \cdots \hat{P}_{2n} \gamma_5) 
& = & 
 \cp{1}{2} \sp{2}{3} \cdots \sp{2n}{1} 
~-~ \sp{1}{2} \cp{2}{3} \cdots \cp{2n}{1} \nonumber \\
&=& -2i ~\sqrt{S_{12}S_{23} \dots S_{2n1}}
~ \sin(\phi_{12}-\phi_{32}+\phi_{34}-  \dots -\phi_{1~2n} ).
\ea
If $m$ of the vectors in the string 
have negative energy, multiple these vectors by $(-1)$ and 
use all the resulting positive energy vectors 
to calculate the above trace. The original trace is obtained 
by multiplying this answer by $(-1)^m$. Traces involving vectors which 
are massive can be also treated by writing the massive vector as 
a sum of two light-like vectors (usually the decay products
\cite{kls85b,gk85b}).

\newpage
\listoffigures
\newpage
\begin{figure}[h]
\centering
\includegraphics[width=4.cm,angle=-90]{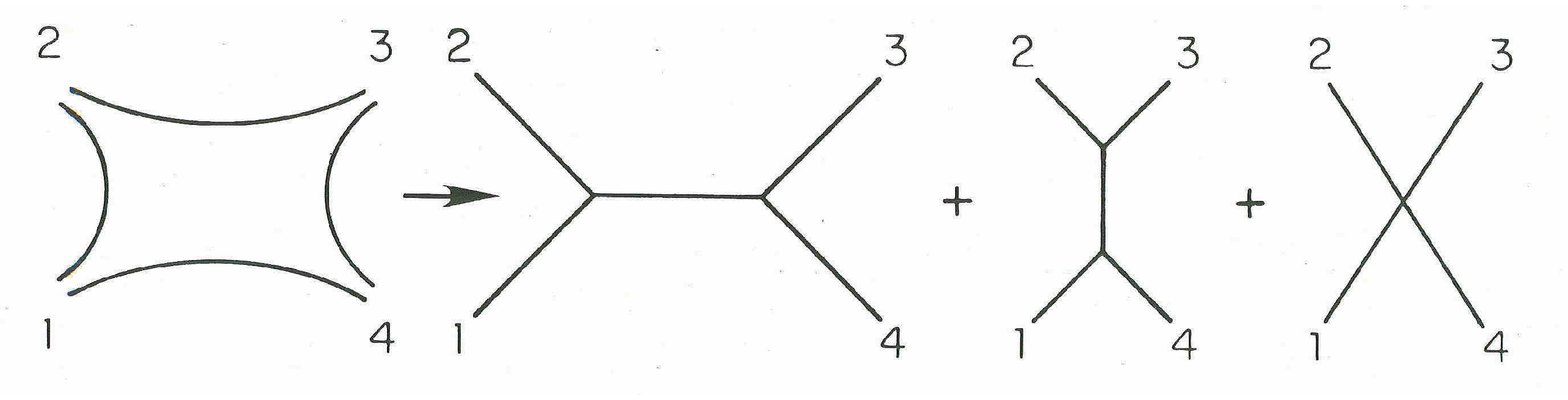}
\caption{The zero-slope limit of the four gluon string diagram in 
terms of Feynman diagrams.}
\label{four.fig}         
\vspace{0.2in}
\end{figure}

\begin{figure}[h]
\centering
\includegraphics[width=8.cm,angle=90]{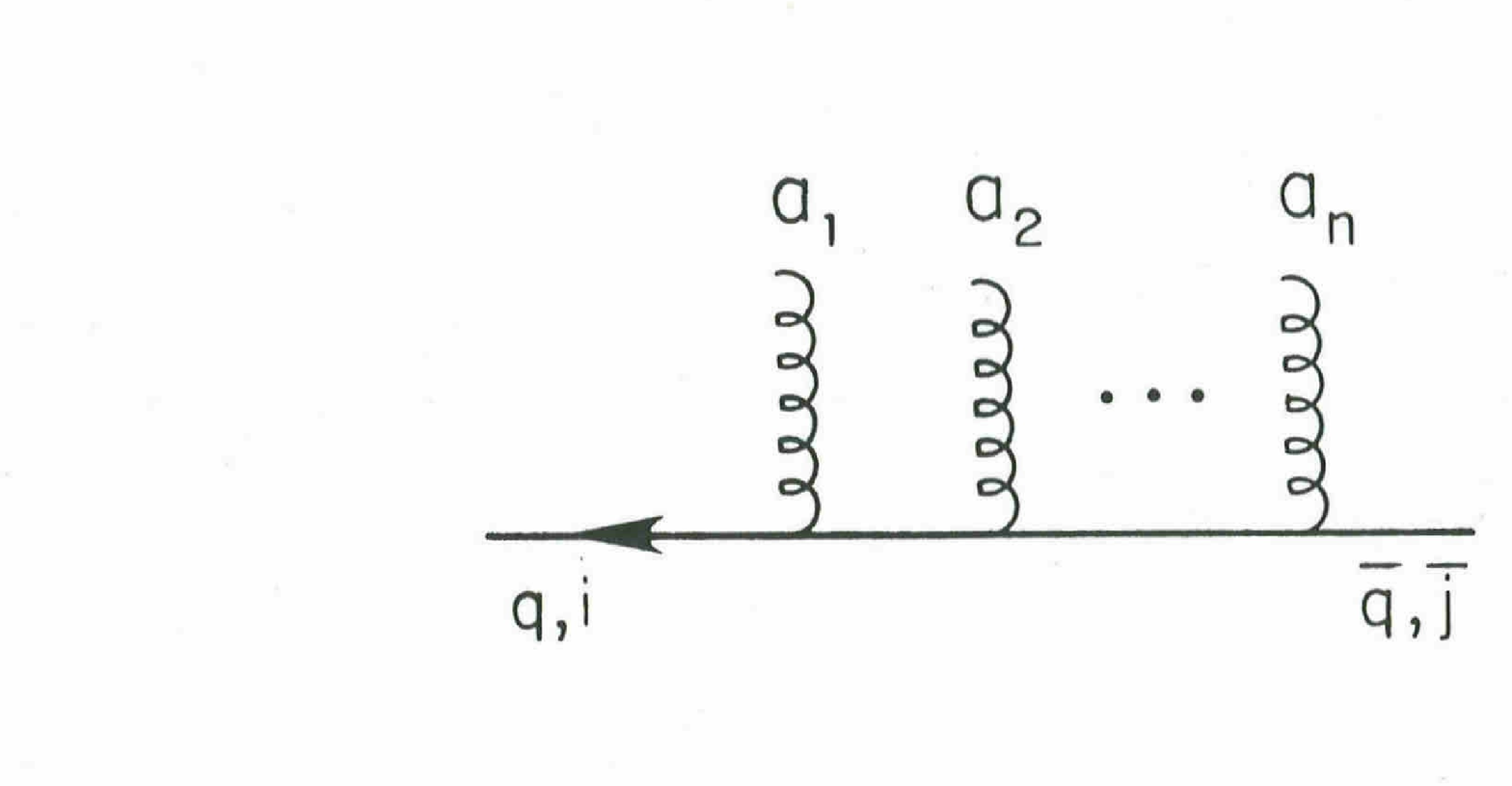}
\caption{QED-type diagrams.}
\label{colfigtwo}         
\vspace{0.2in}
\end{figure}

\begin{figure}[h]
\centering
\includegraphics[width=18.cm]{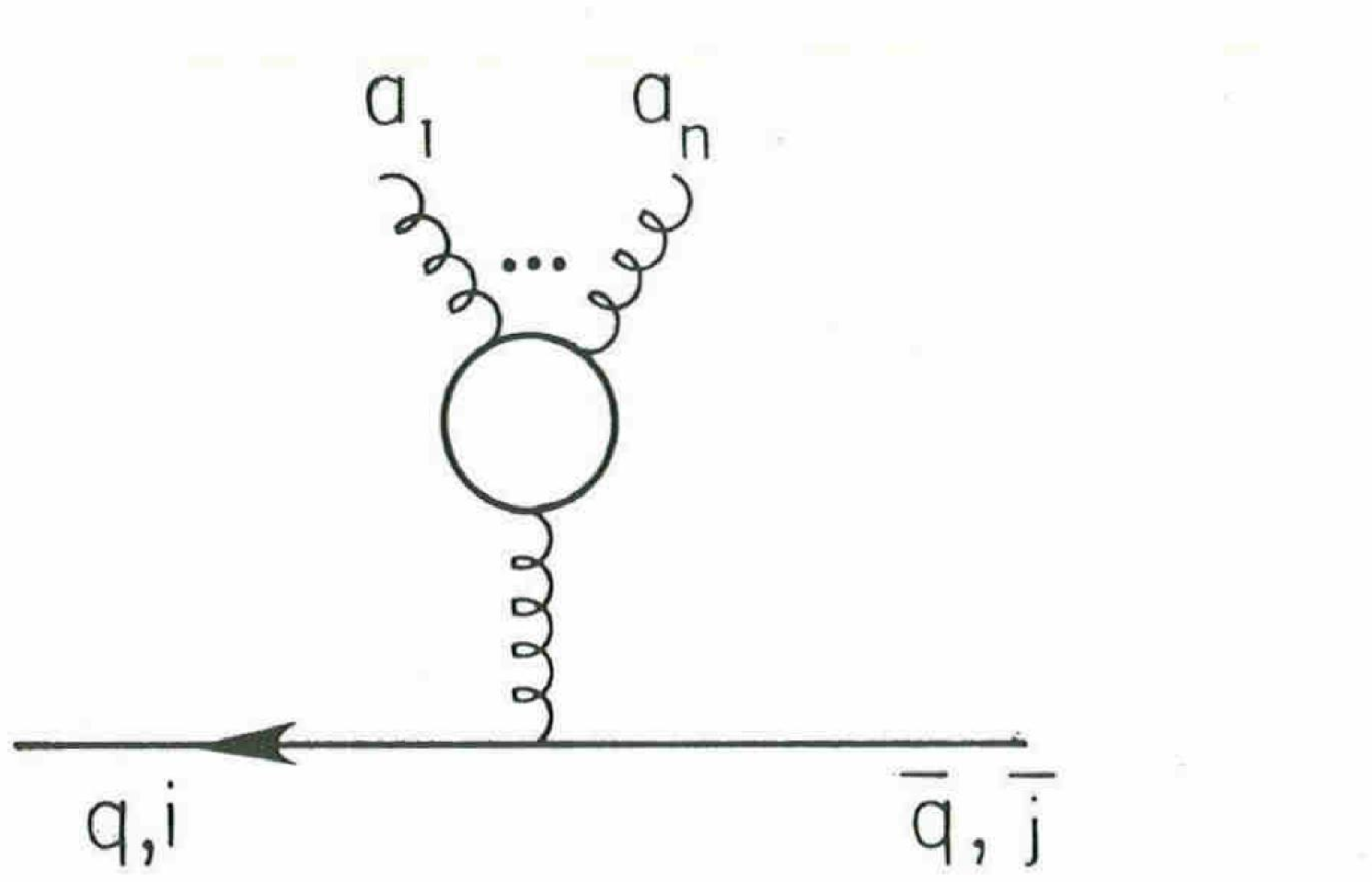}
\caption{Gluon tree off a quark line.}
\label{colfigthree}         
\vspace{0.2in}    
\end{figure}

\begin{figure}[h]
\centering
\includegraphics[width=18.cm]{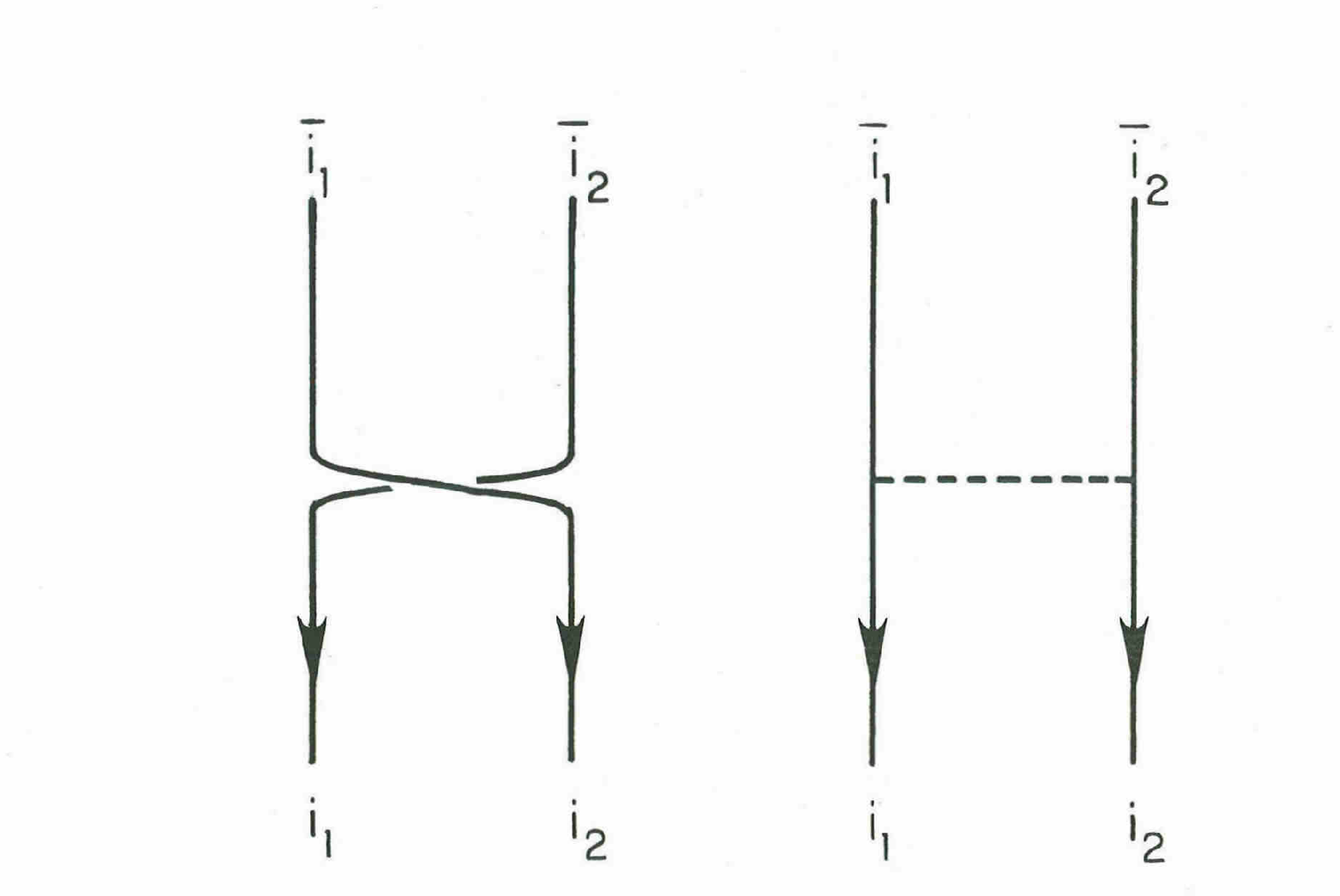}
\caption{The color flows for quark-antiquark scattering.}
\label{colfigfour}                     
\vspace{0.2in}   
\end{figure}

\begin{figure}[h]
\centering
\includegraphics[width=18.cm]{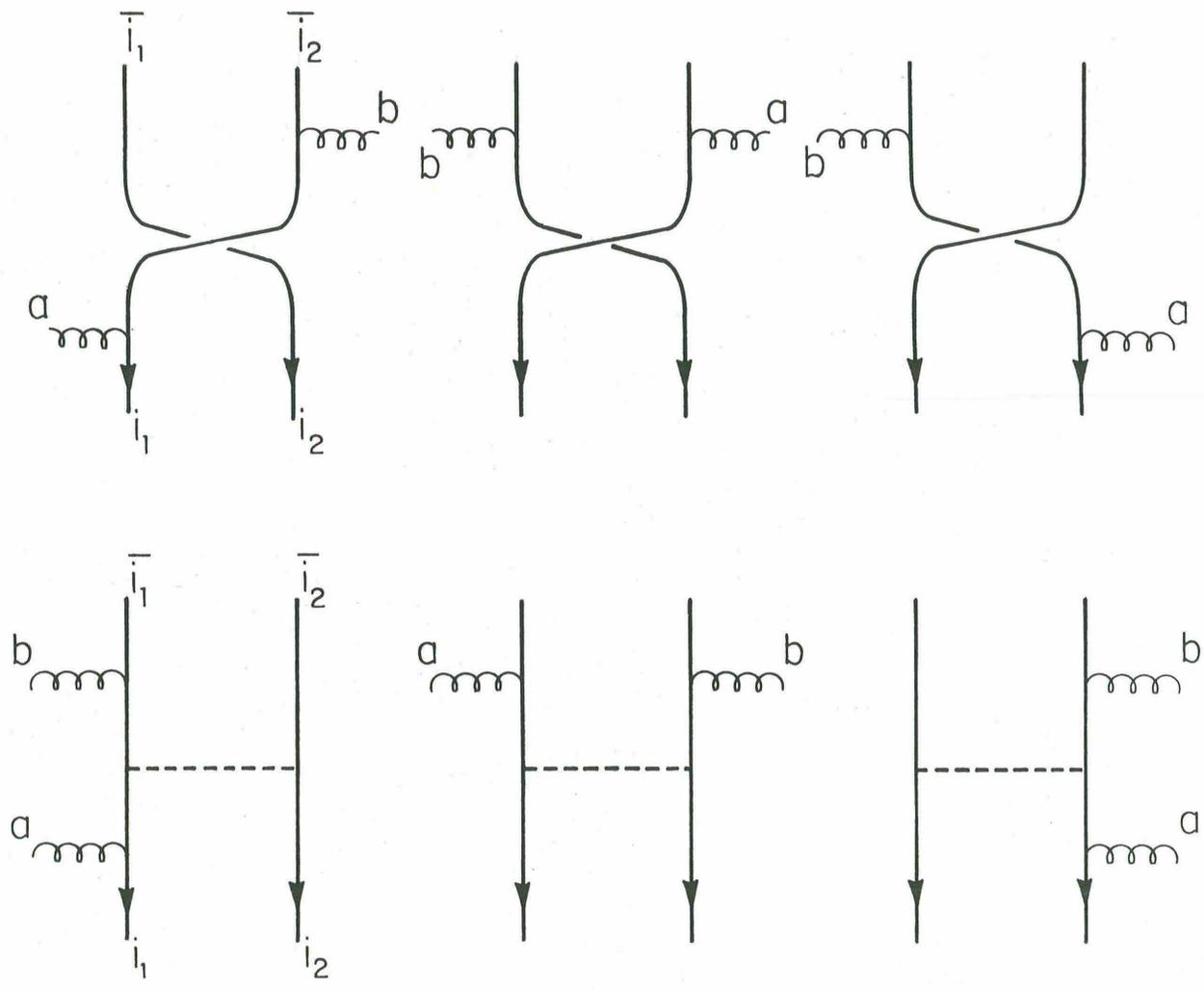}
\caption{The color flows for quark-antiquark scattering with emission of two 
gluons.}
\label{colfigfive}         
\vspace{0.2in}   
\end{figure}

\begin{figure}[h]
\centering
\includegraphics[width=18.cm]{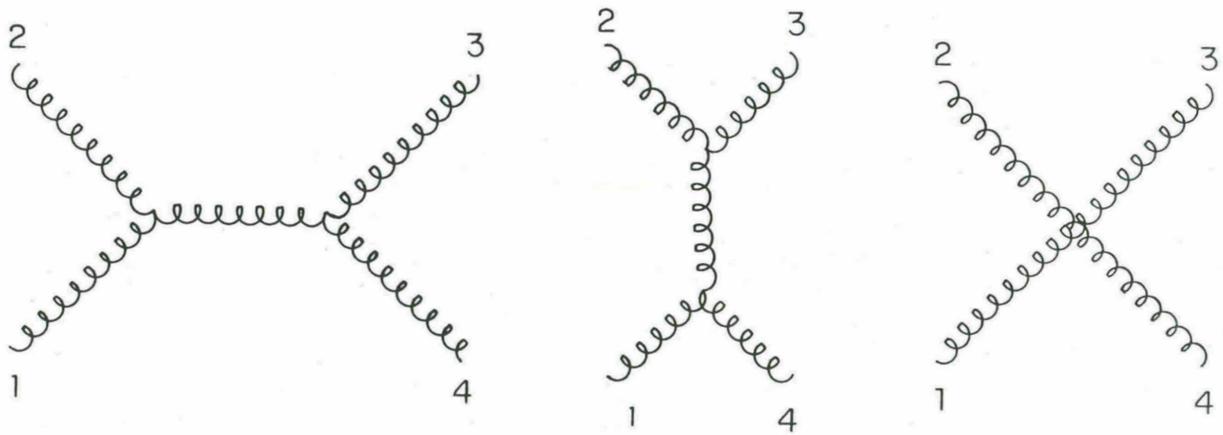}
\caption{The three diagrams contributing to the subamplitude 
$m(1,2,3,4)$.}
\label{exafigone}         
\vspace{0.2in}  
\end{figure}

\begin{figure}[h]
\centering
\includegraphics[width=18.cm]{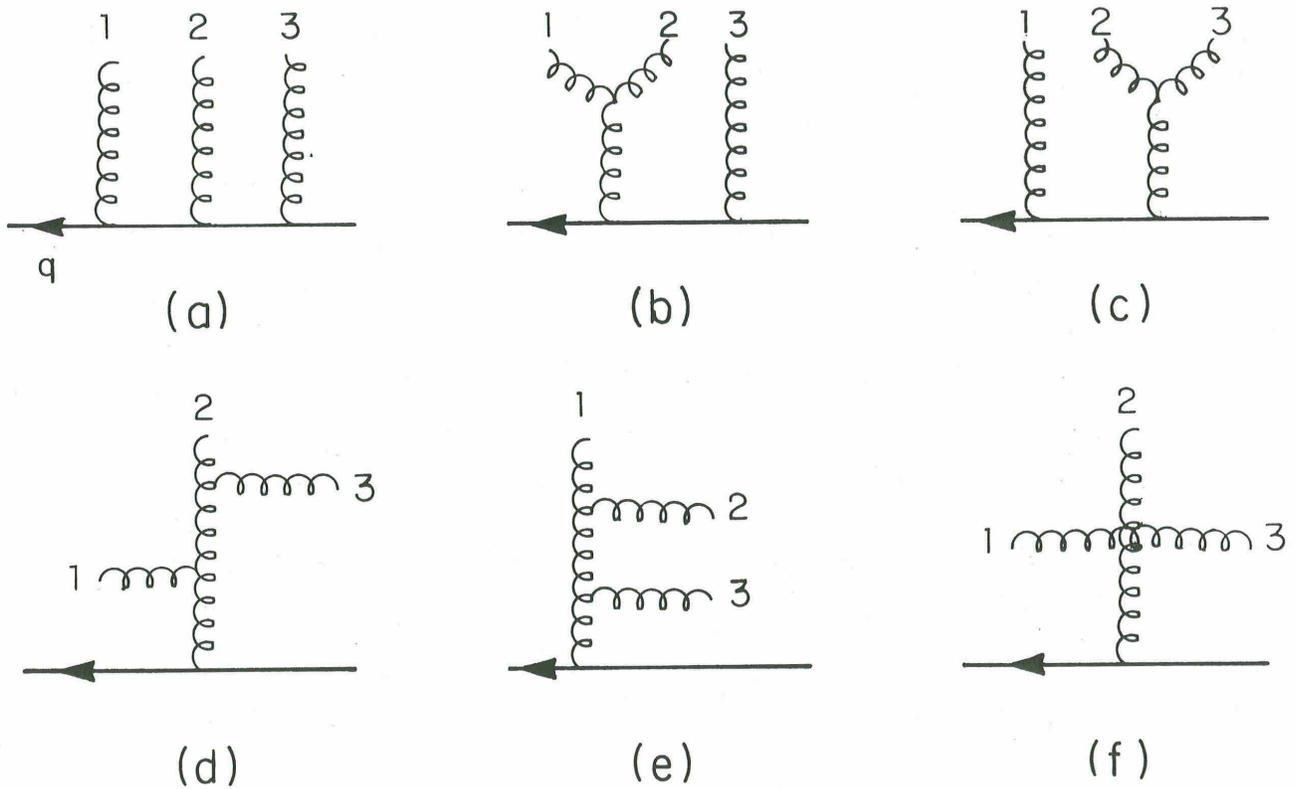}
\caption{The diagrams contributing to the subamplitude $m(q,1,2,3,\qbar)$.}
\label{exafigtwo}                                                      
\vspace{0.2in}  
\end{figure}

\begin{figure}[h]
\centering
\includegraphics[width=18.cm]{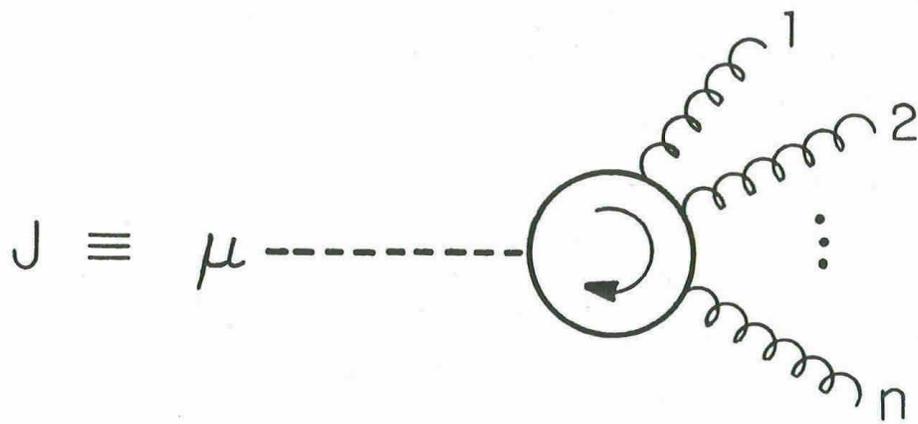}
\caption{The color ordered gluonic current.}
\label{rsqfigone}                  
\vspace{0.2in}
\end{figure}

\begin{figure}[h]
\centering
\includegraphics[width=18.cm]{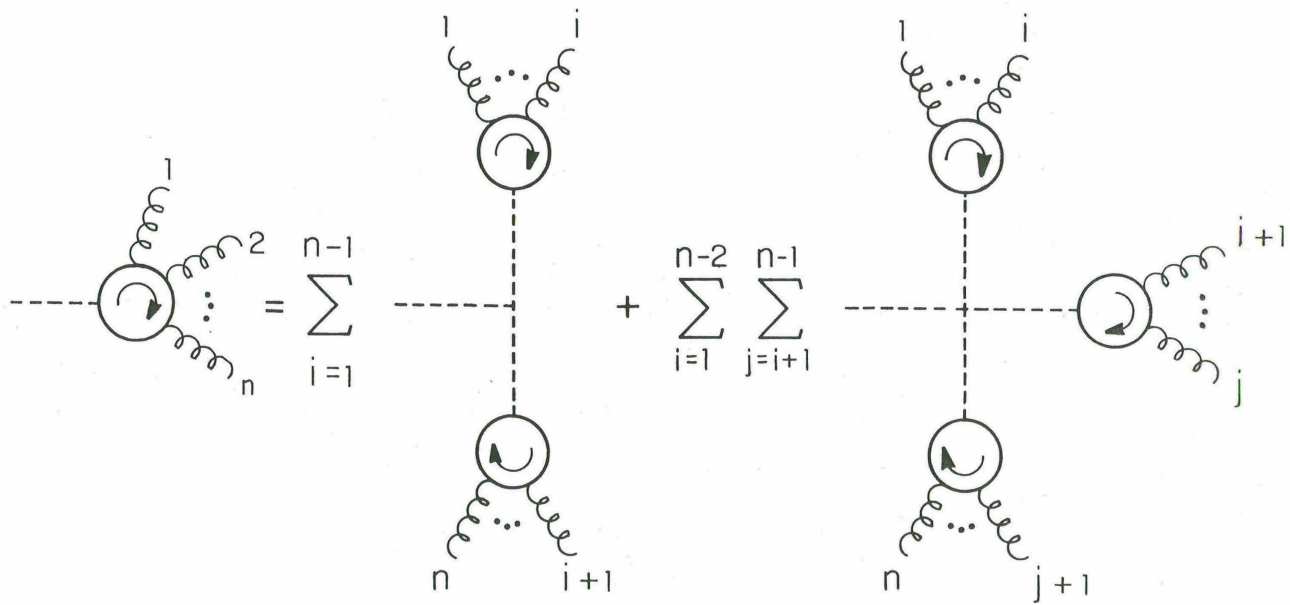}
\caption{A graphical representation for the Berends-Giele gluonic 
recursion relation.}
\label{rsqfigtwo}         
\vspace{0.2in}
\end{figure}

\begin{figure}[h]
\centering
\includegraphics[width=18.cm]{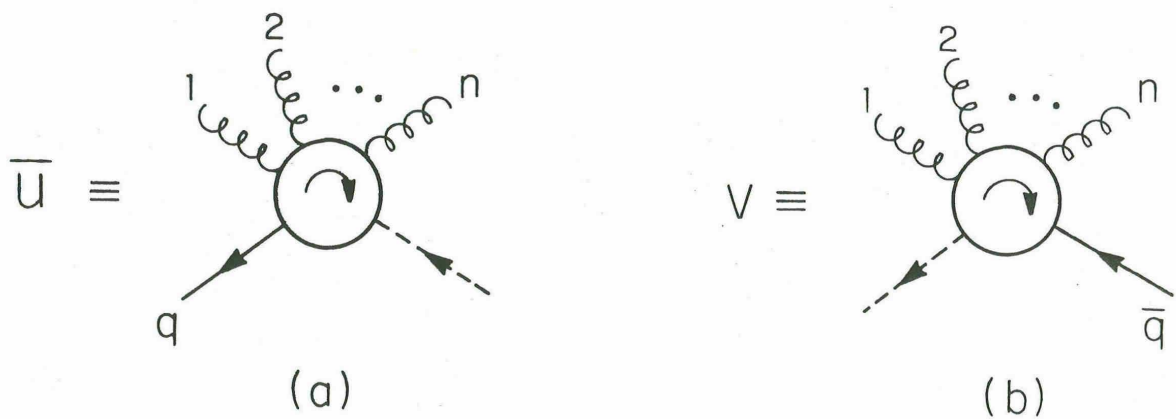}
\caption{The quark, (a), and antiquark, (b),
color ordered spinorial currents.}
\label{rsqfigthree}            
\vspace{0.2in}
\end{figure}

\begin{figure}[h]
\centering
\includegraphics[width=18.cm]{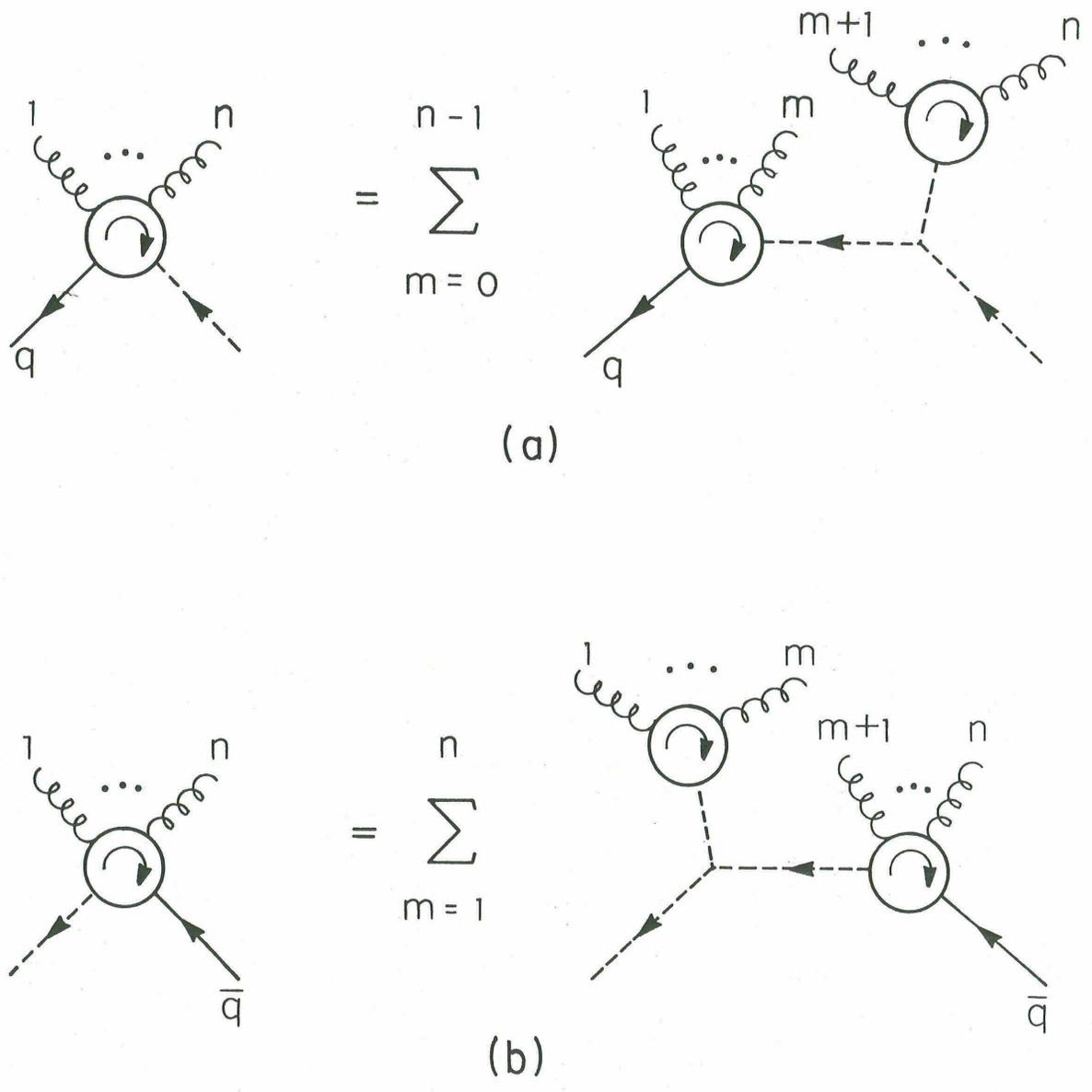}
\caption{Graphical representations of the Berends-Giele quark, (a), and 
antiquark, (b), recursion relations.}
\label{rsqfigfour}         
\vspace{0.2in}   
\end{figure}

\begin{figure}[h]
\centering
\includegraphics[width=18.cm]{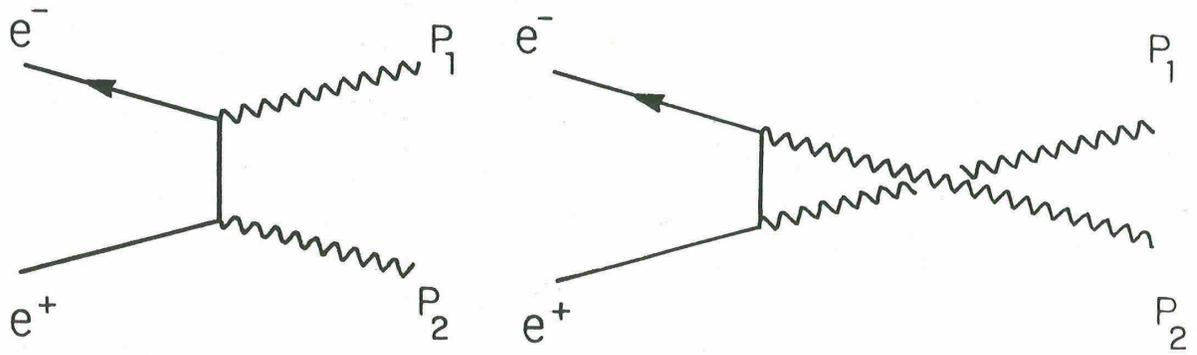}
\caption{Diagrams for $e^+e^- \to \gamma\gamma$ annihilation.}
\label{appfig}         
\vspace{0.2in}
\end{figure}
        
\end{document}